\documentclass{jfp1}

\usepackage{t1enc}
\usepackage{latexsym}
\usepackage{theorem}
\usepackage{amssymb}
\usepackage{hyperref}
\renewcommand\url[1]{\href{#1}{#1}}
\usepackage{prooftree}
\usepackage{mathdots} 
\usepackage{enumerate}
\usepackage{pgf}
\usepackage{tikz}
\usetikzlibrary{arrows}

\newcommand\hide[1]{}

\newcommand\hsp[1][3ex]{\hspace*{#1}}
\newcommand\vsp[1][3mm]{\vspace*{#1}}
\newcommand\moins\setminus
\newcommand\vide\emptyset
\newcommand\eg{{\em e.g.} }
\newcommand\ie{{\em i.e.} }

\newcommand\lub{\mr{lub}}
\newcommand\dom{\mr{dom}}

\newcommand\FV{\mr{FV}}

\newcommand\Pos{\mr{Pos}}

\renewcommand\a\rightarrow
\newcommand\A\Rightarrow
\renewcommand\aa\leftrightarrow
\renewcommand\AA\Leftrightarrow
\newcommand\la\leftarrow
\newcommand\lA\Leftarrow
\newcommand\ad\downarrow
\newcommand\Ad\Downarrow
\renewcommand\au\uparrow 
\newcommand\Au\Uparrow
\renewcommand\to\mapsto

\newcommand\ab{\a_\b}

\newcommand\ar{\a_\cR}

\newcommand\I[1]{[\![#1]\!]}

\renewcommand\N[1]{\|#1\|_\infty}

\newcommand\ex\exists
\newcommand\all\forall
\newcommand\ou\vee
\newcommand\bigou\bigvee
\newcommand\biget\bigwedge
\newcommand\et\wedge
\newcommand\non\neg
\newcommand\st\star
\newcommand\B\Box
\renewcommand\th\vdash

\newcommand\sle\subseteq
\newcommand\sge\supseteq
\newcommand\slt\subset
\newcommand\sgt\supset
\newcommand\tle\unlhd
\newcommand\tge\unrhd
\newcommand\tlt\lhd
\newcommand\tgt\rhd
\newcommand\cle\preceq
\newcommand\cge\succeq
\newcommand\clt\prec
\newcommand\cgt\succ
\newcommand\qle\sqsubseteq
\newcommand\qge\sqsupseteq
\newcommand\qlt\sqsubset
\newcommand\qgt\sqsupset

\newcommand\lex{\mr{lex}}

\renewcommand\prod{\mr{prod}}

\newcommand\h[1]{{\widehat{#1}}}
\renewcommand\o[1]{{\overline{#1}}}

\newcommand\al\alpha
\renewcommand\b\beta
\newcommand\g\gamma
\newcommand\G\Gamma
\renewcommand\d\delta
\newcommand\D\Delta
\newcommand\ep\epsilon
\newcommand\vep\varepsilon
\newcommand\z\zeta
\renewcommand\t\theta
\newcommand\T\Theta
\newcommand\io\iota
\newcommand\kap\kappa
\renewcommand\l\lambda
\renewcommand\L\Lambda
\renewcommand\r\rho
\newcommand\s\sigma
\renewcommand\S\Sigma
\newcommand\up\upsilon
\newcommand\Up\Upsilon
\newcommand\vphi\varphi
\newcommand\w\omega
\newcommand\W\Omega

\newcommand\bs\boldsymbol
\newcommand\mi\mathit
\newcommand\mc\mathcal
\newcommand\mt\mathtt
\newcommand\mr\mathrm
\newcommand\mb\mathbb 
\newcommand\mg\mathbf
\newcommand\mk\mathfrak
\newcommand\ms\mathsf
\DeclareMathAlphabet\mz{OT1}{pzc}{m}{it} 

\newcommand\bB{\mb{B}}
\newcommand\bC{\mb{C}}
\newcommand\bD{\mb{D}}
\newcommand\bE{\mb{E}}
\newcommand\bF{\mb{F}}

\newcommand\bH{\mb{H}}
\newcommand\bI{\mb{I}}
\newcommand\bJ{\mb{J}}

\newcommand\bL{\mb{L}}

\newcommand\bN{\mb{N}}

\newcommand\bP{\mb{P}}
\newcommand\bQ{\mb{Q}}
\newcommand\bR{\mb{R}}
\newcommand\bS{\mb{S}}
\newcommand\bT{\mb{T}}

\newcommand\bV{\mb{V}}

\newcommand\bZ{\mb{Z}}

\newcommand\cD{\mc{D}}

\newcommand\cR{\mc{R}}
\newcommand\cS{\mc{S}}
\newcommand\cT{\mc{T}}
\newcommand\cU{\mc{U}}
\newcommand\cV{\mc{V}}

\newcommand\cX{\mc{X}}

\newcommand\tc{\mt{c}}
\newcommand\td{\mt{d}}

\newcommand\tf{\mt{f}}
\newcommand\tg{\mt{g}}

\newcommand\ts{\mt{s}}

\newcommand\tx{\mt{x}}

\newcommand\tA{\mt{A}}
\newcommand\tB{\mt{B}}
\newcommand\tC{\mt{C}}

\newcommand\tF{\mt{F}}

\newcommand\tH{\mt{H}}

\newcommand\tN{\mt{N}}

\newcommand\tV{\mt{V}}

\newcommand\ka{\mk{a}}
\newcommand\kb{\mk{b}}
\newcommand\kc{\mk{c}}
\newcommand\kd{\mk{d}}

\newcommand\kh{\mk{h}}

\newcommand\ra{\mr{a}}

\newcommand\rd{\mr{d}}

\newcommand\ri{\mr{i}}

\newcommand\rl{\mr{l}}

\newcommand\rp{\mr{p}}
\renewcommand\rq{\mr{q}}
\newcommand\rr{\mr{r}}
\newcommand\rs{\mr{s}}

\newcommand\rx{\mr{x}}

\newcommand\sa{\ms{a}}
\renewcommand\sb{\ms{b}}
\newcommand\msc{\ms{c}} 

\renewcommand\sf{\ms{f}}
\newcommand\sg{\ms{g}}

\renewcommand\sl{\ms{l}}

\let\latexss\ss
\renewcommand\ss{\ms{s}}
\renewcommand\st{\ms{t}}

\newcommand\sB{\ms{B}}
\newcommand\sC{\ms{C}}
\newcommand\sD{\ms{D}}
\newcommand\sE{\ms{E}}
\newcommand\sF{\ms{F}}

\newcommand\sL{\ms{L}}

\newcommand\sN{\ms{N}}
\newcommand\sO{\ms{O}}
\newcommand\sP{\ms{P}}

\newcommand\sT{\ms{T}}

\newcommand\va{{\vec{a}}}
\newcommand\vb{{\vec{b}}}
\newcommand\vc{{\vec{c}}}

\newcommand\ve{{\vec{e}}}

\newcommand\vl{{\vec{l}}}

\newcommand\vt{{\vec{t}}}
\newcommand\vu{{\vec{u}}}
\newcommand\vv{{\vec{v}}}
\newcommand\vw{{\vec{w}}}
\newcommand\vx{{\vec{x}}}
\newcommand\vy{{\vec{y}}}

\newcommand\vT{{\vec{T}}}
\newcommand\vU{{\vec{U}}}
\newcommand\vV{{\vec{V}}}

\newenvironment{rul}
  {$\begin{array}{rcl}}
  {\end{array}$}

\newenvironment{rew}[1][~~\a~~]
  {$\begin{array}{r@{#1}l}}
  {\end{array}$}
\newenvironment{rewc}[1][~~\a~~]
  {\begin{center}\begin{rew}[#1]}
  {\end{rew}\end{center}}

\hide{
\newcounter{dfnnum}
\newcounter{lemnum}
\newcounter{thmnum}
\newcounter{cornum}
\newcounter{remnum}
\newcounter{propnum}
\newcounter{hypnum}
\newcounter{conjnum}
\newcounter{asmnum}

{\theorembodyfont{\rmfamily} 
  \newtheorem{dfn}[dfnnum]{Definition}
  \newtheorem{lem}[lemnum]{Lemma}
  \newtheorem{thm}[thmnum]{Theorem}
  \newtheorem{cor}[cornum]{Corollary}
  \newtheorem{rem}[remnum]{Remark}
  \newtheorem{prop}[propnum]{Proposition}
  \newtheorem{conj}[conjnum]{Conjecture}
  \newtheorem{hyp}[hypnum]{Hypothesis}
  \newtheorem{asm}[asmnum]{Assumption}
  
}}

\newcommand\cqfd{\hfill$\blacksquare$}
\newenvironment{prf}{Proof. }{}

\newenvironment{ded}
   {\begin{center}\begin{prooftree}}
   {\end{prooftree}\end{center}}

\newenvironment{lstgeneric}[2]
  {\begin{list}{#1}{\topsep=.5mm\itemsep=.5mm\parsep=0mm%
    \itemindent=-3ex\labelsep=1ex\labelwidth=0ex #2}}
  {\end{list}}

\newcommand\nf[1]{{#1\!\!\ad}}
\newcommand\thfphi[1][\sf]{\th^{#1}_\vphi}
\newcommand\thf[1][\sf]{\th^{#1}}

\newcommand\tlea{\tle_{\ra}}
\newcommand\tlta{\tlt_{\ra}}

\newcommand\lea{\le_\tA}
\newcommand\gea{\ge_\tA}
\newcommand\lta{<_\tA}
\newcommand\lna{\lneq_\tA}

\newcommand\leai{\lea^\infty}

\newcommand\eqf{\simeq_\bF}
\newcommand\ltf{<_\bF}
\newcommand\lef{\le_\bF}

\newcommand\arity{\mr{ar}}

\newcommand\SN{\mr{SN}}
\newcommand\Mon{\mr{Mon}}
\newcommand\Var{\mr{Var}}
\newcommand\An{\mr{Annot}}
\newcommand\mgs{\mr{mgs}}

\newcommand\Sub{\mr{Sub}}
\newcommand\Sol{\mr{Sol}}

\newcommand\ext{\mr{ext}}

\newcommand\range{\mr{range}}

\newcommand\rec{\ms{rec}}
\newcommand\last{\ms{last}}
\newcommand\lesseq{\ms{le}}
\newcommand\si{\ms{if}}
\newcommand\newsi{\ms{newif}}
\newcommand\fst{\ms{fst}}
\newcommand\snd{\ms{snd}}
\newcommand\znat{\ms{0}}
\newcommand\zord{\ms{zero}}
\newcommand\snat{\ms{s}}
\newcommand\sord{\ms{succ}}
\renewcommand\lim{\ms{lim}}
\newcommand\pivot{\ms{pivot}}
\newcommand\qs{\ms{qs}}
\newcommand\qsort{\ms{qsort}}
\newcommand\pair{\ms{pair}}
\newcommand\nil{\ms{nil}}
\newcommand\cons{\ms{cons}}
\newcommand\true{\ms{true}}
\newcommand\false{\ms{false}}
\newcommand\nm{\ms{nm}}
\newcommand\at{\ms{at}}
\newcommand\rev{\ms{rev}}
\newcommand\revremlast{\ms{revremlast}}
\newcommand\leaf{\ms{leaf}}
\newcommand\node{\ms{node}}
\newcommand\inj{\ms{inj}}
\newcommand\sub{\ms{sub}}
\renewcommand\div{\ms{div}}
\newcommand\map{\ms{map}}
\newcommand\filter{\ms{filter}}
\renewcommand\empty{\ms{empty}}
\newcommand\add{\ms{add}}
\newcommand\cond{\ms{cond}}
\newcommand\app{\ms{app}}
\newcommand\lam{\ms{lam}}

\newcommand\cfrac[2]{\begin{array}{c}#1\\\hline #2\vsp[5mm]\end{array}}
\newcommand\qed\cqfd

\newcounter{dfnnum}
\newcounter{lemnum}
\newcounter{thmnum}
\newcounter{cornum}

\newcounter{propnum}

\newcounter{exanum}
{\theorembodyfont{\rmfamily} 
  \newtheorem{dfn}[dfnnum]{Definition}
  \newtheorem{lem}[lemnum]{Lemma}
  \newtheorem{thm}[thmnum]{Theorem}
  \newtheorem{cor}[cornum]{Corollary}
  
  \newtheorem{prop}[propnum]{Proposition}

  \newtheorem{exa}[exanum]{Example}
}

\begin{document}

\title{Size-based termination of higher-order rewriting}

\author[F. Blanqui]{Fr\'ed\'eric Blanqui\\INRIA\\[3mm]
  ENS / Universit\'e Paris-Saclay\\
  LSV, 61 avenue du Pr\'esident Wilson, 94235 Cachan Cedex, France}


\maketitle

\begin{abstract}
  We provide a general and modular criterion for the termination of
  simply-typed $\l$-calculus extended with function symbols defined by
  user-defined rewrite rules. Following a work of Hughes, Pareto and
  Sabry for functions defined with a fixpoint operator and
  pattern-matching, several criteria use typing rules for bounding the
  height of arguments in function calls. In this paper, we extend this
  approach to rewriting-based function definitions and more general
  user-defined notions of size.
\end{abstract}

\section{Introduction}
\label{sec-intro}

In this paper, we are interested in the termination of Church's
simply-typed $\l$-calculus \cite{church40jsl} extended with function
symbols defined by user-defined rewrite rules
\cite{dershowitz90chapter,terese03book} like the ones of Figure
\ref{fig-div}. Our results could be used to check the termination of
typed functional programs (\eg in OCaml \cite{ocaml} or Haskell
\cite{haskell}), rewriting-based programs (\eg in Maude \cite{maude}),
or function definitions in proof assistants (\eg Coq \cite{coq}, Agda
\cite{agda}, Dedukti \cite{dedukti}). By termination, we mean the
strong normalization property, that is, the absence of infinite
rewrite sequences $t_0\a t_1\a\ldots$ The mere existence of a normal
form is a weaker property called weak normalization. Termination is an
important property in program verification.

The rewrite system of Figure \ref{fig-div} defines the substraction
and division functions on the sort $\sN$ of natural numbers in unary
notation, \ie with the constructors $\znat:\sN$ for zero and
$\snat:\sN\A\sN$ for the successor function. A way to prove the
termination of this system is to show that, in two successive
functions calls, arguments are strictly decreasing wrt some
well-founded order. A natural order, based on the inductive nature of
$\sN$, is to compare the height of terms. More precisely, let the size
of a terminating term $t$ of sort $\sN$ be the number of $\snat$
symbols at the top of the normal form of $t$ (this rewrite system is
weakly orthogonal and thus confluent \cite{oostrom94phd}). While the
termination of $\sub$ (\ie the absence of infinite reductions starting
from a term of the form $\sub~t~u$ with $t$ and $u$ in normal form) is
not very difficult to establish (the size of the first argument is
strictly decreasing in recursive calls), proving the termination of
$\div$ requires the observation that $\sub$ is not size-increasing,
that is, the size of $(\sub~t~u)$ is less than or equal to the size of
$t$.

\begin{figure}
  \figrule
  \caption{Rewrite system defining substraction and division on natural numbers\label{fig-div}}
  \normalsize
\begin{rewc}
\sub~x~\znat & x\\
\sub~\znat~y & \znat\\
\sub~(\ss~x)~(\ss~y) & \sub~x~y\\[3mm]
\div~\znat~(\ss~y) & \znat\\
\div~(\ss~x)~(\ss~y) & \ss~(\div~(\sub~x~y)~(\ss~y))\\
\end{rewc}
\figrule
\end{figure}

The idea of sized types, introduced by Hughes, Pareto and Sabry in
\cite{hughes96popl} for fixpoint-based function definitions, is to
consider an abstract interpretation of this notion of size into an
algebra of symbolic size expressions, and turn the usual typing rules
of simply-typed $\l$-calculus into deduction rules on the size of
terms. This allows one to automatically deduce some information on the
size of terms, and thus prove termination by checking that, for
instance, the size of some given argument decreases in every recursive
call. Hence, termination is reduced to checking typing and abstract
size decreasingness.

In our example, this amounts to saying: the 2nd rule of $\div$ does not
jeopardize termination since, assuming that $x$ is instantiated by a
term $t$ of abstract size $\al$, and $y$ is instantiated by a term $u$
of abstract size $\b$, then $\div~(\ss~t)~(\ss~u)$ terminates because
its first argument is of size $\al+1$ while, in the recursive call
$\div~(\sub~t~u)~(\ss~u)$, the first argument has a size smaller than
or equal to $\al$.

The goal of this work is to automate this kind of inductive reasoning,
and check the information given by the user (here, the fact that
$\sub$ is not size-increasing). However, when considering type
constructors taking functions as arguments (\eg Sellink's model of
$\mu$CRL \cite{sellink93sosl}, Howard's constructive ordinals in
Example \ref{ex-rec}), the size of a term is generally not a finite
natural number but a transfinite ordinal number. However, abstract
size expressions can also handle transfinite sizes.

Before explaining our contributions and detailing the outline of the
paper, we give hereafter a short survey on the use of ordinals for
proving termination since this is at the heart of our work though, in
the end, we provide an ordinal-free termination criterion.

\subsection{Ordinal-based termination}

A natural (and trivially complete) method for proving the termination
of a relation $\a$ consists in considering a well-founded domain
$(\bD,<_\bD)$, \eg some ordinal $(\kh,<_\kh)$, assigning a ``size''
$\|t\|\in\bD$ to every term $t$, and checking that every rewrite step
(including $\b$-reduction) makes the ``size'' strictly decrease:
$\|t\|>_\bD\|u\|$ whenever $t\a u$.

In theory, it is enough to take $\bD=\w$ (the first infinite ordinal)
when the rewrite relation is finitely branching. However, after
G\"odel's incompleteness theorem \cite{godel31mmp}, defining $\|~\|$
and proving that $\|t\|>_\bD\|u\|$ whenever $t\a u$, may require the
use of much bigger ordinals. For instance, the termination of
cut-elimination in Peano arithmetic (PA) requires induction up to the
ordinal $\vep_0=\w^{\w^{\iddots}}$ but PA cannot prove the
well-foundedness of $\vep_0$ itself \cite{gentzen35ma}. Yet, there is
a function $\|~\|$ from the terms of G\"odel's system T
\cite{godel58dialectica} (which extends PA) to $\w$ such that
$\|t\|>_\bD\|u\|$ whenever $t\a u$ \cite{weiermann98jsl}.

An equivalent approach is finding a well-founded relation containing
$\a$. For instance, Dershowitz's recursive path ordering (RPO)
\cite{dershowitz79focs,dershowitz82tcs} or its extension to the
higher-order case by Jouannaud and Rubio
\cite{jouannaud99lics,jouannaud07jacm,blanqui15lmcs}. But, in this
paper, we will focus on the explicit use of size functions. For a
connection between RPO and ordinals, see for instance
\cite{dershowitz88lics}.

Early examples of this approach are given by Ackermann's proof of
termination of second-order primitive recursive arithmetic functions
using $\kh=\w^{\w^\w}$ \cite{ackermann25ma}, Gentzen's proof of
termination of cut elimination in Peano arithmetic using $\kh=\vep_0$
\cite{gentzen35ma,howard68ipt,wilken12lmcs}, Turing's proof of weak
normalization of Church's simply-typed $\l$-calculus
\cite{turing42unpub}, and Howard's proof of termination of his system
V (an extension of G\"odel's system T with an inductive type for
representing ordinals) using Bachmann's ordinal
\cite{howard72jsl}. This approach developed into a whole area of
research for measuring the logical strength of axiomatic theories,
involving ever growing ordinals, that can hardly be automated. See for
instance \cite{rathjen06icm} for some recent survey. Instead, Monin
and Simonot developed an algorithm for trying to find size assignments
in $\kh=\w^\w$ \cite{monin01tcs}.

But, up to now, there has been no ordinal analysis for powerful
theories like second-order arithmetic: the termination of cut
elimination in such theories is based on another approach introduced
by Girard \cite{girard72phd,girard88book}, which consists in
interpreting types by so-called computability predicates and typing by
the membership relation.

In the first-order case, \ie when there is no rule with abstraction or
applied variables, size-decreasingness can be slightly relaxed by
conducting a finer analysis of the possible sequences of function
calls. This led to the notions of dependency pair in the theory of
first-order rewrite systems
\cite{arts96caap,arts00tcs,hirokawa05ic,giesl06jar}, and size-change
principle for first-order functional programs \cite{lee01popl}. These
two notions are thoroughly compared in \cite{thiemann05aaecc}. In both
cases, it is sufficient to define a measure on the class of terms
which are arguments of a function call only. Various extensions to the
higher-order case have been developed
\cite{sakai01ieice,wahlstedt07phd,jones08lmcs,kusakari09ieice,kop11rta},
but no general unifying theory yet.

The present paper is not concerned with this problem but with defining
a practical notion of size for simply-typed $\l$-terms inhabiting
inductively defined types.

Note by the way that the derivational complexity of a rewrite system,
\ie the function mapping every term $t$ to the maximum number of
successive rewrite steps one can do from $t$ \cite{hofbauer89rta},
does not seem to be related, at least in a simple way, to the ordinal
necessary to prove its termination: there are rewrite systems whose
termination can be proved by induction up to $\w$ only and yet have
huge derivational complexities \cite{moser14jlc}, unless perhaps one
bounds the growth rate of the size of terms (measured here as the
number of symbols) \cite{schmitz14rp}. The notion of runtime
complexity, \ie the function mapping every $n\in\bN$ to the maximum
number of successive rewrite steps one can do from a term whose
subterms are in normal form and whose size is smaller than $n$, seems
to provide a better (Turing related) complexity model
\cite{avanzini10rta}.

\subsection{Model-based termination}

In \cite{manna70hicss}, Manna and Ness proposed to interpret every
term whose free variables are $x_1,\ldots,x_n$ by a function from
$\bE^n$ to $\bE$, where $(\bE,<_\bE)$ is a well-founded domain. That
is, $\bD$ is the set of all the functions from some power of $\bE$ to
$\bE$ and $<_\bD$ is the pointwise extension of $<_\bE$, \ie
$f:\bE^n\a\bE<_\bD g:\bE^n\a\bE$ if, for all $x_1,\ldots,x_n\in\bE$,
$f(x_1,\ldots,x_n)<_\bE g(x_1,\ldots,x_n)$.

In the first-order case, this can be done in a structured way by
interpreting every function symbol $\sf$ of arity $n$ by a function
$\sf_\bE:\bE^n\a\bE$ and every term by composing the interpretations
of its symbols, \eg $\|\sf\,(\sg\,x)\|$ is the function mapping $x$ to
$\sf_\bE(\sg_\bE(x))$. If moreover these interpretation functions are
monotone in each argument, then checking that rewriting is
size-decreasing can be reduced to checking that every rule is
size-decreasing.

A natural domain for $(\bE,<_\bE)$ is of course $(\bN,<_\bN)$. In this
case, both monotony and size-decreasingness can be reduced to absolute
positivity. Indeed,\\\hsp[1cm]$f(x_1,\ldots,x_p)>g(x_1,\ldots,x_q)$ is
equivalent to $f(x_1,\ldots,x_p)-g(x_1,\ldots,x_q)-1\ge 0$\\and
monotony is equivalent to checking that, for all $i$,
$f(\ldots,x_i+1,\ldots)-f(\ldots,x_i,\ldots)-1\ge 0$. By restricting
the class of functions, \eg to polynomials of bounded degree, one can
develop heuristics for trying to automatically find monotone
polynomial interpretation functions making rules size-decrease
\cite{cherifa87scp,lucas05ita,contejean05jar,fuhs07sat}. Unfortunately,
polynomial absolute positivity is undecidable on $\bN$ since it is
equivalent to the solvability of Diophantine equations (Proposition
6.2.11 in \cite{terese03book}), which is undecidable
\cite{matiyasevich70,matiyasevich93book}. Yet, these tools get useful
results in practice by restricting degrees and coefficients to small
values, \eg $2$.

A similar approach can be developed for dense sets like $\bQ^+$ or
$\bR^+$ by ordering them with the (not well-founded!) usual orderings
on $\bQ^+$ and $\bR^+$ if one assumes moreover that the functions
$\sf_\bE$ are strictly extensive (\ie $\sf_\bE(x_1,\ldots,x_n)>x_i$
for all $i$) \cite{dershowitz79ipl}, or with the well-founded relation
$<_\d$ where, for some fixed $\d>0$, $x<_\d y$ if $x+\d\le y$
\cite{lucas05ita,fuhs08aisc}. In the case of $\bR^+$, polynomial
absolute positivity is decidable but of exponential complexity
\cite{tarski48tr,collins75atfl}. Useful heuristics have however been
studied \cite{hong98jar}.

These approaches have also been successfully extended to linear
functions on domains like $\bE=\bB^n$ (vectors of dimension $n$) or
$\bE=\bB^{n\times n}$ (square matrices of dimension $n$)
\cite{endrullis08jar,courtieu10sofsem}, where $\bB$ is a well-founded
domain.

Instead of polynomial functions, Cicho\'n considered the class of
Hardy functions \cite{hardy04qjm} indexed by ordinals smaller than
$\vep_0$ \cite{cichon96caap}. The properties of Hardy functions
(composition is addition of indices, etc.) can be used to reduce the
search of appropriate Hardy functions to solving inequalities on
ordinals.

Manna and Ness' approach has also been extended to the higher-order
case.

In \cite{gandy80chapter}, Gandy remarks that terms of the $\l
I$-calculus (\ie when, in every abstraction $\l xt$, $x$ freely occurs
at least once in $t$) can be interpreted in the set of hereditary
strictly monotone functions on some well-founded set $(\bE,<_\bE)$,
that is, a closed term of base type $\sB$ is interpreted in the set
$\I\sB=\bE$, a closed term of type $T\A U$ is interpreted by a
monotone function from $\I{T}$ to $\I{U}$, and $f:\I{T\A U}<_{\I{T\A
    U}}g:\I{T\A U}$ if, for all $x\in\I{T}$, $f(x)<_{\I{U}}g(x)$ (note
that, in contrast with the first-order case, $x$ itself may be a
function). Then, by taking $\bE=\bN$ and extending the $\l$-calculus
with constants $\ms{0}:o$, $\ss:o\A o$ and $\ms{+}:o\A o\A o$ for each
base type $o$, he defines a size function that makes $\b$-reduction
size-decrease and provide an upper bound to the number of rewrite
steps. An exact upper bound was later computed by de Vrijer in
\cite{devrijer87im}.

Gandy's approach was later extended by van de Pol
\cite{vandepol93hoa,vandepol96phd} and Kahrs \cite{kahrs95rta} to
arbitrary higher-order rewriting {\em \`a la} Nipkow
\cite{nipkow91lics,mayr98tcs}, that is, to rewriting on terms in
$\b$-normal $\eta$-long form with higher-order pattern-matching
\cite{miller89elp}. But this approach has been implemented only
recently \cite{fuhs12rta}.

Interestingly, van de Pol also showed that, in the simply-typed
$\l$-calculus, Gandy's approach can be seen as a refinement of
Girard's proof of termination based on computability predicates
\cite{vandepol95hoa,vandepol96phd}.

Finally, a general categorical framework has been developed by Hamana
\cite{hamana06hosc}, that is complete wrt. the termination of binding
term rewrite systems, a formalism based on Fiore, Plotkin and Turi's
binding algebra \cite{fiore99lics} and close to a typed version of
Klop's combinatory reduction systems \cite{klop93tcs}.

\medskip

To the best of our knowledge, nobody seems to have studied the
relations between Howard's approach based on ordinals
\cite{howard68ipt,wilken12lmcs} and Gandy's approach based on
interpretations \cite{gandy80chapter,devrijer87im,vandepol96phd}.

Note also that the existence of a quasi-interpretation, \ie
$\|t\|\ge_\bD\|u\|$ whenever $t\a u$, not only may give useful
information on the complexity of a rewrite system \cite{bonfante11tcs}
but, sometimes, may also simplify the search of a termination
proof. Indeed, Zantema proved in \cite{zantema95fi} that the
termination of a first-order rewrite system $\cR$ is equivalent to the
termination of ${\mr{lab}(\cR)}\cup{>_\bD}$, where $\mr{lab}(\cR)$ are
all the variants of $\cR$ obtained by annotating function symbols by
the interpretation of their arguments, a transformation called
semantic labeling. Although usually infinite, the obtained labeled
system may be simpler to prove terminating, and some heuristics have
been developed to use this technique in automated termination tools
\cite{middeldorp96cade,koprowski06ijcar,sternagel08rta}. This result
was later extended to the higher-order case by Hamana
\cite{hamana07ppdp}.

\subsection{Termination based on typing with size annotations}

Finally, there is another approach based on the semantics of inductive
types, that has been developed for functions defined with a fixpoint
combinator and pattern-matching \cite{burstall80lfp}.

The semantics of an inductive type $\sB$, $\I\sB$, is usually defined,
following Hessenberg's theorem \cite{hessenberg09crelle}, Knaster and
Tarski's theorem \cite{knaster28aspm} or Tarski's theorem
\cite{tarski55pjm}, as the smallest fixpoint of a monotone function
$\bH^\sB$ on some complete lattice. Moreover, following Kuratowski
\cite{kuratowski22fm,cousot79pjm}, such a fixpoint can be reached by
transfinite iteration of $\bH^\sB$ from the smallest element of the
lattice $\bot$. Hence, every element $t\in\I\sB$ can be given as size
the smallest ordinal $\ka$ such that $t\in \cS^\sB_\ka$, where
$\cS^\sB_\ka$ is the set obtained after $\ka$ transfinite iterations
of $\bH^\sB$ from $\bot$. In particular, terms of a first-order data
type like the type of Peano integers, lists, binary trees, \ldots
always have a size smaller than $\w$.

Mendler used this notion of size to prove the termination of an
extension of G\"odel's system T \cite{godel58dialectica} and Howard's
system V \cite{howard72jsl} to functionals defined by recursion on
higher-order inductive types, \ie types with constructors taking
functions as arguments \cite{mendler87phd,mendler91apal}, in which
case the size of a term can be bigger than $\w$.

In \cite{hughes96popl,pareto00phd}, Hughes, Pareto and Sabry proposed
to internalize this notion of size by extending the type system with,
for each data type $\sB$, new type constants $\sB_0$, $\sB_1$, \ldots
$\sB_\infty=\sB$ for typing the terms of type $\sB$ of size smaller
than or equal to $0$, $1$, \ldots, $\infty$ respectively, and the
subtyping relation induced by the fact that a term of size at most $a$
is also of size at most $b$ whenever $a\le_\bN b$ or $b=\infty$. More
generally, to provide some information on how a function behaves
wrt. sizes, they consider as size annotations not only $0,1,\ldots$
but any first-order term built from the function symbols $\mt{0}$ for
zero, $\ts$ for successor and $\mt{+}$ for addition, and arbitrary
size variables $\al,\b,\ldots$, that is the language of Presburger
arithmetic \cite{presburger29}. So, for instance, the usual list
constructor $\cons$ gets the type $\sN\A\sL_\al\A\sL_{\ts\al}$, and
the usual $\map$ function on lists can be typed by
$(\sN\A\sN)\A\sL_\al\A\sL_\al$, where $\al$ is a free size variable
that can be instantiated by any size expression in a way similar to
type instantiation in ML-like programming languages
\cite{milner78jcss}.

Hughes, Pareto and Sabry do not actually prove the termination of
their calculus but provide a domain-theoretic model
\cite{scott72chapter}. However, following Plotkin \cite{plotkin77tcs},
a closed term of first-order data type terminates iff its
interpretation is not $\bot$. The first termination proof for
arbitrary terms seems to have been given by Amadio and Coupet-Grimal
in \cite{amadio97tr,amadio98fossacs}, who independently developed a
system similar to the one of Hughes, Pareto and Sabry, inspired by
Gim\'enez's work on the use of typing annotations for termination and
productivity \cite{gimenez96phd}. Gim\'enez himself later proposed a
similar system in \cite{gimenez98icalp} but provided no termination
proof. Note that Plotkin's result was later extended to higher-order
types and rewriting-based function definitions by Berger, and Coquand
and Spiwack in \cite{berger05cie,coquand07lmcs,berger08apal}.

Size annotations are an abstraction of the semantic notion of size
that one can use to prove properties on the actual size of terms like
termination (size-decreasingness) or the fact that a function is not
size-increasing (\eg $\map$), which can in turn be used in a
termination proof \cite{walther88cade,giesl97jar}. Following
\cite{cousot97popl}, it could certainly be described as an actual
abstract interpretation.

Hence, termination can be reduced to checking that a term has some
given type in the system with size-annotated type constants and
subtyping induced by the ordering on size annotations, the usual
typing rules being indeed valid deduction rules wrt. the size of terms
(\eg if $t:\sN_a\A\sN_b$ and $u:\sN_a$, then $tu:\sN_b$).

But, in such a system, a term can have infinitely many different types
because of size instantiation or because of subtyping. As already
mentioned, size instantiation is similar to type instantiation in
Hindley-Milner's type system \cite{hindley69tams,milner78jcss} where
the set of types of a term has a smallest element wrt. the
instantiation ordering if it is not empty \cite{huet76hdr}. In this
case, there is a complete type-checking algorithm for $(t,T)$ which
consists of checking that $T$ is an instance of the smallest type of
$t$ \cite{hindley69tams}. Unfortunately, with subtyping, there is no
smallest type wrt. the instantiation ordering (\eg $\l xx$ has type
$\al\A\al$ for all $\al$, and type $\sB\A\sC$ if $\sB<\sC$, but
$\sB\A\sC$ is not an instance of $\al\A\al$), or subtyping composed
with instantiation (\eg $\l f\l xf(fx))$ has type
$(\al\A\al)\A(\al\A\al)$ for all $\al$, and type
$(\sB\A\sC)\A(\sB\A\sC)$ if $\sB<\sC$, but no instance of
$(\al\A\al)\A(\al\A\al)$ is a subtype of $(\sB\A\sC)\A(\sB\A\sC)$)
\cite{fuh90tcs}. To recover a notion of smallest type and
completeness, all the works we know on type inference with subtyping
extend the notion of type to include subtyping constraints.

We will not follow this approach though. One reason is that we
consider Church-style $\l$-terms (\ie with type-annotated
abstractions) instead of Curry-style $\l$-terms and, in this case, as
we will prove it, there is a smallest type wrt. to subtyping composed
with instantiation when size expressions are only built from
variables, the successor symbol and an arbitrary number of constants
(the ``successor'' size algebra). Note moreover that, although
structural (function types and base types are incomparable), subtyping
is not well-founded in this case since, for instance,
$\sN_\al\A\sN>\sN_{\ts\al}\A\sN>\ldots$ However, if we disregard how
size annotations are related to the semantics of inductive types, our
work has important connections with more general extensions of
Hindley-Milner's type system with subtypes
\cite{mitchell84popl,fuh90tcs,pottier01ic}, indexed types
\cite{zenger97tcs}, DML(C) \cite{xi02hosc}, HM(X)
\cite{sulzmann00phd}, or generalized algebraic data types (GADTs)
\cite{xi03popl,cheney03tr}, which are all a restricted form of
dependent types \cite{debruijn68sad,martinlof73lc}.

Hughes, Pareto and Sabry's approach was later extended to higher-order
data types \cite{barthe04mscs}, polymorphic types
\cite{abel04ita,barthe05tlca,abel06phd,abel08lmcs}, rewriting-based
function definitions in the calculus of constructions
\cite{blanqui04rta,blanqui05csl}, conditional rewriting
\cite{blanqui06lpar-sbt}, product types \cite{barthe08csl}, and
fixpoint-based function definitions in the calculus of constructions
\cite{barthe06lpar,gregoire10lpar,sacchini11phd}.

It should be noted that, in contrast with the ordinal-based approach,
not all terms are given a size, but only those of base type. Moreover,
although ordinals are used to define the size of terms, no ordinal is
actually used in the termination criterion since one considers an
abstraction of them. Indeed, when comparing two terms, one does not
need to actually know their size: it is enough to differentiate between
their size. Hence, transfinite computations can be reduced to finite
ones.

Finally, Roux and the author proved in \cite{blanqui09csl} that size
annotations provide a quasi-model, and thus can be used in a semantic
labeling. Terms whose type is annotated by $\infty$ (unknown size) are
interpreted by using a technique introduced by Hirokawa and Middeldorp
in \cite{hirokawa06rta}. Interestingly, semantic labeling allows one
to deal with function definitions using matching on defined symbols,
like in a rule for associativity (\eg $(x+y)+z\a x+(y+z)$), while
termination criteria based on types with size annotations are
restricted to matching on constructor symbols.

Current implementations of termination checkers based on typing with
size annotations include ATS \cite{xi03types,ats}, MiniAgda
\cite{abel10par,miniagda}, Agda \cite{agda}, cicminus
\cite{sacchini11phd,cicminus} or HOT \cite{hot}. Most of these tools
assume given the annotated types of function symbols (\eg to know
whether the size of a function is bounded by the size of one of its
arguments). Heuristics for inferring the annotations of function
symbols have been proposed in \cite{telford99bctcs,chin01hosc}. They
are both based on abstract interpretation techniques
\cite{cousot96csur}.

\subsection{Contributions}

\begin{enumerate}
\item The first contribution of the present paper is to give a rigorous and
detailed account, for the simply-typed $\l$-calculus, of the approach
and results sketched in \cite{blanqui04rta,blanqui05csl}, hence
providing the first complete account of the extension of Hughes,
Pareto and Sabry's approach to rewriting-based function definitions
\cite{dershowitz90chapter,terese03book}.

\item In all the works on size-annotated types, the size algebra is
fixed. In those considering first-order data types only, the size
algebra is usually the language of Presburger arithmetic, the
first-order theory of which is decidable
\cite{presburger29,fischer74sam}. In those considering higher-order
data types, the successor symbol $\ts$ is usually the only symbol
allowed, except in \cite{barthe08csl} which allows addition too. Yet,
there are various examples showing that, within a richer size algebra,
more functions can be proved terminating since types are more precise.

The second contribution of the present paper is to provide a
type-checking algorithm for a general formulation of Hughes, Pareto
and Sabry's calculus parametrized, for size annotations, by a
quasi-ordered first-order term algebra $(\tA,\lea)$ interpreted in
ordinals. In particular, we prove that this algorithm is complete
whenever size function symbols are monotone, the existential fragment
of $(\tA,\lea)$ is decidable and every satisfiable set of size
constraints admits a smallest solution.

\item In all the previous works, the notion of size is also fixed: the
  size of $t$ is the height of the set-theoretical tree representation
  of the normal form of $t$ (an abstraction being represented as an
  infinite set of trees).

The third contribution of the paper is to enable users to define their
own notion of size by annotating the types of constructors. These
annotations generate a stratification of the interpretation of
inductive types. We prove that one can build such a stratification in
the domain of Girard's computability predicates
\cite{girard72phd,girard88book} when annotations form monotone and
extensive functions.

\item The fourth contribution is the proof that, in the successor algebra,
the satisfiability of a finite set of constraints is decidable in
polynomial time, and every satisfiable finite set of constraints has a
smallest solution computable in polynomial time too.
\end{enumerate}

In contrast with \cite{blanqui04rta,blanqui05csl}, the present paper:
\begin{itemize}
\item includes a short survey on the use of ordinals in termination
  proofs;
\item develops a stratification-based notion of size for inhabitants
  of inductive types;
\item introduces the notion of constructor size function;
\item shows how to define a stratification from constructor size
  functions that are monotone and strictly extensive on recursive
  arguments;
\item proves the existence and polynomial complexity of the
  computation of a smallest solution for a solvable set of constraints
  in the successor algebra, using max-plus algebra techniques instead
  of pure linear algebra techniques.
\end{itemize}

\subsection{Organization of the paper}

In Section \ref{sec-terms}, we recall the definitions of types, terms
and rewriting, and the interpretation of types as computability
predicates.
In Section \ref{sec-size}, we introduce the notions of stratification,
size and constructor size functions, and prove properties on the size
of computable terms.
In Section \ref{sec-sn}, we present the termination criterion. The
main ingredient of the termination criterion is a type system with
subtyping, parametrized by a quasi-ordered first-order term algebra
for abstract size expressions. It also requires that annotations of
arguments are minimal in some sense.
In Section \ref{sec-min}, we provide a sufficient syntactic condition
for the minimality property to be satisfied when the size is defined
as the height.
In Section \ref{sec-ex}, we provide various examples of the expressive
power of our termination criterion.
In Section \ref{sec-dec}, we provide a complete algorithm for checking
subject-reduction and size-decreasingness under some general assumptions on
the size algebra.
In Section \ref{sec-sub}, we show how subtyping problems can be
reduced to ordering problems in the size algebra.
Finally, in Section \ref{sec-succ}, we prove that the simplest
possible algebra, the successor algebra, satisfies the required
conditions for the type-checking algorithm to be complete.

\section{Types, terms and computability}
\label{sec-terms}

In this section, we define the set of terms that we consider (Church's
simply-typed $\l$-calculus with constants \cite{church40jsl}), the
operational semantics (the combination of $\b$-reduction and
user-defined rewrite rules \cite{dershowitz90chapter,terese03book}),
and the notion of computability used to prove termination.

Given a set $E$, we denote by $E^*$ the set of words or sequences over
$E$ (\ie the free monoid containing $E$), the empty word by $\vep$,
the concatenation of words by juxtaposition, the length of a word $w$
by $|w|$. We also use $\ve$ to denote a (possibly empty) sequence
$e_1,\ldots,e_{|\ve|}$ of elements of $E$.

Given a partial function $f:A\a B$, $a\in A$ and $b\in B$, let
$[a:b,f]$ be the function mapping $a$ to $b$ and every
$x\in\dom(f)-\{a\}$ to $f(x)$.

We recall that, if $X$ is a bounded set of ordinals, \ie when there is
$b$ such that $x\le b$ for all $x\in X$, then the least upper bound of
$X$, written $\sup X$, exists. In particular, $\sup\vide=0$.

\subsection{Types}

Following Church, we assume given a non-empty countable set $\bS$ of
{\em sorts} $\sB$, $\sC$, $\ldots$ and define the set $\bT$ of
(simple) {\em types} as follows:

\begin{itemize}
\item sorts are types;
\item if $T$ and $U$ are types, then $T\A U$ is a type.
\end{itemize}

Implication associates to the right. So, $T\A U\A V$ is the same as
$T\A(U\A V)$. Moreover, $\vT\A U$ is the same as
$T_1\A T_2\A\ldots\A T_n\A U$ where $n=|\vT|$.

The {\em arity} of a type $T$, $\arity(T)$, is defined as follows:
$\arity(\sB)=0$ and $\arity(T\A U)=1+\arity(U)$.

\subsection{Terms}

Given disjoint countable sets $\bV$, $\bC$ and $\bF$, for variables,
constructors and function symbols respectively, we define the set
of {\em pre-terms} as follows:

\begin{itemize}
\item variables, constructors and function symbols are pre-terms;
\item if $x$ is a variable, $T$ a type and $u$ a pre-term, then $\l
  x^Tu$ is a pre-term;
\item if $t$ and $u$ are pre-terms, then $tu$ is a pre-term.
\end{itemize}

Application associates to the left. So, $tuv$ is the same as
$(tu)v$. Moreover, $t\vu$ is the same as
$(\ldots((tu_1)u_2)\ldots u_{n-1})u_n$ where $n=|\vu|$.

As usual, the set of {\em terms} $\bL$ is obtained by quotienting
pre-terms by $\al$-equivalence, \ie renaming of bound variables,
assuming that $\bV$ is infinite \cite{curry58book}.

As usual, positions in a tree (type or term) are denoted by words on
positive integers. Word concatenation is denoted by juxtaposition and
the empty word by $\vep$. Given a tree $t$ and a position $p$ in $t$,
let $t|_p$ be the subtree of $t$ at position $p$, and $\Pos(u,t)$ be
the set of positions $p$ in $t$ such that $t|_p=u$.

A {\em substitution} $\t$ is a map from variables to terms whose {\em
  domain} $\dom(\t)=\{x\in\bV\mid\t(x)\neq x\}$ is finite. In the
following, any finite map $\t$ from variables to terms is
implicitly extended into the substitution $\t\cup\{(x,x)\mid
x\notin\dom(\t)\}$. Let $\FV(\t)=\bigcup\{\FV(\t(x))\mid
x\in\dom(\t)\}$. The application of a substitution $\t$ to a term $t$
is written $t\t$. We have $x\t=\t(x)$, $(tu)\t=(t\t)(u\t)$ and $(\l
x^Tu)\t=\l x^T(u\t)$ if $x\notin\dom(\t)\cup\FV(\t)$, which can always
be achieved by $\al$-equivalence.

\subsection{Typing}

We assume given a map $\T$ assigning a type to every symbol
$s\in\bC\cup\bF$, and will sometimes write $s:T$ instead of
$(s,T)\in\T$ or $\T(s)=T$.

A {\em typing environment} is a finite map $\G$ from variables to
types. The usual deduction rules assigning a type to a term in a
typing environment are recalled in Figure \ref{fig-typ}. As mentioned
at the beginning of the section, $[x:U,\G]$ is the function mapping
$x$ to $U$ and every $y\in\dom(\G)-\{x\}$ to $\G(y)$.

Given a symbol $s$, let $\rr^s=\arity(\T(s))$ be the
maximum number of terms $s$ can be applied to. For all $s$, there are
types $T_1,\ldots,T_{\rr^s}$ and a sort $\sB$ such that
$\T(s)=T_1\A\ldots\A T_{\rr^s}\A\sB$.

Given $\sB\in\bS$, let
$\bC^\sB=\{(\msc,\vt,\vT)\mid\msc\in\bC,\msc:\vT\A\sB,|\vt|=|\vT|\}$
be the set of tuples $(\msc,\vt,\vT)$ such that $\msc$ is maximally
applied in $\msc\vt$ and $\vT$ are the types declared for the
arguments of $\msc$ (but $t_i$ does not need to be of type $T_i$).

\begin{figure}
  \figrule
\caption{Typing rules\label{fig-typ}}\vsp[3mm]
\normalsize
\begin{center}
$\cfrac{(s,T)\in\T\cup\G}{\G\th s:T}
\quad\cfrac{\G\th t:U\A V\quad\G\th u:U}{\G\th tu:V}
\quad\cfrac{[x:U,\G]\th v:V}{\G\th\l x^Uv:U\A V}$
\end{center}
\figrule
\end{figure}

\subsection{Rewriting}

Given a relation on terms $R$, let $R(t)=\{t'\in\bL\mid tRt'\}$ be the
set of immediate reducts of a term $t$, $R^*$ be the reflexive and
transitive closure of $R$, and $R^{-1}$ be its inverse ($xR^{-1}y$ if
$yRx$). $R$ is {\em finitely branching} if, for all $t$, $R(t)$ is
finite. It is {\em monotone} (or congruent, stable by context,
compatible with the structure of terms) if $tuRt'u$, $utRut'$ and $\l
x^UtR\l x^Ut'$ whenever $tRt'$. It is {\em stable} (by substitution)
if $t\t Rt'\t$ whenever $tRt'$. Given two relations $R$ and $S$, let
$RS$ (or $R\circ S$) be their composition ($tRSv$ if there is $u$ such
that $tRu$ and $uSv$). A relation $R$ is {\em locally confluent} if
$R^{-1}R\sle R^*(R^{-1})^*$, and {\em confluent} if $(R^{-1})^*R^*\sle
R^*(R^{-1})^*$.

The relation of {\em $\b$-rewriting $\ab$} is the smallest monotone
relation containing all the pairs $((\l x^Ut)u,t\{(x,u)\})$.

A {\em rewrite rule} is a pair of terms $(l,r)$, written $l\a r$, such
that there are $\sf\in\bF$, $\vl$, $\D$ and $T$ such that
$l=\sf\,\vl$, $\FV(r)\sle\FV(l)$, $\D\th l:T$ and, (SR) for all $\G$
and $U$, $\G\th r:U$ whenever $\G\th l:U$.

Given a set $\cR$ of rewrite rules, let $\ar$ denote the smallest
monotone and stable relation containing $\cR$. The condition (SR)
implies that $\ar$ preserves typing: if $\G\th t:U$ and $t\ar u$, then
$\G\th u:U$ (subject-reduction property). Note that it is satisfied
if, for instance, $l$ contains no abstraction and no subterm of the
form $x\,t$ \cite{barbanera97jfp}.

All over the paper, we assume given a set $\cR$ of rewrite rules and
let $\SN$ be the set of terms strongly normalizing wrt.:
$${\a}={{\ab}\cup{\ar}}$$

We will assume that $\a$ is finitely branching, which is in particular
the case if $\cR$ is finite.

Given $\sB$ and $t$, let
$\bC^\sB_{\a^*}(t)=\{(\msc,\vt,\vT)\in\bC^\sB\mid t\a^*\msc\,\vt\}$ be
the set of triples $(\msc,\vt,\vT)$ such that $t$ reduces to
$\msc\vt$, $\msc$ is maximally applied in $\msc\vt$, and $\vT$ are the
types of the arguments of $\msc$.

Given a relation $R$, let $\vx~R_\prod~\vy$ if $|\vx|=|\vy|$ and there
is $i$ such that $x_i\,R\,y_i$ and, for all $j\neq i$,
$x_j=y_j$. Given $n$ relations $R_1,\ldots,R_n$, let
$\vx~(R_1,\ldots,R_n)_\lex~\vy$ if $|\vx|\ge n$, $|\vy|\ge n$ and
there is $i$ such that $x_i\,R_i\,y_i$ and, for all $j<i$,
$x_j=y_j$. $R_\prod$ and $(R_1,\ldots,R_n)_\lex$ are well-founded
whenever $R,R_1,\ldots,R_n$ so are.

\subsection{Computability}
\label{sec-comp}

Following Tait \cite{tait67jsl}, Girard
\cite{girard72phd,girard88book}, Mendler \cite{mendler87phd}, Okada
\cite{okada89issac}, Breazu-Tannen and Gallier \cite{tannen89icalp},
and Jouannaud and Okada \cite{jouannaud91lics,blanqui02tcs}, \ldots
termination of a rewrite relation on simply-typed $\l$-terms can be
obtained by interpreting types by {\em computability predicates} and
checking that function symbols are computable, that is, map computable
terms to computable terms.

However, to handle matching on constructors taking functions as
arguments (or matching on function symbols), one needs to modify
Girard's definition of computability. In the following, we recall the
definition that we will use and some of its basic properties, and
refer the reader to \cite{blanqui16tcs,riba09tlca} for more details on
the theory of computability predicates with rewriting.

\begin{dfn}[Computability predicates]\label{def-cp}
  A term $t$ is {\em neutral} if it is of the form $x\vv$,
  $(\l{}xt)u\vv$ or $\sf\vt$ with $|\vt|\ge\sup\{|\vl|\mid\ex
  r,\sf\,\vl\a r\in\cR\}$\footnote{The supremum exists since, by
    assumption, for all $\sf\,\vl\a r\in\cR$, $\sf\,\vl$ is typable and
    thus $|\vl|\le\rr^\sf$.}. A {\em computability predicate} is a set
  of terms $\cS$ satisfying the following properties:
\begin{itemize}
\item $\cS\sle\SN$;
\item $\a\!(\cS)\sle\cS$;
\item if $t$ is neutral and $\a\!(t)\sle\cS$, then $t\in\cS$.
\end{itemize}
Let $\bP$ be the set of all the computability predicates.
An element of a computability predicate is said to be {\em computable}.
\end{dfn}

In our definition of neutral terms, not every redex is neutral as it
is the case in Girard's definition. However, the following key
property is preserved: application preserves neutrality, that is, if
$t$ is neutral, then $tu$ is neutral too. This definition also works
with polymorphic and dependent types. It only excludes infinite
rewrite systems where the number of arguments to which a function
symbol is applied is unbounded (at the top of rule left-hand sides
only, not in every term).

Computability predicates enjoy the following properties:
\begin{itemize}
\item the set $\bV$ of variables is included in every computability
  predicate;
\item given a computability predicate $\cS$, $(\l x^Uv)u\in\cS$ iff
  $v\{(x,u)\}\in\cS$ and $u\in\SN$;
\item $\bP$ is a complete lattice wrt. inclusion.
\end{itemize}

The greatest lower bound of a set $\bQ\sle\bP$ is $\bigcap\bQ$ if
$\bQ\neq\vide$, and $\SN$ (the greatest element of $\bP$)
otherwise. Note however that the lowest upper bound of $\bQ$, written
$\lub(\bQ)$, is not necessarily the union. For instance, with the
non-confluent system $\cR=\{\sf\a\sa,\sf\a\sb\}$, if $\bP(\cX)$
denotes the smallest computability predicate containing $\cX$, then
$\bP(\{\sa\})\cup\bP(\{\sb\})$ is not a computability predicate since
it does not contain $\sf$. There are a number of cases where the union
of two computability predicates is known to be a computability
predicate, but this is for a different notion of neutral term:
\begin{itemize}
\item In \cite{riba07fossacs,riba08cie}, Riba proves that his set of
  computability predicates is stable by union if $\cR$ is an
  orthogonal constructor rewrite system.
\item In \cite{werner94phd} (Lemma 4.14 p. 96), Werner proves that his
  set of computability predicates is stable by well-ordered union.
\end{itemize}

Luckily, Werner's proof does not depend on the definition of neutral
terms:

\begin{lem}
\label{lem-wo-union-cp}
If $\a$ is finitely branching and $\bQ$ is a non-empty set of computability
predicates well-ordered wrt. inclusion, then $\bigcup\bQ$ is a
computability predicate.
\end{lem}

\begin{prf}
\begin{itemize}
\item Let $t\in\bigcup\bQ$. Then, there is $\cS\in\bQ$ such that
  $t\in\cS$. Since $\cS\sle\SN$, we have $t\in\SN$.
\item Let $t\in\bigcup\bQ$ and $u$ such that $t\a u$. Then, there is
  $\cS\in\bQ$ such that $t\in\cS$. Since $\a(\cS)\sle\cS$, we have
  $u\in\cS$ and thus $u\in\bQ$.
\item Let $t$ be a neutral term such that $\a\!(t)\sle\bigcup\bQ$. If
  $\a\!(t)=\vide$, then $t$ belongs to every element of
  $\bQ$. Therefore, $t\in\bigcup\bQ$. Otherwise, since $\a$ is finitely
  branching, we have $\a\!(t)=\{t_1,\ldots,t_n\}$ with $n\ge 1$. For
  every $i\in\{1,\ldots,n\}$, there is $\cS_i\in\bQ$ such that
  $t_i\in\cS_i$. Since $\bQ$ is well-ordered wrt. inclusion, there is
  $k\in\{1,\ldots,n\}$ such that $\cS_k$ is the biggest element of
  $\{\cS_1,\ldots,\cS_n\}$ wrt. inclusion. Hence, $\a\!(t)\sle\cS_k$
  and $t\in\cS_k$. Therefore, $t\in\bigcup\bQ$.\qed
\end{itemize}
\end{prf}  

The interpretation of arrow types is defined as usual, in order to
ensure the termination of $\b$-reduction:

\begin{dfn}[Interpretation of arrow types]\label{def-cp-arr}
  A (partial) interpretation of sorts, that is, a (partial) function
  $\bI:\bS\a\wp(\bL)$ (powerset of $\bL$), is extended into a
  (partial) interpretation of types $\tilde\bI:\bT\a\wp(\bL)$ as follows:
\begin{itemize}
\item $\tilde\bI(\sB)=\bI(\sB)$;
\item $\tilde\bI(U\A V)=\tilde\bI(U)\tilde\A\tilde\bI(V)$
  where $\cU\tilde\A\cV=\{t\in\bL\mid\all u\in\cU,tu\in\cV\}$.
\end{itemize}
\end{dfn}

Note that $\tilde\bI(T)$ is defined whenever $\bI$ is defined on every
sort occurring in $T$, and $\tilde\bI(T)=\tilde\bJ(T)$ whenever $\bI$ and
$\bJ$ are defined and equal on every sort occurring in $T$.

Note also that $\cU\A\cV$ is a computability predicate whenever $\cU$
and $\cV$ so are. Hence, $\tilde\bI(T)$ is a computability predicate
whenever $\bI(\sB)$ so is for every sort $\sB$ occurring in $T$.

For interpreting sorts, one could take the computability predicate
$\SN$. But this interpretation does not allow one to prove the
computability of functions defined by induction on types with
constructors taking functions as arguments.

Moreover, a computable term may have non-computable subterms. Consider
for instance $\msc:(\sB\A\sC)\A\sB$, $\sf:\sB\A(\sB\A\sC)$,
$\cR=\{\sf(\msc~x)\a x\}$ and $t=\l x^\sB\sf xx$. Then, assuming that
$\bI(\sB)=\SN$, we have $(\msc~t)\in\bI(\sB)$, but
$t\notin\bI(\sB)\A\bI(\sC)$ since $t(\msc~t)\notin\SN$ because
$t(\msc~t)\ab\sf(\msc~t)(\msc~t)\ar t(\msc~t)$. It is however possible
to enforce that a direct subterm of type $T$ of a computable term of
sort $\sB$ is computable if $\sB$ occurs in $T$ at positive positions
only \cite{mendler87phd}:

\begin{dfn}[Positive and negative positions in a type]\label{def-pos}
  The subsets of {\em positive} ($s=+$) and {\em negative} ($s=-$)
  {\em positions} in a type $T$, $\Pos^s(T)$, are defined as follows:
\begin{itemize}
\item $\Pos^s(\sB)=\{\vep\mid s=+\}$,
\item $\Pos^s(U\A V)=\{1p\mid p\in\Pos^{-s}(U)\}\cup\{2p\mid p\in\Pos^s(V)\}$,
\end{itemize}
\noindent where $--=+$ and $-+=-$.
\end{dfn}

Note that the sets of positive and negative positions of a type are
disjoint. However, in a type, a sort can have both positive and
negative occurrences. For instance, $\Pos^+(\sB,\sB\A\sB)=\{2\}$ and
$\Pos^-(\sB,\sB\A\sB)=\{1\}$.

\begin{dfn}[Accessible arguments]
  \label{def-acc-arg}
  We assume given a well-founded {\em ordering} on sorts $<_\bS$. The
  $i$-th argument of a constructor $\msc:\vT\A\sB$ is:
  \begin{itemize}
  \item {\em recursive} if $\Pos(\sB,T_i)\neq\vide$;
  \item {\em accessible} if $T_i$ is {\em positive wrt.} $\sB$, that is:
    \begin{itemize}
    \item every sort occurring in $T_i$ is smaller than or equal to
      $\sB$:\\for all $\sC$, $\Pos(\sC,T)=\vide$ or $\sC\le_\bS\sB$,
      where $\le_\bS$ is the reflexive closure of $<_\bS$;
    \item $\sB$ occurs only positively in $T_i$: $\Pos(\sB,T_i)\sle\Pos^+(T_i)$.
    \end{itemize}
  \end{itemize}

In the following, we will assume wlog\footnote{Arguments can be permuted if
  needed.} that there are $0\le\rp^\msc\le\rq^\msc$ such that:
\begin{itemize}
\item the arguments $1$ to $\rp^\msc$ are accessible and recursive,
\item the arguments $\rp^\msc+1$ to $\rq^\msc$ are accessible and not recursive:
\end{itemize}
\noindent
$$\T(\msc)=\underbrace{T_1\A\ldots\A
  T_{\rp^\msc}}_{\mbox{rec. acc. args}}\A\underbrace{T_{\rp^\msc+1}\A\ldots\A
  T_{\rq^\msc}}_{\mbox{non-rec. acc. args}}\A\underbrace{T_{\rq^\msc+1}\A\ldots\A
  T_{\rr^\msc}}_{\mbox{non-acc. args}}\A\sB$$
\end{dfn}

For instance, for the sort $\sN$ of natural numbers with the
constructors $\znat:\sN$ and $\ss:\sN\A\sN$ (successor)
\cite{peano89book}, we can take $\rp^\znat=\rq^\znat=0$ and
$\rp^\ss=\rq^\ss=1$ since $\sN$ occurs only positively in
$\sN$. Similarly, for the sort $\sO$ of Howard's constructive ordinals
with the constructors $\zord:\sO$, $\sord:\sO\A\sO$ (successor) and
$\lim:{(\sN\A\sO)}\A\sO$ (limit) \cite{howard72jsl}, we can take
$\rp^\zord=\rq^\zord=0$, $\rp^\sord=\rq^\sord=1$ since $\sO$ occurs
only positively in $\sO$, and $\rp^\lim=\rq^\lim=1$ since $\sO$ occurs
only positively in $\sN\A\sO$ if one takes $\sN<_\bS\sO$. Now, for the
sort $\sL$ of lists of natural numbers with the constructors
$\nil:\sL$ and $\cons:\sL\A\sN\A\sL$, we can take $\rp^\cons=1$ and
$\rq^\cons=2$ if one takes $\sN<_\bS\sL$.

Non-accessible arguments are usually forbidden by requiring all the
arguments to be positive, or even strictly positive\footnote{The
  $i$-th argument of $\msc$ is {\em strictly positive} if
  $\Pos(\sB,T_i)=\vide$, or $T_i=\vU\A\sB$ and $\Pos(\sB,\vU)=\vide$.}
as it is the case in the Coq proof assistant
\cite{coquand88colog}. Here, we do not forbid non-positive arguments
and do not require arguments to be strictly positive. Hence, one can
have a sort $\sD$ with the constructors $\app:\sD\A\sD\A\sD$ and
$\lam:(\sD\A\sD)\A\sD$, for which we must have $\rp^\lam=\rq^\lam=0$
since the first argument of $\lam$ is not positive. However, the
termination conditions will enforce that, although one can use in a
rule left-hand side $(\lam~x)$ as a pattern, $x$ cannot be used in the
corresponding rule right-hand side: in a rule, constructors with
non-positive arguments can be pattern-matched in the left-hand side,
but only their positive arguments can be used by themselves in the
right-hand side.

For the sake of simplicity, we consider an ordering instead of a
quasi-ordering, although a quasi-ordering might a priori be necessary
for dealing with mutually defined inductive types (\eg the types of
trees and forests with the constructors $\empty:\sF$,
$\add:\sF\A\sT\A\sF$ and $\node:\sF\A\sT$). The results described in
this paper can however still be applied if one identifies mutually
defined inductive types, because a term typable with mutually defined
inductive types is a fortiori typable in the type system where they
are identified. This abstraction is correct but not necessarily
complete since more terms get typable when two types are identified
(\eg $\add~\empty~\empty$ is typable if $\sT=\sF$).

\smallskip

Since $<_\bS$ is well-founded, we can define an interpretation $\bI$
for every sort by well-founded induction on it as follows. Let $\sB$
be a sort and assume that $\bI$ is defined for every sort smaller than
$\sB$. Then, let $\bI(\sB)$ be the least fixpoint of the monotone
function $\bH^\sB$ on the complete lattice $\wp(\bL)$ such that:
$$\bH^\sB(\cX)=\{t\in\SN\mid\all(\msc,\vt,\vT)\in\bC^\sB_{\a^*}(t),
\all k\in\{1,\ldots,\rq^\msc\},t_k\in\widetilde{[\sB:\cX,\bI]}(T_k)\}.$$

\noindent
where $\widetilde{[\sB:\cX,\bI]}$ is introduced in Definition \ref{def-cp-arr}.

That such a least fixpoint exists follows from Knaster and Tarski's
fixpoint theorem \cite{knaster28aspm,tarski55pjm} and the following
fact:

\begin{prop}[\cite{blanqui05mscs}]\label{prop-mon}
Let $\sB$ be a sort, $\bI$ be an interpretation for every sort smaller
than $\sB$, and $T$ be a type positive wrt. $\sB$. Then,
$\widetilde{[\sB:\cX,\bI]}(T)$ is monotone wrt. $\cX$.
\end{prop}

Moreover, one can easily check that $\bH^\sB(\cX)$ is a computability
predicate whenever $\cX$ so is. Hence, for every type $T$, $\tilde\bI(T)$
is a computability predicate.

In the following, for the sake of simplicity, we will not mention
$\bI$ anymore and simply write $t\in T$ instead of $t\in\tilde\bI(T)$, and
$t\in[\sB:\cX]T$ instead of $t\in\widetilde{[\sB:\cX,\bI]}(T)$.

\section{Size of computable terms}
\label{sec-size}

In this section, we study a general way of attributing an ordinal size
to computable terms of base type by defining, for each sort, a
stratification of computable terms of this sort using a size function
for each constructor, and assuming that $\a$ is finitely branching.

By Hartogs' theorem \cite{hartogs15ma}, there is an ordinal the
elements of which cannot be injected into $\wp(\bL)$, where $\bL$ is
the set of terms (note that this theorem does not require the axiom of
choice). Let $\kh$ be the smallest such ordinal. Since $\bV$ is
countably infinite and $\bC$ and $\bF$ are countable, $\kh$ is the
successor cardinal of $|\wp(\bL)|=2^\w$ \cite{hrbacek99book}.

\subsection{Stratifications}

\begin{dfn}[Stratification of a sort]
  Given a family $(\cS_\ka)_{\ka<\kh}$ of computability predicates,
  let $\cS_\kh=\lub\{\cS_\ka\mid\ka<\kh\}$.

  A {\em stratification} of a computability predicate $\cS$ is a
  monotone sequence of computability predicates $(\cS_\ka)_{\ka<\kh}$
  included in $\cS$ and converging to $\cS$, that is, such that
  $\cS_\kh=\cS$.

  A stratification of a type $T$ is a stratification of $\tilde\bI(T)$.

  Given a stratification $\cS$, the {\em size} of an element $t\in
  \cS_\kh$, written $o_\cS(t)$, is the smallest ordinal $\ka<\kh$ such
  that $t\in\cS_\ka$.

  A stratification is {\em continuous} if, for all limit ordinals
  $0<\ka<\kh$, $\cS_\ka={\lub(\{\cS_\kb\mid\kb<\ka\})}$.
\end{dfn}

Because $\a$ is finitely branching, we immediately remark:

\begin{lem}
  For all continuous stratifications $\cS$ and limit ordinal
  $0<\ka\le\kh$, we have $\cS_\ka=\bigcup(\{\cS_\kb\mid\kb<\ka\})$.
\end{lem}

\begin{prf}
  By definition, $\cS$ is monotone. So, for all $\ka\le\kh$,
  $\{\cS_\kb\mid\kb<\ka\}$ is well-ordered wrt. inclusion. Since $\a$
  is finitely branching, the conclusion follows from Lemma
  \ref{lem-wo-union-cp}.\qed\\
\end{prf}

We now prove some properties of $o_\cS(t)$:

\begin{lem}
  \label{lem-strat-red}
  \label{lem-strat-cont}
Let $\cS$ be a stratification and $t\in\cS_\kh$.
\begin{itemize}
\item If $t\a t'$, then $t'\in\cS_\kh$ and $o_\cS(t)\ge o_\cS(t')$.
\item If $\cS$ is continuous, then either $o_\cS(t)=0$ or $o_\cS(t)=\kb+1$
  for some ordinal $\kb$.
\end{itemize}
\end{lem}

\pagebreak
\begin{prf}
\begin{itemize}
\item Since $\cS_{o_\cS(t)}$ is stable by reduction,
  $t'\in\cS_{o_\cS(t)}$. Therefore, $o_\cS(t')\le o_\cS(t)$.
\item Assume that $o_\cS(t)$ is a limit ordinal $\ka>0$. Since $\cS$
  is continuous, we have
  $\cS_\ka={\bigcup(\{\cS_\kb\mid\kb<\ka\})}$. Therefore, $t\in\cS_\kb$
  for some $\kb<\ka$. Contradiction.\qed
\end{itemize}
\end{prf}

By Proposition \ref{prop-mon}, $[\sB:\cX](T)$ is monotone wrt. $\cX$
whenever $T$ is positive wrt. $\sB$. Hence, any stratification $\cS$
of $\sB$ provides a way to define a stratification of $T$:

\begin{dfn}[Stratification of a positive type]
  Given a stratification $\cS$ of a sort $\sB$ and a type $T$ positive
  wrt. $\sB$, let $[\sB:\cS](T)$ denote the stratification $\cT$ of
  $T$ obtained by taking $\cT_\ka=[\sB:\cS_\ka](T)$.
\end{dfn}

Note that $[\sB:\cS]T$ is not continuous in general (see Example
\ref{ex-omega-size} below).

\begin{lem}
  \label{lem-size-strict-pos}
If $\cS$ is a stratification of $\sB$, $v\in\vU\A\sB$ and
$\Pos(\sB,\vU)=\vide$, then
$o_{[\sB:\cS](\vU\A\sB)}(v)=\sup\{o_\cS(v\vu)\mid\vu\in\vU\}$.
\end{lem}

\begin{prf}
Let $\ka=o_{[\sB:\cS](\vU\A\sB)}(v)$ and
$\kb=\sup\{o_\cS(v\vu)\mid\vu\in\vU\}$. By definition of $\ka$, we
have ${v}\in{\vU\A\cS_\ka}$. So, for all $\vu\in\vU$, $v\vu\in\cS_\ka$
and $o_\cS(v\vu)\le\ka$. Thus, $\kb\le\ka$. We now prove that
$\ka\le\kb$. To this end, it suffices to prove that
$v\in\vU\A\cS_\kb$. Let $\vu\in\vU$. By definition of $\kb$,
$o_\cS(v\vu)\le\kb$. So, $v\vu\in\cS_\kb$.\qed\\
\end{prf}

A continuous stratification of a sort $\sB$ can be obtained by the
transfinite iteration of $\bH^\sB$ from the smallest computability
predicate $\bot$ \cite{kuratowski22fm,cousot79pjm}:

\begin{itemize}
\item $\cD^\sB_0=\bot$;
\item $\cD^\sB_{\ka+1}=\bH^\sB(\cD^\sB_\ka)$;
\item $\cD^\sB_\ka=\lub(\{\cD^\sB_\kb\mid\kb<\ka\})$ if $\ka$ is an
  infinite limit ordinal.
\end{itemize}

The fact that $\cD^\sB$ is monotone follows from the facts that
$\cD^\sB_0\sle\cD^\sB_1$ and $\bH^\sB$ is monotone
\cite{cousot79pjm}. Now, by definition of $\kh$, $\cD^\sB$ is not
injective. Therefore, there are $\kc<\kd<\kh$ such that
$\cD^\sB_\kc=\cD^\sB_\kd$. Since $\cD^\sB$ is monotone,
$\cD^\sB_\kc=\cD^\sB_{\kc+1}=\cD^\sB_\kd=\cD^\sB_\kh=\sB$
\cite{rubin63book}.

We call this stratification the {\em default} stratification. It is
the one used in all the previous works on sized types, except in
\cite{abel12fics} where, after \cite{sprenger03fossacs}, Abel uses a
stratification having better properties, namely
$\cS^\sB_\ka=\lub(\{\bH^\sB(\cS^\sB_\kb)\mid\kb<\ka\})$.

The size wrt. the default stratification of a term $t$ is the
set-theoretical height of the tree representation of $t$ when
abstractions are interpreted as set-theoretical functions. If no
constructor of $\sB$ has accessible {\em functional} arguments and
$\a$ is finitely branching, then every element of $\sB$ has a size
smaller than $\w$. Hence, when considering first-order data types only
(\eg natural numbers, lists, binary trees) and a finitely branching
rewrite relation $\a$, one can in fact take $\kh=\w$.

On the other hand, when one wants to consider constructors with
accessible functional arguments, then one can get terms of size bigger
than $\w$:

\begin{exa}
  \label{ex-omega-size}
  Take the sort $\sO$ of Howard's constructive ordinals mentioned in
  the previous section and let $\inj:\sN\A\sO$ be the usual injection
  from $\sN$ to $\sO$ defined by the rules $\inj~\znat\a\zord$ and
  $\inj~(\ss~x)\a\sord~(\inj~x)$. Let us prove that
  $o_{\cD^\sO}(\lim~\inj)=\w+1$. By definition,
  $o_{\cD^\sO}(\lim~\inj)$ is the smallest ordinal $\ka$ such that
  $\lim~\inj\in\cD^\sO_\ka$. By definition of $\cD$,
  $\ka=o_\cS(\inj)+1$ where $\cS=[\sO:\cD^\sO](\sN\A\sO)$. By Lemma
  \ref{lem-size-strict-pos},
  $o_\cS(\inj)=\sup\{o_{\cD^\sO}(\inj~t)\mid t\in\sN\}$. Now, a term
  of the form $(\inj~t)$ can only reduce to a term of the form
  $(\inj~u)$, $\zord$ or $(\sord~u)$. Hence,
  $o_{\cD^\sO}(\inj~t)<\w$. Finally, one can easily prove that, for
  all $n<\w$, $o_{\cD^\sO}(\inj(\ss^n\znat))=n+1$. Therefore,
  $o_\cS(\inj)=\w$ and $o_{\cD^\sO}(\lim~\inj)=\w+1$. Moreover, $\cS$
  is not continuous since $\inj\in\cS_\omega-\bigcup\{\cS_n\mid
  n<\omega\}$.\qed
\end{exa}

One can also get terms of size bigger than $\w$ by considering
infinitely branching and non-confluent rewrite relations: with
$\cR=\{\sf\a\ss^i\znat\mid i\in\bN\}$, one gets
$o_{\cD^\sN}(\sf)=\w+1$.

\subsection{Stratifications based on constructor size functions}

We now introduce a general way of defining a stratification:

\begin{dfn}[Constructor size function]\label{def-size-fun}
A {\em size function} for $\msc:\vT\A\sB$ is given by:
\begin{itemize}
\item a function $\S^\msc:\kh^{\rq^\msc}\a\kh$ for computing the size
  of a term of the form $\msc\,\vt$ from the sizes of its accessible
  arguments;
\item for every non-recursive accessible argument
  $k\in\{\rp^\msc+1,\ldots,\rq^\msc\}$, a sort $\sB^\msc_k<_\bS\sB$
  occurring in $T_k$, only positively, and with respect to which we
  will measure the size of the $k$-th argument of $\msc$ (in the
  following, we let $\sB^\msc_k=\sB$ if $k\in\{1,\ldots,\rp^\msc\}$).
\end{itemize}
\end{dfn}

In practice, there is usually no choice for $\sB^\msc_k$. For
having a choice, the order of $T_k$ must be greater than or equal to
2. For instance, if $T_k=(\sC\A\sD)\A\sE$, then one can choose between
$\sC$ and $\sE$ if both are different from $\sD$.

On the other hand, there are many possible choices for $\S^\msc$. For
instance, consider the type $\sT$ of labeled binary trees with the
constructors $\leaf:\sB\A\sT$ and $\node:\sT\A\sT\A\sB\A\sT$, where
$\sB<_\bS\sT$ is a sort for labels. We can take $\rp^\leaf=0$,
$\rq^\leaf=1$, $\rp^\node=2$, $\rq^\node=3$, $\S^\leaf(\ka)=0$ and
$\S^\node(\ka,\kb,\kc)=\ka+\kb+1$, so that the size of a tree is not
its height as in the default stratification but the number of its
nodes.

Interestingly, $\S^\msc$ may depend on all accessible arguments,
including the non-recursive ones. For instance, one can measure the
size of a pair of natural numbers by the sum of their sizes: given a
type $\sP$ for pairs of natural numbers with the constructor
$\pair:\sN\A\sN\A\sP$, one can take $\rp^\pair=0$, $\rq^\pair=2$,
$\sB^\pair_1=\sB^\pair_2=\sN$ and $\S^\pair(\ka,\kb)=\ka+\kb$.

Finally, $\S^\msc$ can be defined by combining of size of recursive
and non-recursive arguments. For instance, the size of a list of
natural numbers can be defined as the sum of the sizes of its
components. With this notion of size, a list with only one big element
can be greater than a list with many small elements.

\begin{dfn}[Stratification defined by size functions]\label{def-size-fun-strat}
  Assume that $\a$ is finitely branching.
  Given a size function $\S^\msc$ for every constructor $\msc$, we
define a continuous stratification $\cS^\sB$ for every sort $\sB$ by
induction on $>_\bS$ as follows, where, given
$(\msc,\vt,\vT)\in\bC^\sB_{\a^*}(t)$, $o_{\cS^\msc}(\vt)$ denotes the
sequence $o_{\cS^{\msc,1}}(t_1)$, \ldots, $o_{\cS^{\msc,n}}(t_n)$ with
$n=\rq^\msc$ and $\cS^{\msc,k}_\ka=[\sB^\msc_k:\cS^{\sB^\msc_k}_\ka]T_k$, that
is, $o_{\cS^{\msc,k}}(t_k)$ is the size of $t_k$ in $T_k$ wrt.
$\sB^\msc_k$ (which is $\sB$ if $k\in\{1,\ldots,\rp^\msc\}$):

\begin{itemize}
\item $\cS^\sB_0$ is the set of terms $t\in\SN$ such that, for all
  $(\msc,\vt,\vT)\in\bC^\sB_{\a^*}(t)$:
  \begin{itemize}
  \item $\rp^\msc=0$\quad(\ie $\msc\,$ has no recursive argument),
  \item $\all k\in\{\rp^\msc+1,\ldots,\rq^\msc\}$, $t_k\in T_k$,
  \item $\S^\msc(o_{\cS^\msc}(\vt))\le 0$.
  \end{itemize}
\item $\cS^\sB_{\ka+1}$ is the set of terms $t\in\SN$ such that, for
  all $(\msc,\vt,\vT)\in\bC^\sB_{\a^*}(t)$:
  \begin{itemize}
    \item $\all k\in\{1,\ldots,\rp^\msc\}$, $t_k\in[\sB:\cS^\sB_\ka]T_k$
    \item $\all k\in\{\rp^\msc+1,\ldots,\rq^\msc\}$, $t_k\in T_k$
    \item $\S^\msc(o_{\cS^\msc}(\vt))\le\ka+1$.
  \end{itemize}
\item $\cS^\sB_\ka=\lub(\{\cS^\sB_\kb\mid\kb<\ka\})$ if $\ka$ is an
  infinite limit ordinal.
\end{itemize}
\end{dfn}

Note that $\cS$ is well-defined because:
\begin{itemize}
\item in the case of $\cS^\sB_0$:
  \begin{itemize}
  \item $\rp^\msc=0$ and thus, for all $k\in\{1,\ldots,\rq^\msc\}$,
    $o_{\cS^{\msc,k}}(t_k)=o_{[\sB^\msc_k:\cS^{\sB^\msc_k}]T_k}(t_k)$ is
    well-defined since $t_k\in T_k$ and $\sB^\msc_k<_\bS\sB$.
  \end{itemize}
\item in the case of $\cS^\sB_{\ka+1}$:
  \begin{itemize}
  \item $\all k\le\rp^\msc$,
    $o_{\cS^{\msc,k}}(t_k)=o_{[\sB:\cS^\sB]T_k}(t_k)$ is well-defined
    and $\le\ka$ since $t_k\in[\sB:\cS^\sB_\ka]T_k$;
  \item $\all k\in\{\rp^\msc+1,\ldots,\rq^\msc\}$,
    $o_{\cS^{\msc,k}}(t_k)$ is well-defined since $t_k\in T_k$ and
    $\sB^\msc_k<_\bS\sB$.
  \end{itemize}
\end{itemize}

The definition of $\cS^\sB$ is similar to the definition of the
default stratification except that the size functions $\S^\msc$ are
used to enforce lower bounds on the size of terms. Hence, if one takes
for every $\S^\msc$ the constant function equal to $0$, then one
almost gets the default stratification. To get the default
stratification one has to slightly change the definition of $\cS^\sB$
by taking $\cS^\sB_0=\bot$. The current definition has the advantage
that both variables and nullary constructors whose size function is
$0$ have size $0$. Hence, if one takes $\S_{\ms{0}}=\S_\ss(\ka)=0$,
then $o_{\cS^\sN}(\ss^ix)=o_{\cS^\sN}(\ss^i\ms{0})=i$ while, in the
default stratification, $o_{\cD^\sN}(\ts^ix)=i$ and
$o_{\cD^\sN}(\ss^i\ms{0})=i+1$ (nullary constructors do not belong to
$\bot$).

We now check that $\cS^\sB$ is indeed a stratification of $\sB$.

\begin{lem}\label{lem-strat-correct}
For every sort $\sB$ and ordinal $\ka<\kh$, $\cS^\sB_\ka\sle\sB$.
\end{lem}

\begin{prf}
We proceed by induction on $<_\bS$ and $\ka$.
\begin{itemize}
\item Let $t\in\cS^\sB_0$. Then, $t\in\SN$. Let now
  $(\msc,\vt,\vT)\in\bC^\sB_{\a^*}(t)$ and
  $k\in\{1,\ldots,\rq^\msc\}$. Then, $\rp^\msc=0$ and $t_k\in
  T_k$. Hence, $t\in\sB$ since $\sB=\bH^\sB(\sB)$.

\item Let $t\in\cS^\sB_{\ka+1}$. Then, $t\in\SN$. Let now
  $(\msc,\vt,\vT)\in\bC^\sB_{\a^*}(t)$ and
  $k\in\{1,\ldots,\rq^\msc\}$. If $k\le\rp^\msc$, then
  $t_k\in[\sB:\cS^\sB_\ka]T_k$. By induction hypothesis,
  $\cS^\sB_\ka\sle\sB$. Since $\sB$ occurs only positively in $T_k$,
  Proposition \ref{prop-mon} gives
  $[\sB:\cS^\sB_\ka]T_k\sle[\sB:\sB]T_k=T_k$. Therefore, $t_k\in
  T_k$.
  Now, if $k\in\{\rp^\msc+1,\ldots,\rq^\msc\}$, then $t_k\in T_k$
  too. Therefore, $t\in\sB$ since $\sB=\bH^\sB(\sB)$.

\item Let $\ka$ be an infinite limit ordinal. Then,
  $\cS^\sB_\ka=\lub\{\cS^\sB_\kb\mid\kb<\ka\}$. For every $\kb<\ka$,
  by induction hypothesis, $\cS^\sB_\kb\sle\sB$. Therefore,
  $\cS^\sB_\ka\sle\sB$.\qed
\end{itemize}
\end{prf}

\begin{lem}
For every sort $\sB$ and ordinal $\ka<\kh$, $\cS^\sB_\ka$ is a
computability predicate.
\end{lem}

\begin{prf}
  We proceed by induction on $<_\bS$ and $\ka$. If $\ka$ is an
  infinite limit ordinal, then $\cS^\sB_\ka$ is a computability
  predicate by definition of $\lub$ since, by induction hypothesis,
  for all $\kb<\ka$, $\cS^\sB_\ka$ is a computability predicate.

  We are left with the cases of $0$ and successor ordinals. Given a
  predicate $P$ on triples $(\msc,\vt,\vT)$, let
  $\SN^\sB(P)=\{t\in\SN\mid\bC^\sB_{\a^*}(t)\sle P\}$. We have
  $\cS^\sB_0=\SN^\sB(P_0)$ for some predicate $P_0$, and
  $\cS^\sB_{\ka+1}=\SN^\sB(P_{\ka+1})$ for some predicate
  $P_{\ka+1}$. However, for all predicates $P$, $\SN^\sB(P)$ is a
  computability predicate:
  \begin{itemize}
  \item $\SN^\sB(P)\sle\SN$ by definition.
  \item If $t\in\SN^\sB(P)$ and $t\a t'$, then $t'\in\SN^\sB(P)$ since
    $t'\in\SN$ and $\bC^\sB_{\a^*}(t')\sle\bC^\sB_{\a^*}(t)$.
  \item Assume now that $t$ is neutral and
    $\a\!(t)\sle\SN^\sB(P)$. Then, $t\in\SN$. Assume moreover that
    $(\msc,\vt,\vT)\in\bC^\sB_{\a^*}(t)$. Since $t$ is neutral, there
    is $t'$ such that $t\a t'$ and
    $(\msc,\vt,\vT)\in\bC^\sB_{\a^*}(t')$. Therefore,
    $(\msc,\vt,\vT)\in P$ and $t\in\SN^\sB(P)$.\qed
  \end{itemize}
\end{prf}

\begin{lem}
\label{lem-strat-mon}
For every sort $\sB$, $\cS^\sB$ is monotone.
\end{lem}

\begin{prf}
  We prove that, for all $(\ka,\kb,\kc)$, if $\kb\le\kc\le\ka$, then
  (1) $\cS^\sB_\kb\sle\cS^\sB_\kc$, hence $\cS^\sB|_\ka$ is monotone,
  (2) $\cS^\sB_\kc\sle\cS^\sB_\ka$, and (3)
  $\cS^\sB_\ka\sle\cS^\sB_{\ka+1}$, by induction on $\ka$. There are 3
  cases:
  \begin{itemize}
  \item $\ka=0$. Then, $\kb=\kc=0$ and (1) and (2) hold trivially. We now
    prove (3). Let $t\in\cS^\sB_0$. We prove that $t\in\cS^\sB_1$:
    \begin{itemize}
    \item $t\in\SN$ since, by definition, $\cS^\sB_0\sle\SN$.
    \end{itemize}
    Let now $(\msc,\vt,\vT)\in\bC^\sB_{\a^*}(t)$.
    \begin{itemize}
    \item We have to prove that, for all $k\in\{1,\ldots,\rp^\msc\}$,
      $t_k\in[\sB:\cS^\sB_0]T_k$. Since $t\in\cS^\sB_0$, we have
      $\rp^\msc=0$. Therefore, the property holds since there is no
      $k\in\{1,\ldots,\rp^\msc\}$.
    \item We have to prove that, for all
      $k\in\{\rp^\msc+1,\ldots,\rq^\msc\}$, $t_k\in T_k$. This holds
      since $t\in\cS^\sB_0$.
    \item Finally, we have to prove that
      $\S^\msc(o_{\cS^\msc}(\vt))\le 1$. This holds since
      $t\in\cS^\sB_0$ and thus $\S^\msc(o_{\cS^\msc}(\vt))\le 0$.
    \end{itemize}

  \item $\ka=\ka'+1$.
    \begin{enumerate}
    \item If $\kc\le\ka'$ then (1) holds by induction hypothesis (1)
      on $(\ka',\kb,\kc)$. Otherwise $\kc=\ka'+1$. If $\kb=\kc$, then
      (1) holds trivially. Otherwise $\kb\le\ka'$ and (1) holds by
      induction hypothesis (1) and (3) on $(\ka',\kb,\ka')$, and
      transitivity of $\le$.
    \item If $\kc\le\ka'$ then (2) holds by induction hypothesis (2)
      and (3) on $(\ka',\kb,\kc)$, and transitivity of
      $\le$. Otherwise (2) holds trivially.
    \item Let $t\in\cS^\sB_{\ka'+1}$. We prove that
      $t\in\cS^\sB_{\ka'+2}$:
      \begin{itemize}
      \item $t\in\SN$ since, by definition, $\cS^\sB_{\ka'+1}\sle\SN$.
      \end{itemize}
      Let now $(\msc,\vt,\vT)\in\bC^\sB_{\a^*}(t)$ and
      $k\in\{1,\ldots,\rq^\msc\}$.
      \begin{itemize}
      \item Assume that $k\le\rp^\msc$. Since $t\in\cS^\sB_{\ka'+1}$,
        we have $t_k\in[\sB:\cS^\sB_{\ka'}]T_k$. Therefore,
        $t_k\in[\sB:\cS^\sB_{\ka'+1}]T_k$ since $\sB$ occurs only
        positively in $T_k$ and $\cS^\sB_{\ka'}\sle\cS^\sB_{\ka'+1}$
        by induction hypothesis (3) on $(\ka',\ka',\ka')$.
      \item Assume that $k>\rp^\msc$. Then, $t_k\in T_k$ since
        $t\in\cS^\sB_{\ka'+1}$.
      \item Since $t\in\cS^\sB_{\ka'+1}$, we have
        $\S^\msc(o_{\cS^\msc}(\vt))\le\ka'+1$. Therefore,
        $\S^\msc(o_{\cS^\msc}(\vt))\le\ka'+2$.
      \end{itemize}
    \end{enumerate}
    
  \item $\ka$ is an infinite limit ordinal. Then,
    $\cS^\sB_\ka=\lub\{\cS^\sB_\kb\mid\kb<\ka\}$.
    \begin{enumerate}
    \item If $\kc<\ka$, then (1) follows by induction hypothesis (1)
      on $(\kc,\kb,\kc)$. Otherwise, $\kc=\ka$. If $\kb=\kc$ then (1)
      holds trivially. Otherwise, $\kb<\kc$ and (1) holds by
      definition of $\lub$.
    \item (2) holds by definition of $\lub$.
    \item Let $t\in\cS^\sB_\ka$. We have to prove that
      $t\in\cS^\sB_{\ka+1}$.

      After (1), $\cS^\sB|_\ka$ is monotone. Therefore, by Lemma
      \ref{lem-wo-union-cp},
      $\cS^\sB_\ka=\bigcup\{\cS^\sB_\kb\mid\kb<\ka\}$ and
      $t\in\cS^\sB$ for some $\kb<\ka$.
      Now, since $\ka$ is a limit ordinal, $\kb+1<\ka$. Therefore, by
      induction hypothesis (2) on $(\kb+1,\kb,\kb)$,
      $\cS^\sB_\kb\sle\cS^\sB_{\kb+1}$ and $t\in\cS^\sB_{\kb+1}$.
      We now prove that $t\in\cS^\sB_{\ka+1}$:
      \begin{itemize}
      \item $t\in\SN$ since $\cS^\sB_\kb$ is a computability predicate.
      \end{itemize}
      Let now $(\msc,\vt,\vT)\in\bC^\sB_{\a^*}(t)$ and
      $k\in\{1,\ldots,\rq^\msc\}$.
      \begin{itemize}
      \item Assume that $k\le\rp^\msc$. Since $t\in\cS^\sB_{\kb+1}$,
        we have $t_k\in[\sB:\cS^\sB_\kb]T_k$. Therefore,
        $t_k\in[\sB:\cS^\sB_\ka]T_k$ since $\sB$ occurs only
        positively in $T_k$ and $\cS^\sB_\kb\sle\cS^\sB_\ka$.
      \item Assume that $k>\rp^\msc$. Then, $t_k\in T_k$ since
        $t\in\cS^\sB_{\kb+1}$.
      \item Since $t\in\cS^\sB_{\kb+1}$, we have
        $\S^\msc(o_{\cS^\msc}(\vt))\le\kb+1$. Therefore,
        $\S^\msc(o_{\cS^\msc}(\vt))\le\ka+1$.\qed
      \end{itemize}
    \end{enumerate}
  \end{itemize}
\end{prf}

\begin{lem}
For every sort $\sB$, $\cS^\sB_\kh=\sB$.
\end{lem}

\begin{prf}
By Lemma \ref{lem-strat-correct}, $\cS^\sB_\ka\sle\sB$. Now, since
$\sB=\cD^\sB_\kh$,where $\cD^\sB$ is the default stratification, it
suffices to prove that, for all $\ka$, $\cD^\sB_\ka\sle\cS^\sB_\kh$,
that is, for all $\ka$, there is $\kb<\kh$ such that
$\cD^\sB_\ka\sle\cS^\sB_\kb$. We proceed by induction on $<_\bS$ and
$\ka$.
\begin{itemize}
\item $\cD^\sB_0=\bot\sle\cS^\sB_0$.

\item Let $\ka$ be an infinite limit ordinal smaller than $\kh$. By
  induction hypothesis, for all $\kb<\ka$,
  $\cD^\sB_\kb\sle\cS^\sB_\kh$. Therefore,
  $\cD^\sB_\ka=\lub\{\cD^\sB_\kb\mid\kb<\ka\}\sle\cS^\sB_\kh$.

\item Let now $\ka+1<\kh$. By induction hypothesis, $\cD^\sB_\ka\sle\cS^\sB_\kh$.

  Since $\kh$ is a successor cardinal, it is regular, that is, it is
  equal to its cofinality. And since it is uncountable, it is
  $\w$-complete, that is, every countable subset of $\kh$ has a least
  upper bound in $\kh$.

  Let $\kc=\sup(X)$ where $X=\{o_{\cS^\sB}(t)\mid
  t\in\cD^\sB_\ka\}$. Since $|X|\le|\cD^\sB_\ka|\le|\bL|\le\w$, we
  have $\kc<\kh$ and $\cD^\sB_\ka\sle\cS^\sB_\kc$.

  Let now $\kd=\sup(X\cup Y)$ where $Y$ is the set of the ordinals
  $\S^\msc(o_{\cS^\msc}(\vt))$ such that there are $t\in\cD^\sB_\ka$
  and $(\msc,\vt,\vT)\in\bC^\sB_{\a^*}(t)$. Since $|Y|\le\w$
  ($\cD^\sB_\ka\sle\SN$ and $\a$ is finitely branching), we have
  $\sup(Y)<\kh$ and thus $\kd<\kh$. Since $\kh$ is a limit ordinal,
  $\kd+1<\kh$.

  We now prove that $\cD^\sB_{\ka+1}\sle\cS^\sB_{\kd+1}$. Let
  $t\in\cD^\sB_{\ka+1}$. Then, $t\in\SN$. Let now
  $(\msc,\vt,\vT)\in\bC^\sB_{\a^*}(t)$ and
  $k\in\{1,\ldots,\rq^\msc\}$. If $k>\rp^\msc$, then $t_k\in
  T_k$. Otherwise, $t_k\in[\sB:\cD^\sB_\ka]T_k$. Since $\sB$ occurs
  only positively in $T_k$, we have
  $[\sB:\cD^\sB_\ka]T_k\sle[\sB:\cS^\sB_\kc]T_k$. Since $\kc\le\kd$
  and $\cS^\sB$ is monotone by Lemma \ref{lem-strat-mon}, we have
  $[\sB:\cS^\sB_\kc]T_k\sle[\sB:\cS^\sB_\kd]T_k$. Finally,
  $\S^\msc(o_{\cS^\msc}(\vt))\le\kd$. Therefore,
  $t\in\cS^\sB_{\kd+1}$.\qed
\end{itemize}
\end{prf}

This ends the proof that $\cS^\sB$ is a stratification of
$\sB$. We now see some of its properties:

\begin{lem}
  \label{lem-in-strat}\hfill
\begin{itemize}
\item $t\in\cS^\sB_0$ iff $t\in\sB$ and, for all
  $(\msc,\vt,\vT)\in\bC^\sB_{\a^*}(t)$,
  $\S^\msc(o_{\cS^\msc}(\vt))=\rp^\msc=0$.
\item $t\in\cS^\sB_{\ka+1}$ iff $t\in\sB$ and, for all
  $(\msc,\vt,\vT)\in\bC^\sB_{\a^*}(t)$, $\S^\msc(o_{\cS^\msc}(\vt))\le\ka+1$ and,
  for all $k\in\{1,\ldots,\rp^\msc\}$, $o_{\cS^{\msc,k}}(t_k)\le\ka$.
\end{itemize}
\end{lem}

\pagebreak
\begin{prf}
  \begin{itemize}
  \item Immediate.
  \item Assume that $t\in\cS^\sB_{\ka+1}$. Then, $t\in\sB$. Assume
    moreover that $(\msc,\vt,\vT)\in\bC^\sB_{\a^*}(t)$. Then,
    $\S^\msc(o_{\cS^\msc}(\vt))\le\ka+1$ and, for all
    $k\in\{1,\ldots,\rp^\msc\}$,
    $t_k\in[\sB:\cS^\sB_\ka]T_k=\cS^{\msc,k}_\ka$. Hence,
    $o_{\cS^{\msc,k}}(t_k)\le\ka$. Conversely, if
    $o_{\cS^{\msc,k}}(t_k)\le\ka$, then
    $t_k\in[\sB:\cS^\sB_\ka]T_k$.\qed
  \end{itemize}
\end{prf}

\begin{lem}
  \label{lem-constr-in-strat}
  If $(\msc,\vt,\vT)\in\bC^\sB$ and $\msc\,\vt\in\sB$, then:
\begin{itemize}
\item $o_{\cS^\sB}(\msc\,\vt)\ge\S^\msc(o_{\cS^\msc}(\vt))$.
\item $o_{\cS^\sB}(\msc\,\vt)>o_{\cS^{\msc,k}}(t_k)$ for all
  $k\in\{1,\ldots,\rp^\msc\}$.
\end{itemize}
\end{lem}

\begin{prf}
Let $\ka=o_{\cS^\sB}(\msc\,\vt)$. Since $\cS^\sB$ is continuous, by
Lemma \ref{lem-strat-cont}, either $\ka=0$ or $\ka=\kb+1$ for some
$\kb$.
\begin{itemize}
\item If $\ka=0$, then $\msc\,\vt\in\cS^\sB_0$ and
  $\S^\msc(o_{\cS^\msc}(\vt))\le\ka$ by definition of
  $\cS^\sB_0$. Otherwise, $\msc\,\vt\in\cS^\sB_{\kb+1}$ and
  $\S^\msc(o_{\cS^\msc}(\vt))\le\ka$ by definition of
  $\cS^\sB_{\kb+1}$.
\item If $\ka=0$, then $\msc\,\vt\in\cS^\sB_0$ and $\rp^\msc=0$. So,
  there is no $k\in\{1,\ldots,\rp^\msc\}$. Otherwise,
  $\msc\,\vt\in\cS^\sB_{\kb+1}$ and
  $t_k\in[\sB:\cS^\sB_\kb]T_k$. Thus,
  $o_{\cS^{\msc,k}}(t_k)\le\kb<\ka$.\qed
\end{itemize}
\end{prf}

\begin{lem}
\label{lem-ord}
If $t\in\sB$, then $o_{\cS^\sB}(t)=\d\sup(R\cup S\cup T)$ where:
\begin{itemize}
\item $\d\ka=\ka+1$ if $\ka$ is an infinite limit ordinal, and $\d\ka=\ka$
  otherwise;
\item $R=\{o_{\cS^\sB}(t')\mid t\a t'\}$;
\item $S=\{o_{\cS^{\msc,k}}(t_k)+1\mid(\msc,\vt,\vT)\in\bC^\sB,~t=\msc\,\vt,~k\in\{1,\ldots,\rp^\msc\}\}$;
\item $T=\{\S^\msc(o_{\cS^\msc}(\vt))\mid(\msc,\vt,\vT)\in\bC^\sB,~t=\msc\,\vt\}$.
\end{itemize}
\end{lem}

\begin{prf}
  Let $\ka=\sup(R\cup S\cup T)$ and $\kb=o_{\cS^\sB}(t)$.

  We first prove that $\kb\ge\d\ka$. Let $t'$ such that $t\a t'$.
  Then, $\kb\ge o_{\cS^\sB}(t')$ by Lemma \ref{lem-strat-red}. Assume
  now that $(\msc,\vt,\vT)\in\bC^\sB$ and $t=\msc\,\vt$. By Lemma
  \ref{lem-constr-in-strat}, $\kb\ge\S^\msc(o_{\cS^\msc}(\vt))$ and,
  for all $k\in\{1,\ldots,\rp^\msc\}$,
  $\kb>o_{\cS^{\msc,k}}(t_k)$. Therefore, $\kb\ge\ka$.

  Since $\cS^\sB$ is continuous, $\kb$ cannot be an infinite limit
  ordinal. So, if $\ka$ is an infinite limit ordinal, then $\kb>\ka$
  and $\kb\ge\ka+1=\d\ka$. Otherwise, $\d\ka=\ka$ and $\kb\ge\d\ka$.

  Now, to have $\kb\le\d\ka$, we prove that $t\in\cS^\sB_{\d\ka}$
  using Lemma \ref{lem-in-strat}:
\begin{itemize}
\item Case $\d\ka=0$. Then, $\ka=0$. Let $(\msc,\vt,\vT)\in\bC^\sB_{\a^*}(t)$.
\begin{itemize}
\item Case $t=\msc\,\vt$. Then, $S=\vide$, $\rp^\msc=0$ and
  $\S^\msc(o_{\cS^\msc}(\vt))=0$. Therefore, $t\in\cS^\sB_0$.
\item Case $t\a t'\a^*\msc\,\vt$. Then, $o_{\cS^\sB}(t')=0$. So,
  $\rp^\msc=0$, $\S^\msc(o_{\cS^\msc}(\vt))=0$ and
  $t\in\cS^\sB_0$.
\end{itemize}

\item Case $\d\ka=\ka'+1$. Let $(\msc,\vt,\vT)\in\bC^\sB_{\a^*}(t)$.
\begin{itemize}
\item Case $t=\msc\,\vt$. First,
  $\S^\msc(o_{\cS^\msc}(\vt))\le\sup(T)\le\ka\le\d\ka=\ka'+1$. Second,
  if $k\in\{1,\ldots,\rp^\msc\}$, then
  $o_{\cS^{\msc,k}}(t_k)<o_{\cS^{\msc,k}}(t_k)+1\le\sup(S)\le\ka\le\d\ka$.
  Therefore, $o_{\cS^{\msc,k}}(t_k)\le\ka'$ and $t\in\cS_{\d\ka}$.

\item Case $t\a t'\a^*\msc\,\vt$. First,
  $\S^\msc(o_{\cS^\msc}(\vt))\le\ka'+1$ since, by Lemma
  \ref{lem-constr-in-strat}, $\S^\msc(o_{\cS^\msc}(\vt))\le
  o_{\cS^\sB}(\msc\,\vt)$ and, by Lemma \ref{lem-strat-red},
  $o_{\cS^\sB}(\msc\,\vt)\le o_{\cS^\sB}(t')\le\sup(S)\le\ka\le\d\ka$.
  Second, if $k\in\{1,\ldots,\rp^\msc\}$, then
  $o_{\cS^{\msc,k}}(t_k)\le\ka'$ since, by Lemma
  \ref{lem-constr-in-strat},
  $o_{\cS^{\msc,k}}(t_k)<o_{\cS^\sB}(\msc\,\vt)$ and, by Lemma
  \ref{lem-strat-red}, $o_{\cS^\sB}(\msc\,\vt)\le
  o_{\cS^\sB}(t')\le\sup(R)\le\ka\le\d\ka=\ka'+1$. So,
  $t\in\cS_{\d\ka}$.\qed
\end{itemize}
\end{itemize} 
\end{prf}

Note that taking
$\S^\msc(\vec\ka)\le\sup\{\ka_k+1\mid k\in\{1,\ldots,\rp^\msc\}\}$
gives the same notion of size as taking $\S^\msc(\vec\ka)=0$. On the
other hand, if
$\S^\msc(\vec\ka)\ge\sup\{\ka_k+1\mid k\in\{1,\ldots,\rp^\msc\}\}$,
then $\S^\msc$ gives the size of irreducible terms of the form
$\msc\,\vt$:

\begin{cor}\label{cor-size-constr-nf}
  Assume that $\S^\msc$ is strictly extensive wrt. recursive arguments
  (\ie $\ka_k<\S^\msc(\vec\ka)$ if $k\in\{1,\ldots,\rp^\msc\}$) and
  $\S^\msc(\vec\ka)$ is never an infinite limit ordinal. Then, for all
  $(\msc,\vt,\vT)\in\bC^\sB$ such that $\msc\,\vt\in\sB$ and
  $\msc\,\vt$ is irreducible, we have
  $o_{\cS^\sB}(\msc\,\vt)=\S^\msc(o_{\cS^\msc}(\vt))$.
\end{cor}

\begin{prf}
  Since $\msc\,\vt$ is irreducible, $R=\vide$. Let
  $\ka=\S^\msc(o_{\cS^\msc}(\vt))$. Since $\ka>o_{\cS^{\msc,k}}(t_k)$
  whenever $k\in\{1,\ldots,\rp^\msc\}$,
  $o_{\cS^\sB}(\msc\,\vt)=\d\ka$. Since $\ka$ is not an infinite
  limit,
  $\d\ka=\ka$.\qed
\end{prf}


\begin{cor}\label{cor-size-constr}
  Assume that $\S^\msc$ is monotone wrt. every argument, strictly
  extensive wrt. recursive arguments and never returns an infinite
  limit ordinal. Then, for all $(\msc,\vt,\vT)\in\bC^\sB$ with
  $\msc\,\vt\in\sB$, we have
  $o_{\cS^\sB}(\msc\,\vt)=\S^\msc(o_{\cS^\msc}(\vt))$.
\end{cor}

\begin{prf}
  We proceed by induction on $\vt$ with $\la_\prod$ as well-founded
  relation. Assume that $\msc\,\vt\a u$. Then, there are $\vu$ such
  that $u=\msc\,\vu$ and $\vt\a_\prod\vu$. Hence,
  $o_{\cS^\msc}(\vu)\le_\prod o_{\cS^\msc}(\vt)$ and, by induction
  hypothesis, $o_{\cS^\sB}(\msc\,\vu)=\S^\msc(o_{\cS^\msc}(\vu))$. So,
  $o_{\cS^\sB}(\msc\,\vu)\le\S^\msc(o_{\cS^\msc}(\vt))$ since
  $\S^\msc$ is monotone. Therefore,
  $o_{\cS^\sB}(\msc\,\vt)=\S^\msc(o_{\cS^\msc}(\vt))$.\qed\\
\end{prf}


Finally, we are going to prove that, if $\a$ is locally confluent,
hence confluent on strongly normalizing terms \cite{newman42am}, then
the size of a term is equal to the size of its normal form when its
type is a strictly positive sort:

\begin{dfn}[Strictly positive sorts]\label{def-strict-pos}
  A sort $\sB$ is {\em strictly positive} if, for every constructor
  $\msc:\vT\A\sB$ and argument $k\in\{1,\ldots,\rq^\msc\}$, $T_k$ is
  positive wrt.  $\sB$ and either $T_k$ is a strictly positive
  sort\footnote{This is a restriction wrt. the definition given in
    \cite{coquand88colog} where $T_k$ can be any type where $\sB$ does
    not occur.} $\sC<_\bS\sB$ or $T_k$ is of the form $\vU\A\sB$ with
  $\Pos(\sB,\vU)=\vide$.
\end{dfn}

Examples of strictly positive sorts are Peano numbers and
Howard constructive ordinals.

\begin{lem}\label{lem-size-eq-size-nf}
  Assume that $\a$ is locally confluent and, for every constructor
  $\msc$, $\S^\msc$ is monotone wrt. every argument, strictly
  extensive wrt. recursive arguments and never returns an infinite
  limit ordinal. Then, for every strictly positive sort $\sB$ and term
  $t\in\sB$, $o_{\cS^\sB}(t)=o_{\cS^\sB}(\nf{t})$, where $\nf{t}$ is
  the normal form of $t$.
\end{lem}

\begin{prf}
  First note that $o_{\cS^\sB}(\nf{t})\le o_{\cS^\sB}(t)$ since
  $t\a^*\nf{t}$. We now prove that, for all strictly positive $\sB$,
  for all $t\in\sB$, $o_{\cS^\sB}(t)\le o_{\cS^\sB}(\nf{t})$, hence
  that $o_{\cS^\sB}(t)=o_{\cS^\sB}(\nf{t})$, by induction on
  $(\sB,o_{\cS^\sB}(t),t)$ with ${(<_\bS,<,\la)_\lex}$ as well-founded
  relation. By Lemma \ref{lem-ord}, $o_{\cS^\sB}(t)=\d\sup(R\cup S\cup
  T)$. Since $\S^\msc$ is strictly extensive,
  $o_{\cS^\sB}(t)=\d\sup(R\cup T)$. Since $o_{\cS^\sB}(\nf{t})$ cannot
  be an infinite limit ordinal, it is sufficient to prove that
  $\sup(R\cup T)\le o_{\cS^\sB}(\nf{t})$.

  Assume that $t\a u$. Then, $o_{\cS^\sB}(u)\le o_{\cS^\sB}(t)$.
  Hence, by induction hypothesis on the 2nd or 3rd component,
  $o_{\cS^\sB}(u)\le o_{\cS^\sB}(\nf{u})=o_{\cS^\sB}(\nf{t})$.

  Assume now that $(\msc,\vt,\vT)\in\bC^\sB$ and $t=\msc\,\vt$. By
  Corollary \ref{cor-size-constr},
  $o_{\cS^\sB}(t)=\S^\msc(o_{\cS^\msc}(\vt))$ and
  $o_{\cS^\sB}(\nf{t})=\S^\msc(o_{\cS^\msc}(\nf\vt))$. Since $\S^\msc$
  is monotone, it suffices to prove that, for all
  $k\in\{1,\ldots,\rq^\msc\}$,
  $o_{\cS^{\msc,k}}(t_k)\le o_{\cS^{\msc,k}}(\nf{t_k})$. Since $\sB$
  is strictly positive, there are two cases:
  \begin{itemize}
  \item $T_k$ is a strictly positive sort $\sC<_\bS\sB$. Then, by
    induction hypothesis on the 1st component,
    $o_{\cS^{\msc,k}}(t_k)=o_{\cS^\sC}(t_k)\le
    o_{\cS^\sC}(\nf{t_k})=o_{\cS^{\msc,k}}(\nf{t_k})$.
  \item There is $\vU$ such that $T_k=\vU\A\sB$ and
    $\Pos(\sB,\vU)=\vide$. Then, by Lemma \ref{lem-size-strict-pos},
    $o_{\cS^{\msc,k}}(t_k)=\sup\{o_{\cS^\sB}(t_k\,\vu)\mid\vu\in\vU\}$
    and
    $o_{\cS^{\msc,k}}(\nf{t_k})=\sup\{o_{\cS^\sB}(\nf{t_k}\,\vu)\mid\vu\in\vU\}$.
    Let $\vu\in\vU$. Since
    $o_{\cS^\sB}(\nf{t_k}\,\vu)\le
    o_{\cS^\sB}(t_k\,\vu)<o_{\cS^\sB}(t)$,
    by induction hypothesis on the 2nd component,
    $o_{\cS^\sB}(t_k\,\vu)=o_{\cS^\sB}(\nf{t_k}\,\vu)$. Therefore,
    $o_{\cS^{\msc,k}}(t_k)=o_{\cS^{\msc,k}}(\nf{t_k})$.\qed
  \end{itemize}
\end{prf}

We end this section by introducing the reflexive and transitive
closure of the notion of accessible argument (Definition
\ref{def-acc-arg}) and prove some properties about it. In order to keep
track of the sort with respect to which the size is measured, we
consider a relation on triples $(t,T,\sB)$ made of a term $t$, its
type $T$ and the sort $\sB$ used to measure the size of $t$ in
$[\sB:\cS^\sB]T$.

\begin{dfn}[Accessible subterm]\label{def-acc}
  We say that $(u,U,\sC)$ is {\em accessible} in $(t,T,\sB)$, written
  $(u,U,\sC)\tlea(t,T,\sB)$, if $(u,U,\sC)=(t,T,\sB)$ or there are
  $(\msc,\vt,\vT)\in\bC^\sB$ and $k\in\{1,\ldots,\rq^\msc\}$ such that
  $t=\msc\,\vt$, $T=\sB$ and $(u,U,\sC)\tlea(t_k,T_k,\sB^\msc_k)$,
  where $\sB^\msc_k=\sB$ if $k\le\rp^\msc$, and $\sB^\msc_k$ is given
  by the size function of $\msc$ if $k>\rp^\msc$ (see Definition
  \ref{def-size-fun}).
\end{dfn}

For example:
\begin{itemize}
\item $(x,\sN,\sN)$ is accessible in $(\ss\,x,\sN,\sN)$ if
  $\ss:\sN\A\sN$;
\item $(f,{\sN\A\sO},\sO)$ is accessible in $(\lim\,f,\sO,\sO)$ if
  $\lim:(\sN\A\sO)\A\sO$;
\item $(x,\sN,\sN)$ is accessible in $(\pair\,(\ss\,x)\,y,\sP,\sP)$ if
  $\pair:\sN\A\sN\A\sP$ and $\ss:\sN\A\sN$.
\item $(x,\sB\A\sC,\sB)$ is not accessible in $(\msc\,x,\sB,\sB)$ if
  $\msc:(\sB\A\sC)\A\sB$, because $\sB$ occurs negatively in
  $\sB\A\sC$ and thus $\rq^\msc=0$.
\end{itemize}

Note that $\tlea$ is stable by substitution, and that $\sC$ occurs
only positively in $U$ whenever $(u,U,\sC)\tlta(t,T,\sB)$, where
$\tlta$ is the strict part of $\tlea$.

\begin{lem}\label{lem-acc}
  If $(u,U,\sC)\tlea(t,T,\sB)$ and $t\in T$, then $u\in U$.
\end{lem}

\begin{prf}
  We proceed by induction on $\tlea$. If
  $(u,U,\sC)=(t,T,\sB)$, this is immediate. Otherwise, there are
  $(\msc,\vt,\vT)\in\bC^\sB$ and $k\in\{1,\ldots,\rq^\msc\}$ such that
  $t=\msc\,\vt$, $T=\sB$ and $(u,U,\sC)\tlea(t_k,T_k,\sB^\msc_k)$. By
  definition of $\tilde\bI(\sB)$, we have $t_k\in T_k$. So, by
  induction hypothesis, $u\in U$.\qed
\end{prf}

\section{Termination criterion}
\label{sec-sn}

In this section, we describe a termination criterion that capitalizes
on the fact that some terms can be assigned an ordinal size. The idea
is simple: if for every rewrite step $\sf l\a r$ and every function
call $\sg m$ in $r$, the size of $m$ is strictly smaller than the size
of $l$, then there cannot be any infinite reduction.

The idea, dating back to Hughes, Pareto and Sabry \cite{hughes96popl},
consists of introducing symbolic expressions representing ordinals and
logical rules for deducing information about the size of terms,
namely, that it is bounded by some expression. Hence, termination is
reduced to checking the decreasingness of symbolic size expressions.

Following these authors, we replace every sort $\sB$ by a pair
$(\sB,a)$, written $\sB_a$, where $a$ is a symbolic expression from an
algebra interpretable in ordinals, so that a term is of size-annotated
type $\sB_a$ if it is of type $\sB$ and of size {\em smaller than or
  equal to} the interpretation of $a$. The typing rules of Figure
\ref{fig-typ} are then easily turned into valid deduction rules on
size annotations. Moreover, the monotony of stratifications naturally
induces a notion of subtyping on size-annotated types: a term of type
$\sB_a$ is also of type $\sB_b$ if $a\le b$.

\subsection{Size-annotated types}

In the previously mentioned works, only two particular algebras have
been considered so far. First, the successor algebra (Definition
\ref{def-succ-alg}). Second, when $\kh$ is restricted to $\w$ (\eg when
inductive types are restricted to first-order data types), the algebra
of Presburger arithmetic generated from the symbols $\mt{0}$, $\ts$
and $\mt{+}$ interpreted by zero, the successor function and the
addition on natural numbers respectively, the first-order theory of
which is decidable \cite{presburger29}.

Other algebras are however interesting as we shall see in some
examples. For instance, the max-successor algebra, that is, the
successor algebra extended by a $\mt{max}$ operator, and the max-plus
algebra, that is, the algebra generated by the symbols $\mt{0}$,
$\mt{1}$, $\mt{+}$ and $\mt{max}$.

So, in the following, we consider an arbitrary size algebra and prove
general results under some conditions on it. Then, in Section
\ref{sec-succ}, we prove that these conditions are in particular
satisfied by the successor algebra.

\begin{dfn}[Size algebra]\label{def-size-alg}
A {\em size algebra} is given by:
\begin{itemize}
\item a first-order term algebra $\tA$ built from a set $\tV$ of size
  variables $\al,\b,\ldots$ and a set $\tF$ of size function symbols
  $\tf,\tg,\ldots$ of fixed arity, disjoint from $\tV$;
\item a quasi-order $\lea$ on $\tA$ stable by substitution:
  $a\vphi\lea b\vphi$ whenever $a\lea b$ and $\vphi\!:\!\tV\!\a\!\tA$;
\item a strict order ${\lta}\sle{\lea}$ also stable:
  $a\vphi\lta b\vphi$ whenever $a\lta b$ and $\vphi\!:\!\tV\!\a\!\tA$;
\item for each size function symbol $\tf\in\tF$ of arity $n$, a
  function $\tf_\kh:\kh^n\a\kh$ so that, for every valuation
  $\mu:\tV\a\kh$, $a\mu\le b\mu$ ($a\mu<b\mu$ resp.) whenever
  $a\lea b$ ($a\lta b$ resp.) where, as usual, $\al\mu=\mu(\al)$ and
  $(\tf a_1\ldots a_n)\mu=\tf_\kh(a_1\mu,\ldots,a_n\mu)$.
\end{itemize}
\noindent
A size algebra is {\em monotone} if every size function symbol is
monotone wrt $\lea$ in every argument, that is, $\tf\,\va\lea\tf\,\vb$
whenever $\va\,(\lea)_\prod\,\vb$. Given a size substitution $\vphi$
and a set $V$ of variables, let $\vphi|_V=\{(\al,\al\vphi)\mid\al\in
V\}$.
\end{dfn}

Let $a\le_\ext b$ ($a<_\ext b$ resp.) iff, for all $\mu$,
$a\mu\le b\mu$ ($a\mu<b\mu$ resp.). Note that $(\le_\ext,<_\ext)$
satisfies the above conditions and, for every pair of relations
$(\lea,\lta)$ satisfying the above conditions, we have
${\lea}\sle{\le_\ext}$ and ${\lta}\sle{<_\ext}$. So, one can always
take $\le_\ext$ ($<_\ext$ resp.) for $\lea$ ($\lta$ resp.).

As remarked in \cite{giesl02jsc}, the strict part of a stable
quasi-order $\lea$, that is ${\lna}={\lea-\gea}$, is not
necessarily stable. On the other hand, its stable-strict part $\lta$
is stable, where $a\lta b$ iff, for all closed substitution $\vphi$,
$a\vphi\lna b\vphi$.

\smallskip
The simplest size algebra is:

\begin{dfn}[Successor algebra]\label{def-succ-alg}
The {\em successor} size algebra is obtained by taking:
\begin{itemize}
\item $\tF=\tC\cup\{\ts\}$ where $\tC$ is an infinite set of constants
  and $\ts$ a unary symbol interpreted by the successor
  function\footnote{$\kh$ is closed by successor since it is a limit
    ordinal.};
\item $\lta$ is the smallest strict ordering on $\tA$ such that, for
  all $a$, $a\lta\ts\,a$;
\item $\lea$ is the reflexive closure of $\lta$.
\end{itemize}
\end{dfn}

Although this algebra may seem overly simple, it is already sufficient
to overtake the Coq termination checker (see Section \ref{sec-ex} for
various examples using it). We will study the properties of this
algebra in Section \ref{sec-succ}.

\begin{dfn}[Size-annotated types]\label{def-annot-types}
  The set $\bT_\tA$ of annotated types is defined as follows:
  \begin{itemize}
\item if $T$ is a type, then $T\in\bT_\tA$;
\item if $\sB$ is a sort and $a$ a size expression, then $\sB_a\in\bT_\tA$;
\item if $U$ and $V$ belong to $\bT_\tA$, then $U\A V\in\bT_\tA$.
\end{itemize}

Let $\Var(T)$ be the set of size variables occurring in $T$.

Given an annotated type $T$, let $|T|$ be the type obtained by
removing every annotation.

Given a sort $\sB$, a size expression $a$ and a type $T$, let
$\An(T,\sB,a)$ be the annotated type obtained by annotating in $T$
every occurrence of $\sB$ by $a$.

The definition of {\em positive} ($s=+$) and {\em negative} ($s=-$)
{\em positions} in a type (Definition \ref{def-pos}) is extended to
annotated types as follows:
\begin{itemize}
\item $\Pos^s(\sB_b)=\{1p\mid p\in\Pos^s(b)\}$;
\item $\Pos^s(\al)=\{\vep\mid s=+\}$;
\item $\Pos^s(\tf)=\{\vep\mid s=+\}$ if $\tf$ is of arity 0;
\item
  $\Pos^s(\tf\,b_1\ldots b_n)=\{ip\mid
  i\!\in\!\Mon^+(\tf),p\!\in\!\Pos^s(b_i)\} \cup\{ip\mid
  i\!\in\!\Mon^-(\tf),p\!\in\!\Pos^{-s}(b_i)\}$ if $\tf$ is of arity $n>0$,
\end{itemize}
where $\Mon^+(\tf)$ ($\Mon^-(\tf)$ resp.) is the set of arguments in
which $\tf$ is monotone (anti-monotone resp.) wrt. $\lea$.
\end{dfn}

In order to combine terms with annotated and unannotated types, we
extend $\tA$ by a greatest element $\infty$ and identify $\sB_\infty$
with $\sB$:

\begin{dfn}[Top-extension of a size algebra]
  The {\em top-extension} of a size algebra $\tA$ is a set
  $\o\tA=\tA\cup\{\infty\}$ with $\infty\notin\tA$. Given $\sB\in\bS$,
  let $\sB_\infty=\sB$ (we identify $\sB_\infty$ and $\sB$). Given
  size expressions $a,b\in\o\tA$, let $a\leai b$ if $a\lea b$ or
  $b=\infty$. Given $\vphi:\tV\a\o\tA$, let $a\vphi=\infty$ if $a$
  contains a variable $\al$ such that $\vphi(\al)=\infty$, and
  $a\vphi$ be the usual substitution otherwise. Terms distinct from
  $\infty$ are called {\em finite}.
\end{dfn}

We now propose to users a syntactic way to specify their own notions
of size through the annotation of constructor types. We assume that
every constructor type is annotated in a way that allows us to define
a size function, hence a stratification for every sort, and thus an
interpretation of every annotated type in computability predicates. To
this end, we use notations similar to the ones of Definition
\ref{def-size-fun}:

\begin{dfn}[Annotated types of constructors]\label{def-cons-annot-typ}
  We assume that every $\msc\in\bC$ with $\T(\msc)=T_1\A\ldots\A
  T_{\rr^\msc}\A\sB$ is equipped with an annotated type
  $\o\T(\msc)=\o{T_1}\A\ldots\A\o{T_{\rr^\msc}}\A\sB_{\s^\msc}$ with:
\begin{itemize}
\item for all $i\in\{1,\ldots,\rq^\msc\}$,
  $\o{T_i}=\An(T_i,\sB^\msc_i,\al^\msc_i)$;
\item for all $i\in\{\rq^\msc+1,\ldots,\rr^\msc\}$, $\o{T_i}=T_i$;
\item $\al^\msc_1,\ldots,\al^\msc_{\rp^\msc}\in\tV$;
\item $\al^\msc_{\rp^\msc+1},\ldots,\al^\msc_{\rq^\msc}\in\tV\cup\{\infty\}$;
\item the variables of $\{\al^\msc_1,\ldots,\al^\msc_{\rq^\msc}\}$ are
  either pairwise equal or pairwise distinct;
\item for all $i\in\{1,\ldots,\rp^\msc\}$, $\sB^\msc_i=\sB$;
\item for all $i\in\{\rp^\msc+1,\ldots,\rq^\msc\}$ with
  $\al^\msc_i\in\tV$, $\sB^\msc_i$ occurs in $T_i$;
\item for all $i\in\{\rp^\msc+1,\ldots,\rq^\msc\}$ with
  $\al^\msc_i\in\tV$, $\Pos(\sB^\msc_i,T_i)\sle\Pos^+(T_i)$;
\item $\s^\msc\in\o\tA$;
\item for all $i\in\{1,\ldots,\rq^\msc\}$,
  $\Pos(\al^\msc_i,\s^\msc)\sle\Pos^+(\s^\msc)$ ($\s^\msc$ is monotone
  wrt. every $\al^\msc_i$);
\item for all $i\in\{1,\ldots,\rp^\msc\}$, $\al^\msc_i\lta\s^\msc$
  ($\s^\msc$ is strictly extensive wrt. recursive arguments).
\end{itemize}
\end{dfn}

The semantics of these annotations is given by the next
definition. The intuition is that the size of a term of the form
$\msc\,\vt$ will be given by the interpretation in ordinals of
$\s^\msc$ with each $\al^\msc_i$, the abstract size of the $i$-th
argument of $\msc$, interpreted by the actual size of $t_i$ in
$[\sB^\msc_i:\cS^{\sB^\msc_i}]T_i$.

We now extend the interpretation of types in computability predicates
to annotated types, by defining a size function $\S^\msc$ for each
constructor $\msc$:

\begin{dfn}[Interpretation of size-annotated types]\label{def-int-annot-typ}
  First, for each constructor $\msc$ with $\o\T(\msc)$ as in Definition
  \ref{def-cons-annot-typ}, we define a size function $\S^\msc$
  (see Definition \ref{def-size-fun}) as follows:
  $$\S^\msc(\vec\ka)=\left\{\begin{array}{l}
  \mbox{0 if $\s^\msc=\infty$}\\
  \mbox{$\s^\msc\nu$ otherwise where
  $\nu(\al)=\left\{\begin{array}{l}
  \mbox{$\ka_i$ if $\al=\al^\msc_i$ and all the $\al^\msc_i\in\tV$ are distinct}\\
  \mbox{$\sup\{\ka_i\mid i\in\{1,\ldots,\rq^\msc\},\al^\msc_i\in\tV\}$ otherwise}
    \end{array}\right.$}
  \end{array}\right.$$

\noindent
Then, given a valuation $\mu:\tV\a\kh$, we interpret annotated types
as follows:
\begin{itemize}
\item $\sB\mu=\sB$,
\item $\sB_a\mu=\cS^\sB_{a\mu}$ if $a\in\tA$, where $\cS$ is the
  stratification defined by $\S$ (see Definition \ref{def-size-fun-strat}),
\item $(U\A V)\mu=U\mu\A V\mu$.
\end{itemize}
\end{dfn}

Note that $\S^\msc$ is monotone wrt. every argument and strictly
extensive wrt. recursive arguments since $\s^\msc$ so is.

Note also that, by definition of $\sup$, if $\al$ is distinct from
every $\al^\msc_i$, then $\nu(\al)=0$.

In the successor algebra, a constructor $\msc$ can always be annotated
as in Definition \ref{def-cons-annot-typ} above by taking:

\begin{exa}[Canonical annotations in the successor algebra]
  \label{ex-annot-succ}
  The {\em canonical type of a constructor $\msc$ in the successor
    algebra} is obtained by taking:
\begin{itemize}
  \item $\al^\msc_1=\ldots=\al^\msc_{\rp^\msc}$,
  \item $\al^\msc_{\rp^\msc+1}=\ldots=\al^\msc_{\rq^\msc}=\infty$,
  \item $\s^\msc\in\tV$ if $\rp^\msc=0$,
  \item $\s^\msc=\ss\,\al^\msc_1$ otherwise.
\end{itemize}
\end{exa}

In this case, we get
$\S^\msc(\ka_1,\ldots,\ka_{\rq^\msc})=\sup\{\ka_1+1,\ldots,\ka_{\rp^\msc}+1\}$,
that is, the size is the constructor height, the size of a constant being 0.

For the constructors of the sort $\sO$ of Howard's constructive
ordinals, we get:
\begin{itemize}
\item $\zord:\sO_\al$, $\s^\zord=\al$ and $\S^\zord=0$;
\item $\sord:\sO_\al\A\sO_{\ts\al}$, $\sB^\sord_1=\sO$,
  $\al^\sord_1=\al$, $\s^\sord=\ts\,\al$ and $\S^\sord(\ka)=\ka+1$;
\item $\lim:(\sN\A\sO_\al)\A\sO_{\ts\al}$, $\sB^\lim_1=\sO$,
  $\al^\lim_1=\al$, $\s^\lim=\ts\,\al$ and $\S^\lim(\ka)=\ka+1$.
\end{itemize}

Remark that we could have $\zord$ of size $2$ by simply taking
$\zord:\sN_{\ts(\ts\,\al)}$ instead.

For the constructors of the sort $\sT$ of binary trees with labels in
a sort $\sB<_\bS\sT$, we get:
\begin{itemize}
\item $\leaf:\sB\A\sT_\al$, $\sB^\leaf_1=\sB$, $\al^\leaf_1=\infty$,
  $\s^\leaf=\al$, and $\S^\leaf(\ka)=0$;
\item $\node:\sT_\al\A\sT_\al\A\sB\A\sT_{\ts\al}$,
  $\sB^\node_1=\sB^\node_2=\sT$, $\al^\node_1=\al^\node_2=\al$,
  $\s^\node=\ts\,\al$ and $\S^\node(\ka,\kb,\kc)=\sup\{\ka+1,\kb+1\}$.
\end{itemize}

Note that, in the successor algebra, constructors with at least two
accessible arguments (\eg $\node$) cannot have functional annotated
types (because there is only one non-nullary symbol, namely $\ts$).

\subsection{Termination conditions}

An important ingredient of the termination criterion is the way the
sizes of function arguments are compared. In frameworks where
functions are defined by fixpoint and case analysis, exactly one
argument must decrease at a time. Here, we allow the comparison of
various arguments simultaneously, possibly through some interpretation
functions $\z$.

Since not every term can be assigned a notion of size, and since two
function calls can have different numbers of arguments, we first need
to specify what arguments have to be taken into account and how their
sizes are compared:

\begin{dfn}[Order on function calls]\label{def-call-ord}
  We assume given:
\begin{itemize}
\item a well-founded quasi-ordering $\lef$ on $\bF$ (precedence)
  that we extend into a well-founded quasi-ordering on
  $\bV\cup\bC\cup\bF$ by taking $s\ltf\sf$ whenever $s\in\bV\cup\bC$
  and $\sf\in\bF$;
\item for every $\sf:T_1\A\ldots\A T_{\rr^\sf}\A\sB$:
\begin{itemize}
\item a number $\rq^\sf$ such that, for all
  $i\in\{1,\ldots,\rq^\sf\}$, $T_i$ is a sort $\sB^\sf_i$ (the first
  $\rq^\sf$ arguments of $\sf$ are the arguments that will be taken
  into account for proving termination);
\item an annotated type
  $\o\T(\sf)=\o{T_1}\A\ldots\A\o{T_{\rr^\sf}}\A\sB_{\s^\sf}$ such
  that:
  \begin{itemize}
  \item for all $i\in\{1,\ldots,\rq^\sf\}$,
    $\o{T_i}=\An(\sB^\sf_i,\sB^\sf_i,\al^\sf_i)$;
  \item for all $i\in\{\rq^\sf+1,\ldots,\rr^\sf\}$, $\o{T_i}=T_i$;
  \item $\vec\al^\sf$ are distinct variables;
  \item ${\s^\sf}\in{\o\tA}$;
  \item $\Var(\s^\sf)\sle\{\vec\al^\sf\}$;
  \end{itemize}
\item for each $X\in\{\tA,\kh\}$, a set $\bD^\sf_X$, a quasi-order
  $\le^\sf_X$ on $\bD^\sf_X$, a well-founded relation
  ${<^\sf_X}\sle{\le^\sf_X}$ and a map
  $\z_X:X^{\rq^\sf}\a\bD^\sf_X$ such that:
\begin{itemize}
\item ${(\bD^\sf_X,\le^\sf_X,<^\sf_X)}={(\bD^\sg_X,\le^\sg_X,<^\sg_X)}$ whenever
  $\sf\eqf\sg$;
\item $\va\mu<^{\sg,\sf}_\kh\vb\mu$ whenever
  $\va<^{\sg,\sf}_\tA\vb$ and $\mu:\tV\a\kh$;
\item $\va<^{\sg,\sf}_\tA\vc$ whenever $\va\,(\leai)_\prod\,\vb$ and
  $\vb<^{\sg,\sf}_\tA\vc$, that is,
  ${(\leai)_\prod\circ<^{\sg,\sf}_\tA}\sle{<^{\sg,\sf}_\tA}$;
\item $\vec\ka<^{\sg,\sf}_\kh\vec\kc$ whenever
  $\vec\ka<^{\sg,\sf}_\kh\vec\kb$ and $\vec\kb\le_\prod\vec\kc$, that
  is, ${<^{\sg,\sf}_\kh\circ\le_\prod}\sle{<^{\sg,\sf}_\kh}$;
\end{itemize}
\noindent
where
$(x_1,\ldots,x_{\rq^\sg})<^{\sg,\sf}_X(y_1,\ldots,y_{\rq^\sf})$ iff
$\sg\eqf\sf$ and
$\z^\sg_X(x_1,\ldots,x_{\rq^\sg})<^\sf_X\z^\sf_X(y_1,\ldots,y_{\rq^\sf})$.
\end{itemize}
\end{itemize}

The condition ${<^{\sg,\sf}_\kh\circ\le_\prod}\sle{<^{\sg,\sf}_\kh}$
is used in Theorem \ref{thm-sn} (in the case (app-decr)). On the other
hand, the condition
${(\leai)_\prod\circ<^{\sg,\sf}_\tA}\sle{<^{\sg,\sf}_\tA}$ is only
used in Lemma \ref{lem-thc-mon}. Note that, because $<^{\sg,\sf}_\tA$
is only defined on terms of $\tA$, if $\va\,(\leai)_\prod\,\vb$ and
$\vb<^{\sg,\sf}_\tA\vc$, then $\va$ must be in $\tA$ too since, by
definition, $a\leai b$ iff $a\lea b$ or $b=\infty$.

In the following, we may drop the exponent $\sf$ when there is no ambiguity.

In the coming termination criterion, a function call $\sf\,\vt$ will
give rise to a pair $(\sf,\vphi)$ where $\vphi:\{\vec\al^\sf\}\a\o\tA$
maps $\al^\sf_i$ to the size of $t_i$.

We therefore define a quasi-ordering on pairs $(\sf,\vphi)$ as
follows. Given $h\in\bV\cup\bC\cup\bF$,
$\psi:\{\vec\al^h\}\a\o\tA$ (with $\{\vec\al^h\}=\vide$ if
$h\in\bV\cup\bC$), $\sf\in\bF$,
$\vphi:\{\vec\al^\sf\}\a\o\tA$, let
$$(h,\psi)\lta(\sf,\vphi) \mbox{ if $h\ltf\sf$
or $\vec\al^h\psi\lta^{h,\sf}\vec\al^\sf\vphi$}.$$

Its counterpart on pairs $(\sf,\mu)$ is defined similarly as
follows. Given $h\in\bV\cup\bC\cup\bF$,\linebreak $\nu:\{\vec\al^h\}\a\kh$,
$\sf\in\bF$, $\mu:\{\vec\al^\sf\}\a\kh$, let $(h,\nu)<_\kh(\sf,\mu)$
if $h\ltf\sf$ or
$\vec\al^h\nu<^{h,\sf}_\kh\vec\al^\sf\mu$.
\end{dfn}

For the sake of simplicity, we assume that termination arguments come
first. This is not a real restriction since arguments can always be
permuted if needed.

For $\z^\sf_X$, one can often take the identity (assuming that
$\rq^\sf=\rq^\sg$ whenever $\sf\eqf\sg$). In Example \ref{ex-rev}, we
use a different function. When $\z^\sf_X$ is the identity, one can for
instance take for $\le^\sf_\tA$ ($\le^\sf_\kh$ resp.) the
lexicographic or multiset extension \cite{dershowitz79cacm} of $\lea$
($\le$ resp.), or some combination thereof, for which one can easily
prove the compatibility of $\le^\sf_\tA$ ($\le^\sf_\kh$ resp.) with
$\leai$ ($\le$ resp.). Indeed, we have
${(\leai)_\prod\circ(\lta)_\lex}\sle{(\lta)_\lex}$ since
${\leai\circ\lea}\sle{\lea}$, and
${<_\lex\circ\le_\prod}\sle{<_\lex}$.

We can now state our general termination theorem. In Section
\ref{sec-ex}, we will provide many examples of rewrite systems whose
termination is implied by it.

\begin{thm}[Termination criterion]\label{thm-sn}
  Assume that constructor types are annotated as in Definition
  \ref{def-cons-annot-typ}. By Definition \ref{def-int-annot-typ} and
  \ref{def-size-fun-strat}, this provides us with a size function $\S$
  and a stratification $\cS$. Assume moreover that $\ar$ is finitely
  branching and no $\s^\msc$ can be interpreted by an infinite limit
  ordinal.

  Then, the relation ${\a}={{\ab}\cup{\ar}}$ terminates on the set
  $\bT$ of well-typed terms if, for each rule ${l\a
    r}\in\cR\sle\bT^2$, $l$ is of the form $\sf\,\vl$, the type of $\sf$
  is annotated as in Definition \ref{def-call-ord}, $|\vl|\ge\rq^\sf$
  and there are:
  \begin{itemize}
  \item[--] a typing environment $\G:\FV(r)\a\bT_\tA$ with, for every
    $(x,U)\in\G$, an integer $k^x$ such that $x$ occurs in $l_{k^x}$,
    a sort $\sB^x$ occurring only positively in $|U|$ and a size
    variable $\al^x$ such that $U=\An(|U|,\sB^x,\al^x)$, indicating
    how to measure the size of $x$;\footnote{Note that, if we do not
      care about the size of $x$, or if no sort occurs only positively
      in $U$, then we can always take for $\sB^x$ any sort {\em not
        occurring} in $U$.}
  \item[--] {\em finite} symbolic size upper bounds
    $\vphi:\{\vec\al^\sf\}\a\tA$ for $l_1,\ldots,l_\rq$;
  \end{itemize}
  \noindent
  such that:
  \begin{itemize}
  \item{\bf Monotony.} For all $i\in\{1,\ldots,\rq^\sf\}$,
    $\Pos(\al^\sf_i,\s^\sf)\sle\Pos^+(\s^\sf)$;

  \item{\bf Accessibility.}
    For every $(x,U)\in\G$, one of the following holds:
    \begin{itemize}
    \item $x=l_{k^x}$ and $U=\o{T_{k^x}}\vphi$,
    \item $T_{k^x}$ is a sort and $(x,|U|,\sB^x)\tlea(l_{k^x},T_{k^x},T_{k^x})$;
    \end{itemize}

  \item{\bf Minimality.}\footnote{Lemma \ref{lem-min} provides a
    syntactic condition for checking minimality in the successor
    algebra.} For all substitutions $\t$ with $\vl\t\in\vT$, there
    exists a valuation $\nu$ such that:
    \begin{itemize}
    \item for all $(x,U)\in\G$,
      $o_{[\sB^x:\cS^{\sB^x}]|U|}(x\t)\le\al^x\nu$,
    \item for all $i\in\{1,\ldots,\rq^\sf\}$,
      $\al^\sf_i\vphi\nu=o_{\cS^{\sB_i}}(l_i\t)$;
    \end{itemize}

 \item{\bf Subject-reduction and decreasingness.}\\$\G\thfphi
    r:T_{|\vl|+1}\A\ldots\A T_{\rr^\sf}\A\sB_{\s^\sf}\vphi$, where
    $\thfphi$ is defined in Figures \ref{fig-thc} and
    \ref{fig-subtyping}.
  \end{itemize}
\end{thm}

\begin{figure}
  \figrule
  \caption{Computability closure of $(\sf,\vphi)$\label{fig-thc}}\vsp[3mm]
  \normalsize
\begin{center}
  \begin{minipage}{16mm}\vsp[0mm](app-decr)\end{minipage}
  $\cfrac{
    \begin{array}{c}
      (h,\vV\A V)\in\G\cup\o\T\quad
      h\ltf\sf\ou(h\eqf\sf\et|\vV|\ge\rq^h)\\
      \psi:\{\vec\al^h\}\a\o\tA\quad
      (h,\psi)\lta(\sf,\vphi)\quad
      (\all i)\G\thfphi w_i:V_i\psi\\
    \end{array}}{\G\thfphi h\,\vw:V\psi}$\\

  \begin{minipage}{9mm}\vsp[-4mm](lam)\end{minipage}
  $\cfrac{\G,x:U\thfphi w:V}{\G\thfphi\l x^Uw:U\A V}$
  \hsp[1cm]
  \begin{minipage}{9mm}\vsp[-4mm](sub)\end{minipage}
  $\cfrac{\G\thfphi t:U\quad U\le V}{\G\thfphi t:V}$
\end{center}
\end{figure}

\begin{figure}
  \figrule
  \caption{Subtyping rules\label{fig-subtyping}}\vsp[3mm]
  \normalsize
\begin{center}
  \begin{minipage}{9mm}\vsp[-4mm](size)\end{minipage}
  $\cfrac{a\leai b}{\sB_a\le\sB_b}$
  \hsp[1cm]
  \begin{minipage}{10mm}\vsp[-4mm](prod)\end{minipage}
  $\cfrac{U'\le U\quad V\le V'}{U\A V\le U'\A V'}$\\

  \begin{minipage}{8mm}\vsp[-4mm](refl)\end{minipage}
  $\cfrac{}{T\le T}$
  \hsp[1cm]
  \begin{minipage}{11mm}\vsp[-4mm](trans)\end{minipage}
  $\cfrac{T\le U\quad U\le V}{T\le V}$
\end{center}
\figrule
\end{figure}

\begin{prf}
{\bf Computability of constructors.} We first prove that, for all
$(\msc,\mu,\vt)$ with $\T(\msc)=T_1\A\ldots\A T_\rr\A\sB$,
$\o\T(\msc)=\o{T_1}\A\ldots\A\o{T_\rr}\A\sB_\s$ as in Definition
\ref{def-cons-annot-typ} (we drop the $\msc$'s in exponents),
$|\vt|=\rr$ and $(\all{}i)t_i\in\o{T_i}\mu$, we have
$\msc\,\vt\in\sB_\s\mu$. First, we have $\msc\,\vt\in\SN$ since
$\vt\in\SN$ and there is no rule of the form $\msc\,\vl\a r$. Second,
by Proposition \ref{prop-mon}, for every $i\in\{1,\ldots,\rq\}$, we
have $\o{T_i}\mu\sle T_i$ since $\o{T_i}=\An(T_i,\sB_i,\al_i)$ and
$\Pos(\sB_i,T_i)\sle\Pos^+(T_i)$. Therefore, $\msc\,\vt\in\sB$. Now,
if $\s=\infty$, then we are done. Otherwise, we are left to prove that
$o_{\cS^\sB}(\msc\,\vt)\le\s\mu$. By Corollary \ref{cor-size-constr},
$o_{\cS^\sB}(\msc\,\vt)=\S(o_\cS(\vt))$. By definition,
$\S(o_\cS(\vt))=\s\nu$ where $\nu$ is defined in Definition
\ref{def-int-annot-typ}. Since $\s$ is monotone and
$\Var(\s)\sle\{\vec\al\}$, it suffices to prove that, for all $i$ such
that $\al_i\in\tV$, $\al_i\nu\le\al_i\mu$. If all the $\al_i\in\tV$
are distinct, then $\al_i\nu=o_{\cS^i}(t_i)\le\al_i\mu$ since
$t_i\in\o{T_i}\mu$ and $\al_i$ occurs only positively in
$\o{T_i}$. Otherwise, all the $\al_i\in\tV$ are equal. If there is no
$\al_i\in\tV$, then the property holds trivially. Otherwise, all the
$\al_i\in\tV$ are equal to some variable $\al$ and
$\al\nu=\sup\{o_{\cS^i}(t_i)\mid\al_i=\al\}\le\al\mu$ since, for all
$i$ such that $\al_i=\al$, $t_i\in\o{T_i}\mu$ and $\al$ occurs only
positively in $\o{T_i}$.

\smallskip
{\bf Computability of function symbols.} We now prove that, for all
$((\sf,\mu),\vt)$ with $\T(\sf)=T_1\A\ldots\A T_\rr\A\sB$ and
$\o\T(\sf)=\o{T_1}\A\ldots\A\o{T_\rr}\A\sB_\s$ as in Definition
\ref{def-call-ord} (we drop the $\sf$'s in exponents), $|\vt|=\rr$ and
$(\all i)t_i\in\o{T_i}\mu$, we have $\sf\,\vt\in\sB_\s\mu$, by induction
on $((\sf,\mu),\vt)$ with $(<_\kh,\la_\prod)_\lex$ as well-founded
relation (1). Since $\sf\,\vt$ is neutral, it suffices to prove that,
for all $u$ such that $\sf\,\vt\a u$, we have $u\in\sB_\s\mu$. There are
two cases:
  \begin{itemize}
  \item[(a)] $u=\sf\,\vu$ and $\vt\a_\prod\vu$. Since computability is
    preserved by reduction, $(\all i)u_i\in\o{T_i}\mu$. Therefore, by
    induction hypothesis (1), $u\in\sB_\s\mu$.

  \item[(b)] $\vt=\vl\t\vu$, $\sf\,\vl\a r\in\cR$ and $u=r\t\vu$. Let
    $\nu$ be the valuation given by minimality. For all $i\le\rq$,
    $\al_i\vphi\nu=o_{\cS^{\sB_i}}(l_i\t)$. Since $l_i\t\in\o{T_i}\mu$
    and $T_i={\sB_i}_{\al_i}$, we have $\vphi\nu\le\mu$ (*).
    
    \begin{itemize} \item[(i)] {\bf Correctness of the computability
    closure.} We prove that, for all $(\G,t,T,\t)$, if $\G\thfphi t:T$
    and $x\t\in U\nu$ when $(x,U)\in\G$, then $t\t\in T\nu$, by
    induction on $\thfphi$ (2).

      \begin{itemize}
      \item(app-decr) By induction hypothesis (2), $w_i\t\in
        V_i\psi\nu$. There are 3 cases:
        \begin{itemize}
        \item $h\in\bV$. Then, $h\t\vw\t\in V\psi\nu$ since
          $\psi=\vide$ and $h\t\in(\vV\A V)\nu$ by assumption.
        \item $h\in\bC$ and $V=\vU\A\sC_\s$. For all
          $\vu\in\vU\psi\nu$, we have $h\vw\t\vu\in\sC_\s\psi\nu$ by
          computability of constructors. Therefore, by Definition
          \ref{def-cp-arr}, $h\vw\t\in V\psi\nu$.
        \item $h\in\bF$ and $V=\vU\A\sC_\s$. Since
          $(h,\psi)\lta(\sf,\vphi)$ and $\va\nu<^{h,\sf}_\kh\vb\nu$
          whenever $\va<^{h,\sf}_\tA\vb$, we have
          $(h,\psi\nu)<_\kh(\sf,\vphi\nu)$. Since $\vphi\nu\le\mu$ and
          ${<^{h,\sf}_\kh\circ\le_\prod}\sle{<^{h,\sf}_\kh}$, we
          have $(h,\psi\nu)<_\kh(\sf,\mu)$. Hence, for all
          $\vu\in\vU\psi\nu$, we have $h\vw\t\vu\in\sC_\s\psi\nu$ by
          induction hypothesis (1). Therefore, by Definition
          \ref{def-cp-arr}, $h\vw\t\in V\psi\nu$.
        \end{itemize}
      \item(lam) Wlog. we can assume that
        $x\notin\dom(\t)\cup\FV(\t)$. We have $(\l x^Uw)\t=\l
        x^U(w\t)\in U\nu\A V\nu$ because, for all $u\in U\nu$,
        $(w\t)\{(x,u)\}\in V\nu$ (cf. remarks after Definition
        \ref{def-cp}) since $(w\t)\{(x,u)\}=w\t'$ where
        $\t'=\t\cup\{(x,u)\}$ and $w\t'\in V\nu$ by induction hypothesis
        (2).
      \item(sub) We prove that $U\nu\sle V\nu$ whenever $U\le V$ by
        induction on $\le$ (3):
        \begin{itemize}
        \item(size) If $b=\infty$, then $\sB_a\nu\sle\sB$ by
          definition. Otherwise, $a\nu\le b\nu$ and
          $\sB_a\nu=\cS^\sB_{a\nu}\sle\cS^\sB_{b\nu}=\sB_b\nu$ since
          $\cS^\sB$ is monotone.
        \item(prod) Let $t\in U\nu\a V\nu$ and $u'\in U'\nu$. By
          induction hypothesis (3), $U'\nu\sle U\nu$. Hence, $u\in U\nu$
          and $tu\in V\nu$. By induction hypothesis (3), $V\nu\sle
          V\nu'$. Thus, $tu'\in V'\nu$.
        \item(refl) Immediate.
        \item(trans) By induction hypothesis (3) and transitivity of $\sle$.
        \end{itemize}
      \end{itemize}

    \item[(ii)] {\bf Computability of the matching substitution:} for
      all $(x,U)\in\G$, $x\t\in U\nu$. By assumption, there is $k$
      such that $x$ occurs in $l_k$, and $l_k\t\in\o{T_k}\mu$. After
      the accessibility condition, there are two
      cases: \begin{itemize} \item $x=l_k$ and $U=\o{T_k}\vphi$. If
      $k>q$, then $\o{T_k}=T_k$ and $\o{T_k}\mu=U\nu$. Therefore,
      $x\t\in U\nu$ since $l_k\t\in\o{T_k}\mu$. If $k\le q$, then
      $\o{T_k}=\sB_{\al_k}$ for some sort $\sB$. By minimality,
      $\al_k\vphi\nu=o_{\cS^\sB}(l_k\t)$. Therefore, $x\t\in U\nu$
      since $U=\sB_{\al_k}\vphi$.  \item $T_k$ is a sort and
      $(x,|U|,\sB^x)\tlea(l_k,T_k,T_k)$. By Lemma \ref{lem-acc},
      $x\t\in|U|$ since, by assumption, $l_k\t\in T_k$. By assumption,
      $U=\An(|U|,\sB^x,\al^x)$ and
      $\Pos(\sB^x,|U|)\sle\Pos^+(|U|)$. By minimality,
      $o_{[\sB^x:\cS^{\sB^x}]|U|}(x\t)\le\al^x\nu$. Therefore, $x\t\in
      U\nu$.  \end{itemize}

    \item[(iii)] We can now end the proof that $u\in\sB_\s\mu$. Since
      $\G\thfphi r:V\vphi$ with
      $V=\o{T_{|\vl|+1}}\A\ldots\A\o{T_\rr}\A\sB_\s$, and $x\t\in
      U\nu$ whenever $(x,U)\in\G$ by (ii), we have $r\t\in V\vphi\nu$
      by (i). Hence, $u=r\t\vu\in\sB_\s\vphi\nu$. We now prove that
      $\sB_\s\vphi\nu\sle\sB_\s\mu$. If $\s=\infty$, then
      $\sB_\s\vphi\nu=\sB_\s\mu$. Otherwise, we have
      $\vphi\s\neq\infty$ since $\vphi:\{\vec\al\}\a\tA$. Moreover, we
      have seen in (ii) that, for all $k\le q$, $\o{T_k}=\sB_{\al_k}$
      for some sort $\sB$ and
      $\al_k\vphi\nu=o_{\cS^\sB}(l_k\t)$. Since $l_k\t\in\o{T_k}\mu$,
      $\al_k\vphi\nu\le\al_k\mu$. Now, by monotony,
      $\Pos(\al_k,\s)\sle\Pos^+(\s)$. Therefore, by Proposition
      \ref{prop-mon}, $\sB_\s\vphi\nu\sle\sB_\s\mu$.
    \end{itemize}
  \end{itemize}

\smallskip
{\bf Computability of well-typed terms.} Now, it is easy to prove that
every well-typed term is computable by proceeding as for the
correctness of the computability closure: if $\G\th t:T$ and $x\t\in
U$ whenever $(x,U)\in\G$, then $t\t\in T$. We just detail the case of
a function symbol $\sf$ with $\T(\sf)=T_1\A\ldots\A T_\rr\A\sB$ and
$\o\T(\sf)=\o{T_1}\A\ldots\A\o{T_\rr}\A\sB_\s$. After Definition
\ref{def-cp-arr}, $\sf\in\T(\sf)$ iff, for all $\vt\in\vT$ such that
$|\vt|=\rr$, $\sf\,\vt\in\sB$. By assumption, for all
$i\in\{1,\ldots,\rq\}$, $T_i$ is a sort $\sB_i$. Let $\mu$ be the
valuation mapping, for every $i\in\{1,\ldots,\rq\}$, $\al_i$ to the
smallest ordinal $\kh_i<\kh$ such that
$\cS^{\sB_i}_{\kh_i}=\sB_i$. Then, $t_i\in\o{\sB_i}\mu$ and, by
computability of function symbols,
$\sf\,\vt\in\sB_\s\mu\sle\sB$. Finally, we conclude by noting that the
identity substitution is computable (cf. remark after Definition
\ref{def-cp}).\qed\\
\end{prf}

It is worth remarking that this criterion is modular since the above
conditions are for each rule. Hence, if both $\cR_1$ and $\cR_2$
satisfy the criterion with the same parameters, then $\cR_1\cup\cR_2$
satisfies the criterion too.

We now discuss each condition in turn.

\smallskip
{\bf Accessibility.} The accessibility is easy to check. As explained
in Section \ref{sec-comp}, not every subterm of a computable term is
computable. The definition of computability ensures that all
accessible subterms so are (Lemma \ref{lem-acc}). The accessibility
condition ensures that each free variable $x$ of the right hand-side
is accessible in some $l_i$. Hence, every instance of $x$ is
computable if the arguments of $\sf$ so are. Now, when $x$ is
accessible in a termination argument ($k^x\le\rq^\sf$), there must be
a sort $\sB^x$ with respect to which the size of the instances of $x$
are measured. Since $x$ can be instantiated by terms of any size, the
type of $x$ should be of the form $\An(|U|,\sB^x,\al^x)$, that is,
every occurrence of $\sB^x$ should be annotated by some size variable
$\al^x$, and no other sort should be annotated.

\smallskip
{\bf Subject-reduction and decreasingness.} This condition enforces
two properties at once. First, the right hand-side has the same type
as the left hand-side. This subject-reduction property is required
since the interpretation of a type has to be stable by reduction. So,
there should be no rule $\sf\,\vl\a r$ such that the size of $r$ is
strictly bigger than the size of $\sf\,\vl$. Second, by (app-decr), in
every function call $h\vt$, the symbolic upper bounds $\psi$ of the
actual sizes of the termination arguments of $h$ are strictly smaller
than those of $\sf\,\vl$ given by $\vphi$.

In (app-decr), $\psi$ is any size substitution of the size variables
of $\vV$. This rule works like the rule for type instantiation in
Hindley-Milner type system \cite{hindley69tams,milner78jcss} except
that, here, $\psi$ is not a type substitution but a size
substitution. Hence, if $\ss$ is declared of type
$\sN_\al\A\sN_{\ts\al}$ then, by (app-decr),
$\thfphi\ss:\sN_a\A\sN_{\ts a}$ for any size expression $a$. This
means that, in annotated types, size variables are implicitly
universally quantified.

The rule (app-decr) is a compact formulation that subsumes in a single
rule the usual rules of simply-typed $\l$-calculus for variables
($\G\thfphi x:T$ if $(x,T)\in\G$), constructors and function symbols
($\G\thfphi \msc:T\psi$ if $(\msc,T)\in\o\T$ and $\psi$ is any size
substitution), and application ($\G\thfphi tu:V$ if $\G\thfphi t:U\A
V$ and $\G\thfphi u:U$), with the following restrictions on
application and function symbols. First, the head of an application
cannot be an abstraction: $\thfphi$ only accepts terms in $\b$-normal
form since rule right-hand sides usually so are. Second, if an
application is headed by a function symbol $\sg$, then $\sg\ltf\sf$
(note that $h\ltf\sf$ whenever $h\in\bV\cup\bC$), or we have:
$\sg\eqf\sf$, $\sg$ applied to at least $\rq^\sg$ arguments, and the
sizes of the arguments of $\sg$, represented by $\psi$, are smaller
than $\vphi$ in $\lta$.

Hence, in (app-decr), $h$ is either a variable of $\G$, in which case
$\vV\A V$ is the type of $h$ declared in $\G$, or a constructor or
function symbol, in which case $\vV\A V$ is the annotated type of $h$
declared in $\o\T$. In addition, if $h$ is a variable, a constructor
symbol or a function symbol strictly smaller than $\sf$, then $h$ can
be applied to any number of arguments compatible with its type. On the
other hand, if $h$ is a function symbol equivalent to $\sf$, then it
must be applied to at least $\rq^h$ arguments, and the abstract sizes
of these arguments, given by the size substitution $\psi$, must be
strictly smaller than $\vphi$ in $\lta$.

In the examples below, we will however use (var), (cons) and (prec) to
denote the rule (app-decr) when $h$ is variable, a constructor or a
function symbol smaller than $\sf$ respectively.

Note that the typability of $r$ may require two variables $x$ and $y$
to have the same size over-approximation, that is, to have
$\al^x=\al^y$. This will always be the case in the successor algebra
when $x$ and $y$ are two recursive arguments of a constructor because,
in this algebra, the types of constructor arguments are annotated by
the same variable. For instance, if $x$ and $y$ are the first two
arguments of $\node:\sT_\al\A\sT_\al\A\sB\A\sT_{\ts\al}$, we must have
$\al^x=\al^y$.

Note also that the termination conditions do not require $l$ itself to
be typable in $\th^\sf_\vphi$. Hence, for instance, assuming that
$\sB$ has two constructors $\msc:\sB_\al\A\sB_{\ts\al}$ and
$\sb:\sB_\al\A\sB_\al\A\sB_{\ts\al}$, we can handle the rule
$\sf\,(\sb\,x_1\,(\msc\,x_2))\a\sf\,x_2$ by taking
$\G=[x_2:\sB_{\al^{x_2}}]$ and $\al^\sf_1\vphi=\ts\al^{x_2}$.  On the
other hand, we cannot handle the rule
$\sf\,(\sb\,x_1\,(\msc\,x_2))\a\sf\,(\sb\,x_1\,x_2)$. Indeed, in this
case, we can have
$o_{\cS^\sB}(\sb\,x_1\t\,x_2\t)=o_{\cS^\sB}(\sb\,x_1\t\,(\msc\,x_2\t))$
if $o(x_2\t)<o(x_1\t)$: the height is not a decreasing measure in this
case.

The relation $\thfphi$ is similar to the notion of computability closure
introduced in \cite{blanqui02tcs,blanqui16tcs} except that, when
comparing function arguments, it uses the sizes given by the type
system instead of the structure of terms. As already mentioned in the
introduction, using the size information instead of the structure of
terms relates our termination technique to well-founded monotone
algebras \cite{manna70hicss,vandepol96phd,hamana06hosc}, semantic
labeling \cite{zantema95fi,hamana07ppdp} or the notion of size-change
principle \cite{lee01popl,hyvernat14lmcs}. Now, as remarked in
\cite{blanqui06wst-hodp,kusakari07aaecc}, the notion of computability
closure itself has strong connections with the notion of dependency
pair \cite{arts00tcs}. It is also a tool for defining and
strengthening the higher-order recursive path ordering
\cite{blanqui06tr,jouannaud07jacm,blanqui15lmcs}. Finally, some
relations between these notions have been formally established:
size-change principle and dependency pairs \cite{thiemann05aaecc},
semantic labeling and recursive path ordering \cite{kamin80note},
dependency pairs and recursive path ordering \cite{dershowitz13wst},
and size-based termination and semantic labeling \cite{blanqui09csl}.

The decidability of $\thfphi$ will be studied in Section \ref{sec-dec}
and following.

\smallskip
{\bf Monotony.} The monotony condition is easy to check. It requires
the size of terms generated by $\sf$ to be monotone wrt. the sizes of
its termination arguments. It can always be satisfied by taking
$\s^\sf=\infty$. It is also satisfied if $\tA$ is monotone. This
condition also appears in \cite{abel04ita,barthe04mscs}. It is
necessary because, in the rule (app-decr), $\psi$ is not necessarily
minimal: it may be set to a {\em strict} upper bound by using the rule
(sub) beforehand. This could lead to invalid deductions
wrt. sizes. Take for instance the subtraction on natural numbers
defined by the rules of Figure \ref{fig-div} and assume that
$\sub:\sN_\al\A\sN_\b\A\sN_{\al-\b}$ in the size algebra with
${\lea}={\le_\ext}$ and $\mt{0}$, $\ts$ and $-$ interpreted by $0$,
successor and minus respectively. Then, given $\sf:\sN_\al\A\sN$ with
$\sub\ltf\sf$, the rule $\sf~(\ss~x)\a\sf~(\sub~(\ss~x)~x)$ satisfies
the other conditions. Indeed, take $\G=[x:\sN_x]$ and
$\vphi=\{(\al,\ts\,x)\}$. By (var), $\thfphi x:\sN_x$. By (cons),
$\thfphi\ss~x:\sN_{\ts\,x}$. By (sub), $\thfphi x:\sN_{\ts\,x}$. By
(prec), $\G\thfphi\sub~(\ss~x)~x:\sN_{\ts\,x-\ts\,x}$. By (sub),
$\G\thfphi\sub~(\ss~x)~x:\sN_{\mt{0}}$ (while
$o_{\cD^\sN}(\sub~(\ss~x)~x)>0$!). Therefore,
$\G\thfphi\sf~(\sub~(\ss~x)~x):\sN$ since $\mt{0}\lta\ts\,x$, but the
system does not terminate since
$\sf~(\ss~x)\a\sf~(\sub~(\ss~x)~x)\a\sf~(\ss~x)$.

\smallskip
{\bf Minimality.} Since $\vphi$ provides symbolic {\em upper bounds}
only, this does not suffice for getting termination. We also need
$\vphi$ to be minimal. Indeed, consider the rule $\sf~x\a\sf~x$ with
$\sf:\sN_\al\A\sN$ and $\G=[x:\sN_x]$. By taking $\al\vphi=\ts\,x$,
one has $\G\thfphi\sf~x:\sN$ since $\G\thfphi x:\sN_x$ and
$x\lta\ts~x$.

In Theorem \ref{thm-sn}, minimality is expressed in the most general
way by using the interpretation of annotated types in computability
predicates. With some acquaintance, it is not too difficult to check
this condition by hand on simple systems as shown in Example
\ref{ex-div}. In fact, we think that it is always possible to find a
minimal $\vphi$ when the type of every constructor $\msc$ is annotated
in the max-successor algebra (extension of the successor algebra with
a $\mt{max}$ operator) in the canonical way, that is, by taking
$\s^\msc\in\tV$ if $\rp^\msc=0$ and
${\s^\msc}={\ts(\mt{max}\al^\msc_1\ldots\al^\msc_{\rp^\msc})}$ with
distinct variables for $\al^\msc_1,\ldots,\al_{\rp^\msc}$
otherwise. However, in this paper, we want to focus on the successor
algebra and, in this case, minimality may not be satisfiable whatever
$\vphi$ is. This is due to the fact that, in the successor algebra,
one often needs to approximate the sizes of two distinct term
variables by the same size variable. Indeed, in the successor algebra,
there is no function symbol of arity $\ge 2$. Hence, the annotated
type of a binary constructor can only be of the form
$\sB_\al\A\sB_\al\A\sB_\s$ with the same size variable $\al$ for both
arguments.

In the following section, we study in more details the size of
constructor terms when the size is defined as the height like it is
the case with the canonical annotations of constructor types in the
successor and max-successor algebras. Then, we give a syntactic
condition for minimality to be satisfied in the successor algebra.

\section{Minimality property when the size is defined as the height}
\label{sec-min}

In this section, we provide sufficient conditions for the minimality
condition of Theorem \ref{thm-sn} to be satisfied when the notion of
size is the height and the size of constants is $0$, that is, when,
for every constructor $\msc$, we have:
$$\S^\msc(\ka_1,\ldots,\ka_{\rq^\msc})=\sup\{\ka_1+1,\ldots,\ka_{\rp^\msc}+1\}.$$

After Definition \ref{def-int-annot-typ}, this can be achieved in the
successor algebra by taking the canonical annotation for constructor
types (cf. Example \ref{ex-annot-succ}).

To check the minimality condition, we need to know how the size of a
term of the form $t\t$ depends on the sizes of the subterms $x\t$
where $x$ is a variable of $t$. To this end, we introduce a number of
definitions to express what are the subterms that contribute to the
size of a term and how they contribute to it:

\begin{dfn}[Recursive subterms]\label{def-sub}
  Let $\bD$ be the set of triples $(u,U,k)$ made of a term $u$, a type
  $U$ and a number $k\in\bN$. Given a sort $\sB$ and $(u,U,k)\in\bD$,
  let
  $$\Sub_\sB^1(u,U,k)=\left\{\begin{array}{l}
  \{(u_i,U_i,k+1)\mid i\in\{1,\ldots,\rp^\msc\}\}\\
  \quad\mbox{ if there is
    $(\msc,\vu,\vU)\in\bC^\sB$ such that $u=\msc\,\vu$ and $U=\sB$}\\
  \vide\mbox{ otherwise}
  \end{array}\right.$$
  Then, let $\a_\sB$ be the relation on finite sets of triples such
  that $S\a_\sB S'$ if there is $d\in S$ such that
  $\Sub_\sB^1(d)\neq\vide$ and $S'=(S-\{d\})\cup\Sub_\sB^1(d)$ (we
  replace $d$ by $\Sub_\sB^1(d)$). We say that a set $S\sle\bD$ is a
  {\em set of $\sB$-recursive subterms} of a term $t$ if
  $\{(t,\sB,0)\}\a_\sB^*S$.
\end{dfn}

For instance, if $\sa:\sB$, $\msc:\sB\A\sB$, $\rp^\msc=1$,
$\sb:\sB\A\sB\A\sB$, $\rp^\sb=2$ and $t=\sb(\msc(\msc\sa))x$, then
$\{(t,\sB,0)\}\a_\sB\{(\msc(\msc\sa),\sB,1),(x,\sB,1)\}\a_\sB\{(\msc\sa,\sB,2),(x,\sB,1)\}\a_\sB\{(\sa,\sB,3),(x,\sB,1)\}$.

\begin{lem}\label{lem-size-rec}
  If $S$ is a set of $\sB$-recursive subterms of $t\in\sB$, then
  $$o_{\cS^\sB}(t)=\sup\{o_{[\sB:\cS^\sB]U}(u)+k\mid(u,U,k)\in S\}.$$
\end{lem}

\begin{prf}
  Let $M(S)=\{o_{[\sB:\cS^\sB]U}(u)+k\mid(u,U,k)\in S\}$. The lemma
  trivially holds for $S=\{(t,\sB,0)\}$. Hence, if suffices to check
  that, if it holds for $S$ and $S\a_\sB S'$, then it holds for $S'$
  too. So, assume that there is $(\msc\,\vu,\sB,k)\in S$ such that
  $\Sub_\sB^1(\msc\,\vu,\sB,k)\neq\vide$. Then,
  $M(S')=(M(S)-\{o_{[\sB:\cS^\sB]U}(u)+k\})\cup\{o_{[\sB:\cS^\sB]U_i}(u_i)+k+1\mid
  i\in I\}$ where $I=\{1,\ldots,\rp^\msc\}$. But, by Corollary
  \ref{cor-size-constr},
  $o_{[\sB:\cS^\sB]\sB}(\msc\,\vu)=o_{\cS^\sB}(\msc\,\vu)=\S^\msc(o_{\cS^{\msc,1}}(u_1),\ldots,o_{\cS^{\msc,p^\msc}}(u_{p^\msc}))=\sup\{o_{\cS^{\msc,i}}(u_i)+1\mid
  i\in I\}=\sup\{o_{[\sB:\cS^\sB]U_i}(u_i)+1\mid i\in I\}$. Therefore,
  $\sup M(S)=\sup M(S')$.\qed
\end{prf}

\begin{lem}\label{lem-rec-subs}
  If $S$ is a set of $\sB$-recursive subterms of $t$ and $\t$ is a
  substitution, then $S\t=\{(u\t,U,k)\mid(u,U,k)\in S\}$ is a set of
  $\sB$-recursive subterms of $t\t$.
\end{lem}

\begin{prf}
  The lemma holds for $S=\{(t,\sB,0)\}$. Hence, if suffices to check
  that, if it holds for $S$ and $S\a_\sB S'$, then it holds for $S'$
  too. But $\Sub_\sB^1(u\t,U,k)=\Sub_\sB^1(u,U,k)\t$.\qed\\
\end{prf}

Note that $\a_\sB$ terminates (it acts on finite sets and replaces a
term by smaller subterms) and is confluent (it is orthogonal). Hence,
every finite set has a $\a_\sB$-normal form.

\begin{dfn}[Simple terms]\label{def-simple}
  Given a sort $\sB$ and a term $t$, let $\Sub_\sB(t)$ be the
  $\a_\sB$-normal form of $\{(t,\sB,0)\}$. A term $t$ of sort $\sB$ is
  {\em simple} if, for all $(u,U,k)\in\Sub_\sB(t)$, either $u\in\bV$
  or there is $(\msc,\vu,\vU)\in\bC^\sB$ such that $u=\msc\,\vu$,
  $U=\sB$ and $\rp^\msc=0$ ($\msc$ has no recursive argument).
\end{dfn}

\begin{lem}\label{lem-size-simple}
  If $t$ is a simple term of sort $\sB$ and $t\t\in\sB$ then:
  $$o_{\cS^\sB}(t\t)=\sup(\{\rd_\sB(t)\}\cup\{o_{[\sB:\cS^\sB]V}(x\t)+\rd_\sB^x(t)\mid(x,V)\in\Var_\sB(t)\})$$
  \noindent where:
  \begin{itemize}
  \item $\Var_\sB(t)=\{(x,U)\mid\ex k,(x,U,k)\in\Sub_\sB(t)\}$,
  \item $\rd_\sB^x(t)=\sup\{k\mid\ex U,(x,U,k)\in\Sub_\sB(t)\}$,
  \item $\rd_\sB(t)=\sup\{k\mid\ex u,\ex U,(u,U,k)\in\Sub_\sB(t)\}$.
  \end{itemize}
\end{lem}

\begin{prf}
  By Lemma \ref{lem-rec-subs}, $\Sub_\sB(t)\t$ is a set of
  $\sB$-recursive subterms of $t\t$. Hence, by Lemma
  \ref{lem-size-rec},
  $o_{\cS^\sB}(t\t)=\sup\{o_{[\sB:\cS^\sB]U}(u)+k\mid(u,U,k)\in\Sub_\sB(t)\t\}=\sup\{o_{[\sB:\cS^\sB]U}(u\t)+k\mid(u,U,k)\in\Sub_\sB(t)\}$.
  Let $(x,V)\in\Var_\sB(t)$. Since $t$ is well-typed, for all
  $(x,V')\in\Var_\sB(t)$, we have $V'=V$. Hence,
  $\sup\{o_{[\sB:\cS^\sB]U}(u\t)+k\mid(u,U,k)\in\Sub_\sB(t),u=x\}=o_{[\sB:\cS^\sB]V}(x\t)+\rd_\sB^x(t)$.
  Let now $(u,U,k)\in\Sub_\sB(t)$ with $u\notin\bV$. Since $t$ is
  simple, there is $(\msc,\vu,\vU)\in\bC^\sB$ such that $u=\msc\,\vu$,
  $U=\sB$ and $\rp^\msc=0$. By Corollary \ref{cor-size-constr},
  $o_{[\sB:\cS^\sB]U}(\msc\,\vu\t)=o_{\cS^\sB}(\msc\,\vu\t)=\S^\msc(o_{\cS^{\msc,1}}(u_1\t),\ldots,o_{\cS^{\msc,p^\msc}}(u_{p^\msc}\t))=\sup\{o_{\cS^{\msc,i}}(u_i\t)+1\mid
  i\in\{1,\ldots,\rp^\msc\}\}=0$. Therefore,
  $\sup\{o_{[\sB:\cS^\sB]U}(u\t)+k\mid(u,U,k)\in\Sub_\sB(t),u\notin\bV\}=\rd_\sB(t)$
  and
  $o_{\cS^\sB}(t\t)=\sup(\{\rd_\sB(t)\}\cup\{o_{[\sB:\cS^\sB]V}(x\t)+\rd_\sB^x(t)\mid(x,V)\in\Var_\sB(t)\}$.\qed\\
\end{prf}

To carry on with the previous example, $t=\sb(\msc(\msc\sa))x$ is
simple and we have
$o_{\cS^\sB}(t\t)=\sup\{o_{\cS^\sB}(\msc(\msc\sa))+1,o_{\cS^\sB}(x\t)+1\}=\sup\{3,o_{\cS^\sB}(x\t)+1\}=\sup\{\rd_\sB(t),o_{\cS^\sB}(x\t)+\rd_\sB^x(t)\}$.

\medskip
Assume now that we are under the conditions of Theorem \ref{thm-sn}
for some rule $\sf\,\vl\a r\in\cR$, typing environment
$\G=[x_1:U_1,\ldots,x_n:U_n]$ and substitution
$\vphi:\{\vec\al\}\a\tA$. In particular:
$$\o\T(\sf)={\sB_1}_{\al_1}\A\ldots\A{\sB_\rq}_{\al_\rq}\A T_{\rq+1}\A\ldots\A T_{\rr}\A\sB_{\s}$$
\noindent
with $\vec\al$ distinct variables, $\s\in\o\tA$ and
$\Var(\s)\sle\{\vec\al\}$.

Assume moreover that, for all $j\in\{1,\ldots,\rq\}$, $l_j$ is a
simple term of sort $\sB_j$ and there are $n_j\in\bN$ and $\g_j\in\tV$
such that $\al_j\vphi=\ts^{n_j}\g_j$.

Then, after Lemma \ref{lem-size-simple}, the minimality property is
equivalent to the following purely numerical problem on ordinals: for
all $\ka_1,\ldots,\ka_n$ (for the sizes of $x_1\t,\ldots,x_n\t$
respectively), there are $\kb_1,\ldots,\kb_n$ (for
$\al^{x_1}\nu,\ldots,\al^{x_n}\nu$ respectively) and
$\kc_1,\ldots,\kc_{\rq}$ (for $\g_1\nu,\ldots,\g_{\rq}\nu$
respectively) such that:
\begin{enumerate}
\item $(\all j)(\all k)$ $\kb_j=\kb_k$ if $\al^{x_j}=\al^{x_k}$,
\item $(\all j)(\all k)$ $\kc_j=\kc_k$ if $\g_j=\g_k$,
\item $(\all j)(\all k)$ $\kb_j=\kc_k$ if $\al^{x_j}=\g_k$,
\item $(\all j)$ $\ka_j\le\kb_j$,
\item $(\all j)$
  $\kc_j+n_j=\sup(\{\rd_{\sB_j}(l_j)\}\cup\{\ka_m+\rd_{\sB_j}^{x_m}(l_j)\mid
  x_m\in\dom(\Sub_{\sB_j}(l_j))\}$.
\end{enumerate}

The first three constraints are coherence conditions for $\nu$ to be
well defined. The last two correspond to the first and second
conditions of the minimality property respectively.

One of the problems for these inequations to be satisfied is when two
arguments of $\sf$, say $l_1$ and $l_2$, share some variable but have
distinct sets of variables, or when shared variables occur at
different depths. Take for instance $l_1=x_1$ and
$l_2=\sb\,(\msc\,x_1)\,(\msc\,x_2)$ with constructors annotated in the
canonical way in the successor algebra, that is,
$\msc:\sB_\al\A\sB_{\ts\al}$ and
$\sb:\sB_\al\A\sB_\al\A\sB_{\ts\al}$. Then, for having $\sb\,x_1\,x_2$
in the right hand-side, we need to take
$\al^{x_1}=\al^{x_2}=\g_1=\g_2$. In this case, the minimality
condition says that, for all $\ka_1,\ka_2$, there is $\kb$ such that
$\ka_1\le\kb$, $\ka_2\le\kb$, $\ka_1=\kb+n_1$ and
$\sup\{\ka_1+2,\ka_2+2\}=\kb+n_2$, which is not possible. Take now
$l_1=\sb\,(\msc\,x_1)\,(\msc\,x_2)$ and
$l_2=\sb\,(\msc\,(\msc\,x_1))\,(\msc\,x_2)$. Again, for having
$\sb\,x_1\,x_2$ in the right hand-side, we need to take
$\al^{x_1}=\al^{x_2}=\g_1=\g_2$. In this case, the minimality
condition says that, for all $\ka_1,\ka_2$, there is $\kb$ such that
$\ka_1\le\kb$, $\ka_2\le\kb$, $\sup\{\ka_1+2,\ka_2+2\}=\kb+n_1$ and
$\sup\{\ka_1+3,\ka_2+2\}=\kb+n_2$, which is not possible either.

We now give sufficient conditions for the above set of inequations to
be satisfied:

\begin{lem}\label{lem-min}
  Under the conditions of Theorem \ref{thm-sn}, assume that
  constructor types are annotated in the canonical way in the
  successor algebra (cf. Example \ref{ex-annot-succ}). Then, the
  minimality property is satisfied if, for all $j\in\{1,\ldots,\rq\}$:
  \begin{enumerate}[(a)]
  \item $l_j$ is a simple term of sort $\sB_j$;
  \item there are $n_j\in\bN$ and $\g_j\in\tV$ such that
    $\al_j\vphi=\ts^{n_j}\g_j$;
  \item $n_j\le\inf(\range(D_j))$;
  \item for all $k\in\{1,\ldots,\rq\}$, if $\g_j=\g_k$, then
    $n_j=n_k$, $\rd_{\sB_j}(l_j)=\rd_{\sB_k}(l_k)$ and $D_j=D_k$;
  \item for all $x\in\dom(\G)$, if $\g_j=\al^x$ then $x\in\dom(D_j)$;
  \end{enumerate}
  \noindent
  where $D_j=\{(x,\rd_{\sB_j}^x(l_j))\mid
  x\in\dom(\Sub_{\sB_j}(l_j))\}$, $\Sub_\sB(l)$ is introduced in
  Definition \ref{def-simple}, $\rd_\sB$ and $\rd_\sB^x$ are defined
  in Lemma \ref{lem-size-simple}.
\end{lem}

\begin{prf}
  Let $\kc_i=\sup(\{\rd_{\sB_i}(l_i)\}\cup\{\ka_p+d\mid(x_p,d)\in
  D_i\})-n_i$. It is well defined since
  $n_i\le\inf(\range(D_i))\le\rd_{\sB_i}(l_i)$. Now, let $\kb_i=\kc_m$
  if $\al^{x_i}=\g_m$ for some $m$, and
  $\kb_i=\sup\{\ka_p\mid\al^{x_p}=\al^{x_i}\}$ otherwise. It is
  well-defined since, if $\g_j=\g_k$, then $\kc_j=\kc_k$ because
  $n_j=n_k$, $\rd_{\sB_j}(l_j)=\rd_{\sB_k}(l_k)$ and $D_j=D_k$. We now
  prove that the five numerical constraints equivalent to minimality
  are satisfied:
  \begin{enumerate}
  \item Assume that $\al^{x_j}=\al^{x_k}$. If $\al^{x_j}=\g_m$, then
    $\kb_j=\kc_m=\kb_k$.\\Otherwise,
    $\kb_j=\sup\{\ka_m\mid\al^{x_m}=\al^{x_j}\}=\kb_k$.
  \item Assume that $\g_j=\g_k$. Then, $\kc_j=\kc_k$.
  \item Assume that $\al^{x_j}=\g_k$. Then, $\kb_j=\kc_k$.
  \item For all $j$, $\ka_j\le\kb_j$. Indeed, if $\al^{x_j}=\g_m$,
    then $\kb_j=\kc_m\ge\ka_j$ since $x_j\in\dom(D_m)$. Otherwise,
    $\kb_j=\sup\{\ka_p\mid\al^{x_p}=\al^{x_j}\}\ge\ka_j$.
  \item For all $j$,
    $\kc_j+n_j=\sup(\{\rd_{\sB_j}(l_j)\}\cup\{\ka_m+\rd_{\sB_j}^{x_m}(l_j)\mid
    x_m\in\dom(\Sub_{\sB_j}(l_j))\})$ by definition of $\kc_j$.\qed
  \end{enumerate}
\end{prf}

For instance, with the last rule of Figure \ref{fig-div},
$\div~(\ss~x)~(\ss~y)\a\ss~(\div~(\sub~x~y)~(\ss~y))$, if we take
$\div:\sN_\al\A\sN\A\sN_\al$, $\G=[x:\sN_x,y:\sN_y]$, $\al^x=x$,
$\al^y=y$ and $\vphi=\{(\al,\ts\,x)\}$, we have $n_1=1$,
$\g_1=x=\al^x$ and $D_1=\{(x,1)\}$. So, the conditions above are
satisfied.

On the contrary, if $l_1=\msc\,x_1$,
$l_2=\sb\,(\msc\,x_1)\,(\msc\,x_2)$, $\al^{x_1}=\al^{x_2}=\g_1=\g_2$,
$n_1=1$ and $n_2=2$, then (d) is not satisfied because $\g_1=\g_2$ but
$D_1=\{(x_1,1)\}$ and $D_2=\{(x_1,2),(x_2,2)\}$.

\section{Examples}
\label{sec-ex}

In this section, we show various examples whose termination can be
established by using Theorem \ref{thm-sn}. In proofs of $\thfphi$
judgments, (var), (cons) and (prec) will refer to the specialization
of (app-decr) to variables, constructors and function symbols smaller
than $\sf$ respectively.

We will use the following sorts and constructors with $\sN<_\bS\sL$
and $\sN<_\bS\sO$:
\begin{itemize}
\item $\sB$: the sort of booleans with the constructors $\true:\sB$
  and $\false:\sB$;
\item $\sN$: the sort of natural numbers with the constructors
  $\znat:\sN$ and $\ss:\sN\A\sN$;
\item $\sO$: the sort of Howard's constructive ordinals with the
  constructors\\$\zord:\sO$, $\sord:\sO\A\sO$ and
  $\lim:(\sN\A\sO)\A\sO$;
\item $\sL$:
  the sort of lists with the constructors $\nil:\sL$
  and $\cons:\sL\A\sN\A\sL$\footnote{We
    permuted the usual order of the arguments of $\cons$
    so that its type conforms to Definition \ref{def-acc-arg}.};
\end{itemize}

Unless stated otherwise, we always use the successor algebra (Definition
\ref{def-succ-alg}) and, for constructor types, the canonical
annotations (Example \ref{ex-annot-succ}).

\begin{exa}[Division]\label{ex-div}
  Consider the function symbols $\sub$ (substraction) and $\div$
  (division) both of type $\sN\A\sN\A\sN$ defined by the rules of
  Figure \ref{fig-div}.

  We take $\sub\ltf\div$. For annotated types, we take, for each
  $f\in\{\sub,\div\}$, $\o\T(f)=\sN_\al\A\sN\A\sN_\al$, $\rq^f=1$,
  $\sB^f_1=\sN$, $\al^f_1=\al$, which expresses the fact that these
  functions are not size-increasing. And, for $\z^\sub_X$ and
  $\z^\div_X$, we take the identity.

  We now detail the conditions of Theorem \ref{thm-sn} for each rule
  in turn (monotony is trivial).

  \begin{enumerate}
  \item $\sub~x~\znat\a x$. Take $\G=[x:\sN_x]$, $k^x=1$, $\sB^x=\sN$,
    $\al^x=x$ and $\vphi=\{(\al,x)\}$. Then,
    $\sN_x=\An(\sN,\sB^x,\al^x)$ and:
    \begin{itemize}
    \item Accessibility. $x$ is accessible since $x=l_{k^x}$ and
      $\sN_x=\sN_\al\vphi$.
    \item Minimality. One could use Lemma \ref{lem-min}. We give a
      direct proof instead. Let $\t$ be such that $x\t\in\sN$. We have
      to prove that there exists $\nu$ such that
      $o_{\cS^\sN}(x\t)\le\al^x\nu$ and
      $\al\vphi\nu=o_{\cS^\sN}(x\t)$. It suffices to take
      $\nu(x)=o_{\cS^\sN}(x\t)$.
    \item Subject-reduction. By (var),
      $\thfphi[\sub]x:\sN_x=\sN_\al\vphi$.
    \end{itemize}

  \item $\sub~\znat~y\a\znat$. Take $\G=\vphi=\vide$. Then:
    \begin{itemize}
    \item Minimality. Let $\t$ be such that $y\t\in\sN$. We have to
      prove that there exists $\nu$ such that
      $\al\vphi\nu=o_{\cS^\sN}(\znat)$. It suffices to take
      $\nu(\al)=o_{\cS^\sN}(\znat)$.
    \item Subject-reduction. By (cons),
      $\thfphi[\sub]\znat:\sN_\al=\sN_\al\vphi$.
    \end{itemize}

  \item $\sub~(\ss~x)~(\ss~y)\a\sub~x~y$. Take $\G=[x:\sN_x,y:\sN_y]$,
    $k^x=1$, $\sB^x=\sN$, $\al^x=x$, $k^y=2$, $\sB^y=\sN$, $\al^y=y$,
    $\vphi=\{(\al,\ts\,x)\}$. Then, $\sN_x=\An(\sN,\sB^x,\al^x)$,
    $\sN_y=\An(\sN,\sB^y,\al^y)$ and:
    \begin{itemize}
    \item Accessibility. $x$ is accessible since
      $(x,\sN,\sN)\tlea(l_{k^x},\sN,\sN)$. $y$ is accessible since
      $(y,\sN,\sN)\tlea(l_{k^y},\sN,\sN)$.
    \item Minimality. Let $\t$ be such that $\ss\,x\t\in\sN$ and
      $\ss\,y\t\in\sN$. We have to prove that there exists $\nu$ such
      that $o_{\cS^\sN}(x\t)\le\al^x\nu$,
      $o_{\cS^\sN}(y\t)\le\al^y\nu$ and
      $\al\vphi\nu=o_{\cS^\sN}(\ss\,x\t)=o_{\cS^\sN}(x\t)+1$. It
      suffices to take $\nu(x)=o_{\cS^\sN}(x\t)$ and
      $\nu(y)=o_{\cS^\sN}(y\t)$.
    \item Subject-reduction. Let ${\th}={\thfphi[\sub]}$. By (var),
      $\th x:\sN_x$ and $\th y:\sN_y$. By (app-decr),
      $\th\sub\,x\,y:\sN_x$ since $x\lta\ts\,x$. Therefore, by (sub),
      $\th\sub\,x\,y:\sN_{\ts\,x}=\sN_\al\vphi$.
    \end{itemize}
    
  \item $\div~\znat~(\ss~y)\a\znat$. Like for rule (2).

  \item $\div~(\ss~x)~(\ss~y)\a\ss~(\div~(\sub~x~y)~(\ss~y))$. Take
    $\G=[x:\sN_x,y:\sN_y]$, $k^x=1$, $\sB^x=\sN$, $\al^x=x$, $k^y=2$,
    $\sB^y=\sN$, $\al^y=y$ and $\vphi=\{(\al,\ts\,x)\}$. Then,
    $\sN_x=\An(\sN,\sB^x,\al^x)$, $\sN_y=\An(\sN,\sB^y,\al^y)$ and:
    \begin{itemize}
    \item Accessibility and minimality like for rule (3).
    \item Subject-reduction. Let ${\th}={\thfphi[\div]}$. By (var),
      $\G\th x:\sN_x$ and $\G\th y:\sN_y$. By (prec),
      $\G\th\sub~x~y:\sN_x$. By (cons), $\G\th\ss~y:\sN_{\ts\,y}$. By
      (app-decr), $\G\th\div~(\sub~x~y)~(\ss~y):\sN_x$ since
      $x\lta\ts\,x$. Finally, by (cons),
      $\G\th\ss~(\div~(\sub~x~y)~(\ss~y)):\sN_{\ts
      x}=\sN_\al\vphi$.\qed
    \end{itemize}
  \end{enumerate}
\end{exa}

\begin{exa}[Map and filter]
  Consider the function symbols
  $\map:\sL\A(\sN\A\sN)\A\sL$,\footnote{We permuted the usual order of
    the arguments of $\map$ so that its type conforms to Definition
    \ref{def-call-ord}.} $\si:\sL\A\sL\A\sB\A\sL$ and
  $\filter:\sL\A(\sN\A\sB)\A\sL$ defined by the rules:
\begin{rewc}
\map~\nil~f & \nil\\
\map~(\cons~l~x)~f & \cons~(\map~l~f)~(f~x)\\[1mm]
\si~x~y~\true & x\\
\si~x~y~\false & y\\[1mm]
\filter~\nil~f & \nil\\
\filter~(\cons~l~x)~f & \si~(\cons~(\filter~l~f)~x)~(\filter~l~f)~(f~x)\\
\end{rewc}

For annotated types, we could take in the successor algebra,
$\map:\sL_\al\A(\sN\A\sN)\A\sL_\al$, $\rq^\map=1$, $\sB^\map_1=\sL$,
$\al^\map_1=\al$, $\si:\sL_\al\A\sL_\al\A\sB\A\sL_\al$, $\rq^\si=2$,
$\sB^\si_1=\sB^\si_2=\sL$, $\al^\sf_1=\al^\sf_2=\al$,
$\filter:\sL_\al\A(\sN\A\sB)\A\sL_\al$ and $\rq^\filter=1$, expressing
the fact that these functions are not size-increasing.

Unfortunately, the annotated type of $\si$ does not satisfy the
conditions of Definition \ref{def-call-ord} because
$\al^\si_1=\al^\si_2$ (the variables $\al^\si_i$ should be
distinct). There are however two solutions to get around this problem:
\begin{enumerate}
  \item Annotate $\si$ in the max-successor algebra by taking
    $\si:\sL_\al\A\sL_\b\A\sB\A\sL_{\mt{max}\al\b}$.
  \item Introduce a new type $\sC>_\bS\sL$ with constructor
    $\cond:\sL_\al\A\sL_\al\A\sB\A\sC_\al$, a new function symbol
    $\newsi:\sC_\al\A\sL_\al$ with $\rq^\newsi=1$, and define $\newsi$
    and $\filter$ by the following rules instead:
\end{enumerate}

\begin{rewc}
  \newsi~(\cond~x~y~\true) & x\\
  \newsi~(\cond~x~y~\false) & y\\[1mm]
  \filter~\nil~f & \nil\\
  \filter~(\cons~l~x)~f & \newsi~(\cond~(\cons~(\filter~l~f)~x)~(\filter~l~f)~(f~x))\\
\end{rewc}

One can easily check the conditions on annotated types and the
monotony condition.

For the other conditions, we only detail the case of the last rule of
$\filter$ by taking $\G=[f:\sN\A\sB,x:\sN,l:\sL_l]$,
$\vphi=\{(\al,\ts\,l)\}$, $k^f=2$, any sort distinct from $\sN$ and
$\sB$ for $\sB^f$ (we do not care about the size of $f$), $\al^f=f$,
$k^x=1$, any sort distinct from $\sN$ for $\sB^x$ (we do not care
about the size of $x$), $\al^x=x$, $k^l=1$, $\sB^l=\sL$, $\al^l=l$,
$\newsi<_\bF\filter$, $\cond<_\bF\filter$ and the identity for
$\z^\filter$.

One can easily check the accessibility and minimality conditions.

We now check subject-reduction. Let ${\th}={\thfphi[\filter]}$. By
(var), $\G\th x:\sN$ and $\G\th l:\sL_l$. By (var), $\G\th f~x:\sB$.
By (app-decr), $\G\th\filter~l~f:\sL_l$ since $l\lta\ts\,l$. By
(cons), $\G\th\cons~(\filter~l~f)~x:\sL_{\ts\,l}$. By (sub),
$\G\th\filter~l~f:\sL_{\ts\,l}$ since $l\leai\ts\,l$. By
(cons),\linebreak
$\G\th\cond~(\cons~(\filter~l~f)~x)~(\filter~l~f)~(f~x):\sL_{\ts\,l}$. Therefore,
by (prec),\linebreak
$\G\th\newsi~(\cond~(\cons~(\filter~l~f)~x)~(\filter~l~f)~(f~x)):\sL_{\ts\,l}=\sL_\al\vphi$.\qed
\end{exa}

\begin{exa}[G\"odel' system T and Howard' system V]
  \label{ex-rec}
  Consider the recursor on natural numbers $\rec^\sN_T:\sN\A T\A(\sN\A
  T\A T)\A T$ from G\"odel' system T \cite{godel58dialectica}, and the
  recursor on ordinals $\rec^\sO_T:\sO\A T\A(\sO\A T\A
  T)\A((\sN\A\sO)\A(\sN\A T)\A T)\A T$ from Howard' system V
  \cite{howard72jsl} defined by the following rules:

  \begin{rewc}
    \rec^\sN_T~\znat~u~v & u\\
    \rec^\sN_T~(\snat~x)~u~v & v~x~(\rec^\sN_T~x~u~v)\\[3mm]
    \rec^\sO_T~\znat~u~v~w & u\\
    \rec^\sO_T~(\sord~x)~u~v~w & v~x~(\rec^\sO_T~x~u~v~w)\\
    \rec^\sO_T~(\lim~f)~u~v~w & w~f~(\l n^\sN.\rec^\sO_T~(f~n)~u~v~w)\\
  \end{rewc}

For the annotated types of function symbols, take
$\rec^\sN_T:\sN_\al\A T\A(\sN\A T\A T)\A T$ and $\rec^\sO_T:\sO_\al\A
T\A(\sO\A T\A T)\A((\sN\A\sO)\A(\sN\A T)\A T)\A T$.

We now detail the subject-reduction condition for the last rule of
$\sf=\rec^\sO_T$ with $\G=[f:\sN\A\sO_\b,u:T,v:\sO\A T\A
  T,w:(\sN\A\sO)\A(\sN\A T)\A T]$, $\vphi=\{(\al,\ts\,\b)\}$ and the
identity for $\z^\sf$. Let ${\th}={\thfphi[\sf]}$ and
$\D=[n:\sN]\G$. By (var), $\G\th f:\sN\A\sO_\b$ and $\D\th
f:\sN\A\sO_\b$, $\D\th u:T$, $\D\th v:\sO\A T\A T$, $\G\th
w:(\sN\A\sO)\A(\sN\A T)\A T$ and $\D\th w:(\sN\A\sO)\A(\sN\A T)\A
T$. By (var), $\D\th f~n:\sO_\b$. By (app-decr),
$\D\th\rec^\sO_T~(f~n)~u~v~w:T$ since $\b\lta\ts\,\b$. By (lam),
$\G\th\l n^\sN.\rec^\sO_T~(f~n)~u~v~w:\sN\A T$. By (sub), $\G\th
f:N\A\sO$ since $\sN\A\sO_\b\le\sN\A\sO$. Finally, by (var), $\G\th
w~f~(\l n^\sN.\rec^\sO_T~(f~n)~u~v~w):T$.\qed
\end{exa}

\begin{exa}[Quicksort]
  Let $\sP$ be the sort of pairs of lists with the constructor
  $\pair:\sL\A\sL\A\sP$, and $\sC$ be the sort with the constructor
  $\cond:\sP\A\sP\A\sB\A\sC$. Then, let the functions
  $\fst,\snd:\sP\A\sL$, $\lesseq:\sN\A\sN\A\sB$, $\si:\sC\A\sP$,
  $\pivot:\sL\A\sN\A\sP$, $\qs:\sL\A\sL\A\sL$ and $\qsort:\sL\A\sL$ be
  defined by the rules:

\begin{center}
\begin{tabular}{cc}
\begin{rew}
\fst~(\pair~l~m) & l\\
\snd~(\pair~l~m) & m\\
\si~(\cond~\true~p~q) & p\\
\si~(\cond~\false~p~q) & q\\[3mm]
\end{rew}\quad
&\quad\begin{rew}
\lesseq~\znat~y & \true\\
\lesseq~(\ss~x)~\znat & \false\\
\lesseq~(\ss~x)~(\ss~y) & \lesseq~x~y\\[3mm]
\end{rew}
\end{tabular}
\end{center}

\begin{center}
\begin{tabular}{cc}
\end{tabular}
\begin{rul}
\pivot~\nil~y &\a& \pair~\nil~\nil\\
\pivot~(\cons~l~x)~y &\a&
\si~(\cond~(\pair~(\cons~p_1~x)~p_2)~(\pair~p_1~(\cons~p_2~x))~(\lesseq~x~y))\\
&&\mbox{where }p_1=\fst~p, p_2=\snd~p, p=\pivot~l~y\\
\qs~\nil~m &\a& m\\
\qs~(\cons~l~x)~m &\a& \qs~p_1~(\cons~(\qs~p_2~m)~x)\\
&&\mbox{where }p_1=\fst~p, p_2=\snd~p, p=\pivot~l~x\\
\qsort~l &\a& \qs~l~\nil\\
\end{rul}
\end{center}

For the annotated types of constructors, we take the canonical
annotations except for $\pair:\sL_\al\A\sL_\al\A\sP_\al$ and
$\cond:\sP_\al\A\sP_\al\A\sB\A\sC_\al$. Hence, a term of type
$\sP_\al$ is a pair of lists of length smaller than or equal to $\al$.

Now, for function symbols, we take $\fst,\snd:\sP_\al\A\sL_\al$,
$\si:\sC_\al\A\sP_\al$, $\pivot:\sL_\al\A\sN\A\sP_\al$, which
expresses the fact that these functions are not size-increasing, and
$\lesseq:\sN_\al\A\sN\A\sB$, $\qs:\sL_\al\A\sL\A\sL$ and
$\qsort:\sL_\al\A\sL$.

We now detail the subject-reduction condition for the case of the last
rule of $\qs$ by taking $\G=[x:\sN,l:\sL_l,m:\sL]$,
$\vphi=\{(\al,\ts\,l)\}$, $\fst,\snd,\pivot\ltf\qs$ and the identity
for $\z^\qs$. Let ${\th}={\thfphi[\qs]}$. By (var), $\G\th x:\sN$,
$\G\th l:\sL_l$ and $\G\th m:\sL$. By (prec), $\G\th p:\sP_l$, $\G\th
p_1:\sL_l$ and $\G\th p_2:\sL_l$. Since $l\lta\ts\,l$, by (app-decr),
$\G\th\qs~p_2~m:\sL$. By (cons),
$\G\th\cons~(\qs~p_2~m)~x:\sL$. Finally, since $l\lta\ts\,l$, by
(app-decr) again, $\G\th\qs~p_1~(\cons~(\qs~p_2~m)~x):\sL$.

We proved the termination of this system. However, we cannot express
that $\qsort$ is not size-increasing, that is, take
$\qsort:\sL_\al\A\sL_\al$. To do so, we need a more precise type
system with existential quantifiers and constraints on size variables
where $\pivot$ can be given the type:\\\hsp[1cm]
$(\all\al)\sL_\al\A\sN\A(\ex\b)(\ex\g)(\al=\b+\g)\sL_\b\times\sL_\g$
\cite{blanqui06lpar-sbt}.\qed
\end{exa}

We now give an example using interpretation functions $\z^\sf_X$
different from the identity:

\begin{exa}[Reverse]\label{ex-rev}
  List reversal can be defined as follows \cite{huet82jcss}:
\begin{rewc}
\last~\nil~x & x\\
\last~(\cons~l~y)~x & \last~l~y\\
\revremlast~\nil~x & \nil\\
\revremlast~(\cons~l~y)~x & \rev~(\cons~(\rev~(\revremlast~l~y))~x)\\
\rev~\nil & \nil\\
\rev~(\cons~l~x) & \cons~(\revremlast~l~x)~(\last~l~x)\\
\end{rewc}

\noindent
where $\rev:\sL\A\sL$, $\revremlast:\sL\A\sN\A\sL$ and
$\last:\sL\A\sN\A\sN$.

Since we have a first-order data type, we can assume that
$\kh=\w$. Let $\tA$ be the size algebra with the constant $\mt{1}$
interpreted by 1 and the binary function symbol $\mt{+}$ interpreted
by the addition. Let $\lea$ and $\lta$ be $\le_\ext$ and $<_\ext$
respectively (cf. remark after Definition \ref{def-size-alg}).

Consider the 4th rule. Take $\cons:\sL_\al\A\sN\A\sL_{\al+1}$,
$\rev:\sL_\al\A\sL_\al$, $\revremlast:\sL_\al\A\sN\A\sL_\al$,
$\G=[x:\sN,y:\sN_y,l:\sL_l]$ and $\vphi=\{(\al,l+1)\}$. One can easily
check monotony, accessibility and minimality. We now check
subject-reduction. Let ${\th}={\thfphi[\revremlast]}$. For comparing
termination arguments, take $\rev\eqf\revremlast$, $\z^\rev(a)=2a$
(formally $a+a$) and $\z^\revremlast(a)=2a+1$. By (var), $\G\th
x:\sN$, $\G\th y:\sN$ and $\G\th l:\sL_l$. By (app-decr),
$\G\th\revremlast~l~y:\sN_l$ since
$\z^\revremlast(l)=2l+1\lta\z^\revremlast(l+1)=2(l+1)+1=2l+3$. By
(app-decr), $\G\th\rev~(\revremlast~l~y):\sN_l$ since
$\z^\rev(l)=2l<2l+3$. By (cons),
$\G\th\cons~(\rev~(\revremlast~l~y))~x:\sL_{l+1}$. Finally, by
(app-decr), we get $\G\th r:\sL_{l+1}$, where
$r=\rev~(\cons~(\rev~(\revremlast~l~y))~x)$, since
$\z^\rev(l+1)=2l+2<2l+3$.\qed
\end{exa}

We end this series of examples with one using non-standard constructor
size annotations:

\begin{exa}[Normalization of conditionals]
  Let $\sC$ be the sort of conditional expressions with the
  constructors $\at:\sC$ and $\si:\sC^3\A\sC$. Following
  \cite{boyer79book}, one can define a normalization function
  $\nm:\sC\A\sC$ as follows:

\begin{rewc}
\nm~\at & \at\\
\nm~(\si~\at~y~z) & \si~\at~(\nm~y)~(\nm~z)\\
\nm~(\si~(\si~u~v~w)~y~z) & \nm~(\si~u~(\nm~(\si~v~y~z))~(\nm~(\si~w~y~z)))\\
\end{rewc}

In \cite{paulson86jar} is given a measure on terms due to Shostak that
is decreasing in recursive calls. Hence, we can prove the termination
of $\nm$ by using the following annotated types: $\at:\sC_\al$,
$\si:\sC_\al\A\sC_\b\A\sC_\g\A\sC_{(\al+1)(\b+\g+3)}$ and
$\nm:\sC_\al\A\sC_\al$. One can easily check the monotony condition.

Now, for the 3rd rule, let
$\G=[u:\sC_u,v:\sC_v,w:\sC_w,y:\sC_y,z:\sC_z]$, $\vphi=\{(\al,a)\}$
where
$a=((u+1)(v+w+3)+1)(y+z+3))=uvy+uvz+uwy+uwz+3uv+3uw+3uy+3uz+vy+wy+vz+wz+9u+3v+3w+4y+4z+12$,
$\z^\nm$ be the identity, and ${\th}={\thfphi[\nm]}$. One can easily
check monotony, accessibility and minimality. We now check
subject-reduction. By (cons), $\G\th\si\,v\,y\,z:\sC_{(v+1)(y+z+3)}$
and $\G\th\si\,w\,y\,z:\sC_{(w+1)(y+z+3)}$. By (app-decr),
$\G\th\nm\,(\si\,v\,y\,z):\sC_{(v+1)(y+z+3)}$ since
$(v+1)(y+z+3)=vy+vz+y+z+3\lta a$, and
$\G\th\nm\,(\si\,w\,y\,z):\sC_{(w+1)(y+z+3)}$ since
$(w+1)(y+z+3)=wy+wz+y+z+3\lta a$. Finally, by (app-decr),
$\G\th\nm\,(\si\,u\,(\nm\,(\si\,v\,y\,z))\,(\nm\,(\si\,w\,y\,z))):\sC_b$
where $b=(u+1)((v+1)(y+z+3)+(w+1)(y+z+3)+3)$ since
$b=uvy+uvz+uwy+uwz+2uy+2uz+vy+vz+wy+wz+9u+2y+2z+9\lta a$. So, by
(sub),
$\G\th\nm\,(\si\,u\,(\nm\,(\si\,v\,y\,z))\,(\nm\,(\si\,w\,y\,))):\sC_a$.\qed
\end{exa}

\section{Decidability of $\thfphi$}
\label{sec-dec}

In this section, we provide an algorithm for deciding the relation
$\thfphi$ used in Theorem \ref{thm-sn} and defined in Figures
\ref{fig-thc} and \ref{fig-subtyping}, under general conditions on the
size algebra $\tA$. We will prove in Section \ref{sec-succ} that these
conditions are satisfied by the successor algebra.

The differences between $\thfphi$ and the usual typing relation for
simply-typed $\l$-calculus are the following. First, the set of
typable symbols is restricted to those smaller than or equivalent to
$\sf$. Second, the application of $t$ to $u$ is restricted to the
terms $t$ whose head is not an abstraction. Moreover, when the head of
$t$ is a symbol equivalent to $\sf$, the number of arguments must be
bigger than $\rq^\sf$ and the size of the arguments must be
decreasing.

If we remove the decreasingness condition, we get the relation $\thf$
defined by the same rules as those of $\thfphi$ except (app-decr)
replaced by:

  $$\mbox{(app)}\quad\cfrac{\begin{array}{c}
  (h,\vV\A V)\in\G\cup\o\T\quad
  h\ltf\sf\ou(h\eqf\sf\et|\vV|\ge\rq^h)\\
  \psi:\{\vec\al^h\}\a\o\tA\quad
  (\all i)\G\thf w_i:V_i\psi\\
\end{array}}{\G\thf h\,\vw:V\psi}$$

\noindent
that is (app-decr) without the decreasingness condition
$(h,\psi)\lta(\sf,\vphi)$. Hence, deciding $\G\thfphi t:T$ can be reduced
to finding a derivation of $\G\thf t:T$ where, at each (app) node, the
decreasingness condition is satisfied.

In Section \ref{sec-thd}, we provide an algorithm for deciding
$\thf$. Then, in Section \ref{sec-thc}, we show how to use this
algorithm to decide $\thfphi$.

\subsection{Decidability of $\thf$}
\label{sec-thd}

First note that, in a given typing environment $\G$, a typable term
$t$ may have several and even infinitely many types for two
reasons. First, in (app), the size variables of function types can be
instantiated arbitrarily. Second, subtyping is generally not
bounded. For instance, in the successor algebra,
$\sN_\al\le\sN_{\ts\al}\le\ldots$

The relation $\thf$ differs from Curry and Feys' typing relation with
functional characters or type-schemes (a type with type variables)
\cite{curry58book} in two points. First, our type-schemes are not
built from type variables but from size variables. Second, we have a
subtyping relation. We will however see that some techniques developed
for Curry and Feys' type system or, more generally, Milner's type
system \cite{milner78jcss} and its extensions, can be adapted to our
framework.

The decidability of type-checking in Curry and Feys' system has been
proved by Hindley in \cite{hindley69tams}. Hindley's algorithm is
based on the fact the set of types of a typable term has a smallest
element wrt some ordering $\qle$. Hence, to decide whether
$\G\th t:T$, the algorithm proceeds in two steps. First, it computes
the smallest type of $t$, say $U$, and then checks whether $U\qle T$.

In Curry and Feys' system, $\qle$ is the instantiation ordering: a
type-scheme $U$ is an instance of a type-scheme $T$, or $T$ is more
general than $U$, written $T\qle U$, if $T\t=U$ for some substitution
$\t$. In \cite{huet76hdr}, Huet proved that every non-empty set of
terms has a greatest lower bound wrt. $\qle$. So, in particular,
$\{T\mid\G\th t:T\}$ has a greatest lower bound if $t$ is typable in
$\G$.

For computing the most general type, Hindley uses an algorithm based
on unification \cite{herbrand30phd,robinson65jacm}. Unifying two terms
$T$ and $U$ consists in solving the equation $T=U$, that is, in
finding a substitution $\t$ such that $T\t=U\t$. In \cite{huet76hdr},
Huet proved that solving $T=U$ is equivalent to finding an
$\qle$-upper bound to both $T$ and $U$. He also showed that every
non-empty bounded set of terms has a least upper bound wrt.
$\qle$. Hence, every solvable unification problem has a most general
solution.

Hindley's work was later extended in many directions by considering
richer types, more complex constructions or by improving the algorithm
computing the most general type-scheme. One of the most advanced
generalizations seems to be Sulzmann's $\mr{HM}(X)$ system
\cite{sulzmann01flops}, where the type variables of a type-scheme are
required to satisfy a formula of an abstract constraint system
$X$. For his system, Sulzmann provides a generic constrained-type
inference algorithm assuming a procedure for solving constraints in
$X$. It would be interesting to study whether our framework can fit in
this general setting. However, in this paper, we will simply follow
Hindley's approach.

But, since we also have subtyping, we define $\qle$ as follows:

\begin{dfn}[More general type]
  We say that an annotated type $T$ is {\em more general than} another
  annotated type $U$, written $T\qle U$, if there is a substitution
  $\t$ such that $T\t$ is a subtype of $U$, \ie $T\t\le U$.
\end{dfn}

One can easily check that $\qle$ is a quasi-ordering.

\begin{dfn}[Subtyping problem]
  \label{def-subtyp-pb}
  A {\em subtyping problem} $P$ is either $\bot$ or a finite set of
  subtyping constraints, a subtyping constraint being a pair of types
  $(T,U)$ written $T\le^?U$. It has a solution $\vphi:\tV\a\o\tA$ if
  $P\neq\bot$, $\dom(\vphi)\sle\Var(P)$ and, for all $T\le^?U\in P$,
  $T\vphi\le U\vphi$. Let $\Sol_{\o\tA}(P)$ be the set of all the
  solutions of $P$. A solution $\vphi$ is {\em more general than}
  another solution $\psi$, written $\vphi\qle\psi$, if there is $\t$
  such that $\vphi\t\leai\psi$, \ie there is $\t$ such that, for all
  $\al$, $\al\vphi\t\leai\al\psi$. Finally, let ${\equiv}$ be the
  equivalence relation ${\qle\cap\qge}$.
\end{dfn}

Again, one can easily check that the ordering $\qle$ on substitutions
is a quasi-ordering.

In order to compute the most general type of a term, we make the
following assumptions:
\begin{itemize}
\item every solvable subtyping problem $P$ has a most general solution
  $\mgs(P)$;
\item there is an algorithm for deciding whether a subtyping problem
  is solvable and, if so, computing its most general solution.
\end{itemize}

We will see in Section \ref{sec-succ} that these assumptions are
satisfied when types are annotated in the successor algebra. On the
other hand, they are not generally satisfied in an algebra with
addition.

\begin{figure}
  \figrule
\caption{Type inference algorithm\label{fig-th-inf}}\vsp[3mm]
\normalsize
\begin{center}
  \begin{minipage}{15mm}\vsp[-4mm](inf-lam)\end{minipage}
  $\cfrac{\G,x:U\thf v\au V}{\G\thf\l x^Uv\au U\A V}$\\[5mm]

  \begin{minipage}{15mm}\vsp[23mm](inf-app)\end{minipage}
  $\cfrac{\begin{array}{c}
(h,\vV\A V)\in\G\cup\o\T\quad
h\ltf\sf\ou(h\eqf\sf\et|\vV|\ge\rq^h)\\[1mm]

(\all i)\G\thf w_i\au U_i\\[1mm]

\r_1,\ldots,\r_n\mbox{ permutations on $\tV$ } (n=|\vV|=|\vw|)\\
(\all i)\Var(U_i\r_i)\cap\Var(\vV\A V)=\vide\\
(\all i)(\all j)i\neq j\A\Var(U_i\r_i)\cap\Var(U_j\r_j)=\vide\\[2mm]

\eta=\mgs(\{U_1\r_1\le^?V_1,\ldots,U_n\r_n\le^?V_n\})\\
\end{array}}{\G\thf h\,\vw\au V\eta}$
\end{center}
\figrule
\end{figure}

Now, following \cite{hindley69tams}, the computation of the most
general type is defined by the rules of Figure \ref{fig-th-inf} where
$\G\thf t\au U$ means that, in the typing environment $\G$,
the most general type of $t$ is $U$.
In the case of an application $h\,\vw$, the algorithm proceeds as follows:
\begin{enumerate}
\item Check that $h$ is declared. Let $T$ be its declared type.
\item Check that $h$ can take $n=|\vw|$ arguments,
\ie $T$ is of the form $\vV\A V$ with $|\vV|=n$.
\item If $h$ is a function symbol equivalent to $\sf$,
  check that $|\vV|\ge\rq^h$.
\item Try to infer the types of every $w_i$.
\item If this succeeds with $U_i$ for the type of $w_i$, then rename
  the variables of every $U_i$ using a permutation $\r_i$, so that,
  for all $i$, $U_i\r_i$ has no variable in common with $T$ and, for
  all $i\neq j$, $U_i\r_i$ and $U_j\r_j$ have no variable in common.
\item Finally, try to compute the most general solution $\eta$
  of the problem $\{U_1\r_1\le^?V_1,\ldots,\linebreak U_n\r_n\le^?V_n\}$
  and return $V\eta$ in case of success.
\end{enumerate}


\begin{exa}\label{ex-inf-div}
  To carry on with Example \ref{ex-div}, let
  $r=\ss~(\div~(\sub~x~y)~(\ss~y))$ be the right hand-side of
  the last rule of Figure \ref{fig-div}. We would like to infer the
  type of $r$ in $\G=[x:\sN_x,y:\sN]$ when
  $\ss:\sN_\al\A\sN_{\ts\al}$, $\sub:\sN_\al\A\sN\A\sN_\al$ and
  $\div:\sN_\al\A\sN\A\sN_\al$. Let ${\th}={\thf[\div]}$ and assume
  wlog that $x$ is a constant of the successor algebra.

  By (inf-app), we get (1) $\G\th x\au\sN_x$ and (2) $\G\th
  y\au\sN$. From (1) and (2), by (inf-app), we get (3)
  $\G\th\sub~x~y\au\sN_x$ since, as we shall see in Section
  \ref{sec-mgs},
  $\mgs\{\sN_x\le^?\sN_\al,\sN\le^?\sN\}=\{(\al,x)\}$. From (2), by
  (inf-app), we get (4) $\G\th\ss~y\au\sN$ since
  $\mgs\{\sN\le^?\sN_\al\}=\{(\al,\infty)\}$. From (3) and (4), we get
  (5) $\G\th\div~(\sub~x~y)~(\ss~y)\au\sN_x$ since
  $\mgs\{\sN_x\le^?\sN_\al,\sN\le^?\sN\}=\{(\al,x)\}$. From (5), by
  (inf-app), we get $\G\th r\au\sN_{\ts{}x}$ since
  $\mgs\{\sN_x\le^?\sN_\al\}=\{(\al,x)\}$.\qed
\end{exa}

We now prove that this algorithm is correct and complete wrt $\thf$,
when the size algebra is monotone and the algorithm is applied to an
environment $\G$ having no size variables. To extend in the next
section this result to $\thfphi$, we need to make derivations
explicit:

\begin{dfn}[Derivation]
Derivations of $\G\thf t:T$ are defined as follows:
\begin{itemize}
\item If $(h,\vV\A V)\in\G\cup\o\T$ and, for all $i$, $\pi_i$ is a
  derivation of $\G\thf w_i:V_i\psi$, written
  $\pi_i\tgt\G\thf w_i:V_i\psi$, then $\ra(\G,h\,\vw,\psi,\vec\pi)$ is
  the derivation of $\G\thf h\,\vw:V\psi$ whose last rule is (app).
\item If $\pi\tgt\G,x:U\thf v:V$, then $\rl(\pi)$ is the derivation of
  $\G\thf\l x^Uv:U\A V$ whose last rule is (lam).
\item If $\pi\tgt\G\thf t:U$ and $U\le V$, then $\rs(\pi,V)$ is the
  derivation of $\G\thf t:V$ whose last rule is (sub).
\end{itemize}

Simarly, derivations of $\G\thf t\au T$ are defined as follows:
\begin{itemize}
\item If $(h,\vV\A V)\in\G\cup\o\T$, $\vec\r$ are permutations
  satisfying the conditions of (inf-app) and, for all $i$,
  $\pi_i\tgt\G\thf w_i\au U_i$, then $\ri(\G,h\,\vw,\vec\r,\vec\pi)$ is
  the derivation of $\G\thf h\,\vw\au V\eta$ whose last rule is
  (inf-app).
\item If $\pi\tgt\G,x:U\thf v\au V$, then $\rl(\pi)$ is the derivation
  of $\G\thf\l x^Uv\au U\A V$ whose last rule is (inf-lam).
\end{itemize}

Given a derivation $\pi$ for $\pi\tgt\G\thf t:T$, we write
$\pi\tgt\G\thfphi t:T$ if, at every (app) node in $\pi$, the
decreasingness condition of (app-decr) is satisfied, that is, if
$\G\thfphi t:T$.
\end{dfn}

Note that $\G\thf t\au T$ has at most one derivation.

\begin{lem}
\label{lem-subs}
If $\pi\tgt\G\thf t:T$ then, for every size substitution $\vphi$,
$\pi\vphi\tgt\G\vphi\thf t:T\vphi$.
\end{lem}

\begin{prf}
  Straightforward induction using the fact that $\lea$ and thus
  $\leai$ and $\le$ are stable by substitution. In the case of (app),
  by induction hypothesis, we have $\G\vphi\thf w_i:V_i\psi'$ with
  $\psi'=\psi\vphi$. Therefore, by (app), $\G\vphi\thf
  h\,\vw:V\psi'=(V\psi)\vphi$.\qed
\end{prf}

\begin{lem}[Correctness wrt. $\thf$]
\label{lem-correct}
If $\pi\tgt\G\thf t\au T$ and $\Var(\G)=\vide$, then there is
$|\pi|$ such that $|\pi|\tgt\G\thf t:T$. In particular, for (inf-app),
$|\ri(\G,h\,\vw,\vec\r,\vec\pi)|=\ra(\G,h\,\vw,\eta,\vec\upsilon)$
where $\upsilon_i=\rs(|\pi_i|\r_i\eta,V_i\eta)$.
\end{lem}

\begin{prf}
  By induction on $\G\thf t\au T$. We only detail the case
  (inf-app). By induction hypothesis, $\G\thf w_i:U_i$. By Lemma
  \ref{lem-subs}, $\G\thf w_i:U_i\r_i\eta$ since
  $\Var(\G)=\vide$. Since $U_i\r_i\eta\le V_i\eta$, by (sub), $\G\thf
  w_i:V_i\eta$. Therefore, by (app), $\G\thf h\,\vw:V\eta$.\qed
\end{prf}

\begin{lem}
\label{lem-pos}
  If $\vphi\leai\psi$ and, for all $\al$, $\Pos(\al,T)\sle\Pos^+(T)$,
  then $T\vphi\le T\psi$.
\end{lem}

\begin{prf}
  We say that $T\in\bT_\tA\cup\tA$ is $\d\in\{+,-\}$ if, for all
  $\al$, $\Pos(\al,T)\sle\Pos^\d(T)$. We first prove that (*) if $a$
  is $\d$ then $a\vphi(\leai)^\d a\psi$ where, for any relation $R$,
  ${R^+}={R}$ and ${R^-}={R^{-1}}$. We proceed by induction on $a$:
  \begin{itemize}
  \item $a$ is a variable $\al$. Then, $\d=+$ and
    $\al\vphi\leai\al\psi$ since $\vphi\leai\psi$.
  \item $a=\tf a_1\ldots a_n$. Let $i\in\{1,\ldots,n\}$. If
    $i\in\Mon^\vep(\tf)$ (cf. Definition \ref{def-annot-types}), then
    $a_i$ is $\d\vep$ and, by induction hypothesis,
    $a_i\vphi(\leai)^{\d\vep}a_i\psi$. If
    $i\notin\Mon^+(\tf)\cup\Mon^-(\tf)$ then $a_i$ contains no
    variable and $a_i\vphi=a_i\psi$. Therefore, by monotony of $\tf$
    in every $i\in\Mon^+(\tf)$, anti-monotony of $\tf$ in every
    $i\in\Mon^-(\tf)$, and transitivity, we get $a\vphi(\leai)^\d
    a\psi$.
  \end{itemize}
  We now prove that, if $T$ is $\d$, then $T\vphi\le^\d T\psi$, by
  induction on $T$.
  \begin{itemize}
  \item $T=U\A V$. Then, $U$ is $-\d$ and $V$ is $\d$. So, by
    induction hypothesis, $U\vphi\le^{-\d}U\psi$ and $V\vphi\le^\d
    V\psi$. Therefore, by (prod), $(U\a V)\vphi\le^\d(U\A V)\psi$.
  \item $T=\sB_a$. Then, $a$ is $\d$ and, by (*), $a\vphi\leai
    a\psi$. Therefore, by (size), $T\vphi\le T\psi$.\cqfd
  \end{itemize}
\end{prf}

\newcommand\tha{\thf_\ra}

\begin{lem}[Completeness wrt. $\thf$]
\label{lem-complete}
  In monotone algebras, if $\pi\tgt\G\thf t:T$ and
  ${\Var(\G)\!=\!\vide}$, then there are $U$ and $\pi\au$ such that
  $\pi\au\tgt\G\thf t\au U$ and $U\qle T$. In particular,
  ${\ra(\G,h\,\vw,\psi,\vec\pi)\!\au}\linebreak
  =\ri(\G,h\,\vw,\vec\r,\vec\pi\au)$ where $\vec\r$ are permutations
  satisfying the conditions of rule (inf-app).
\end{lem}

\begin{prf}
  We proceed by induction on $\G\thf t:T$. We only detail the case
  (app) when $h\in\bC\cup\bF$. By induction hypothesis, $\G\thf w_i\au
  U_i$ and there is $\chi_i$ such that $U_i\chi_i\le V_i\psi$. Wlog.
  we can assume that $\dom(\chi_i)\sle\Var(U_i)$. Let now
  $\r_1,\ldots,\r_n$ be permutations satisfying the conditions of
  (inf-app), and $\xi=\{(\al,\al\psi)\mid\al\in\Var(\vV\A
  V)\}\cup\{(\al,\al\r_i^{-1}\chi_i)\mid\al\in\Var(U_i\r_i),1\le i\le
  n\}$. Then, for all $i$, $U_i\r_i\xi=U_i\chi_i\le
  V_i\psi=V_i\xi$. Therefore,
  $P=\{U_1\r_1\le^?V_1,\ldots,U_n\r_n\le^?V_n\}$ is solvable,
  $\eta=\mgs(P)$ exists and there is $\chi$ such that
  $\eta\chi\leai\xi$. Hence, by (inf-app), $\G\thf h\,\vw\au
  V\eta$. By the monotony condition, variables occur only positively
  in $V$. Therefore, by Lemma \ref{lem-pos}, $V\eta\chi\le
  V\xi=V\psi$. Hence, $V\eta\qle V\psi$.\qed
\end{prf}

\subsection{Decidability of $\thfphi$}
\label{sec-thc}

We now prove that, when the size algebra is monotone, for checking
$\G\thfphi t:T$, it is sufficient to check whether there are $U$ and
$\chi$ such that $\G\thf t\au U$, $U\chi\le T$ and also that, if one
denotes by $\upsilon$ the (unique) derivation of $\G\thf t\au U$, then
$|\upsilon|\chi\tgt\G\thfphi t:U\chi$, that is, at every (app) node in
$|\upsilon|\chi$, the decreasingness condition is satisfied.

\begin{lem}
\label{lem-thc-mon}
In monotone algebras, if $\pi\tgt\G\thf t:T$, $\pi\xi'\tgt\G\thfphi
t:T\xi'$ and $\xi\leai\xi'$, then $\pi\xi\tgt\G\thfphi t:T\xi$.
\end{lem}

\begin{prf}
  By induction on $\pi\tgt\G\thf t:T$. We only detail the case (app)
  when $h\eqf\sf$. We have $(h,\vV\A V)\in\o\T$, $\pi_i\tgt\G\thf
  w_i:V_i\psi$, $\pi_i\xi'\tgt\G\thfphi w_i:V_i\psi\xi'$ and
  $\vec\al^h\psi\xi'<^{h,\sf}_\tA\vec\al^\sf\vphi$. By induction
  hypothesis, $\pi_i\xi\tgt\G\thfphi w_i:V_i\psi\xi$. Since
  $\xi\leai\xi'$ and the size algebra is monotone, we have
  $\psi\xi\leai\psi\xi'$. Since
  ${(\leai)_\prod\circ<^{h,\sf}_\tA}\sle{<^{h,\sf}_\tA}$
  (cf. Definition \ref{def-call-ord}), we have
  $\vec\al^h\psi\xi<^{h,\sf}_\tA\vec\al^\sf\vphi$. Therefore,
  $\pi\xi\tgt\G\thfphi h\,\vw:V\psi\xi$.\qed
\end{prf}

\begin{lem}[Completeness wrt. $\thfphi$]\label{lem-complete-decr}
  Let $\tA$ be a monotone algebra. Assume that $\pi\tgt\G\thfphi t:T$
  and $\Var(\G)=\vide$. By lemma \ref{lem-complete}, there are $U$ and
  $\chi$ such that $\pi\au\G\thf t\au U$ and $U\chi\le T$. Then,
  $|\pi\au|\chi\tgt\G\thfphi t:U\chi$.
\end{lem}

\begin{prf}
  We prove that if $\pi\tgt\G\thfphi t:T$, $\Var(\G)=\vide$,
  ${\pi\!\au}\tgt{\G\thf t\au U}$ and $U\chi\le T$, then
  $\rs(|{\pi\!\au}|\chi,T)\tgt\G\thfphi t:T$, by induction on
  $\pi\tgt\G\thfphi t:T$. We only detail the case (app-decr) when
  $t=h\,\vw$, $(h,\vV\A V)\in\o\T$, $T=V\psi$,
  $\vec\al^h\psi<^{h,\sf}_\tA\vec\al^\sf\vphi$ and $U=V\eta$ where
  $\eta$ is given by the rule (inf-app). We have
  $\pi=\ra(\G,h\,\vw,\psi,\vec\pi)$,
  ${\pi\!\au}=\ri(\G,h\,\vw,\vec\r,{\vec\pi\!\au})$,
  ${|{\pi\!\au}|}=\ra(\G,h\,\vw,\eta,\vec\upsilon)$ where
  $\upsilon_i=\rs(|{\pi_i\!\au}|\r_i\eta,V_i\eta)$,
  $|{\pi\!\au}|\chi=\ra(\G,h\,\vw,\eta\chi,\vec\upsilon\chi)$ and, for
  all $i$, $\pi_i\tgt\G\thfphi w_i:V_i\psi$, ${\pi_i\!\au}\tgt\G\thf
  w_i\au U_i$ and $U_i\chi_i\le V_i\psi$ for some $\chi_i$. By
  induction hypothesis,
  $\rs(|{\pi_i\!\au}|\chi_i,V_i\psi)\tgt\G\thfphi w_i:V_i\psi$. In
  particular, $|{\pi_i\!\au}|\chi_i\tgt\G\thfphi w_i:U_i\chi_i$, that
  is, $|{\pi_i\!\au}|\r_i\xi\tgt\G\thfphi w_i:U_i\r_i\xi$, where $\xi$
  is defined in the proof of Lemma \ref{lem-complete}. Since
  $\eta\chi\leai\xi$, by Lemma \ref{lem-thc-mon}, we get
  $|{\pi_i\!\au}|\r_i\eta\chi\tgt\G\thfphi
  w_i:U_i\r_i\eta\chi$. Hence, $\upsilon_i\chi\tgt\G\thfphi
  w_i:V_i\eta\chi$. Moreover, since
  $\vec\al^h\eta\chi\leai\vec\al^h\xi=\vec\al^h\psi$ and
  $\vec\al^h\psi<^{h,\sf}_\tA\vec\al^\sf\vphi$, we get
  $\vec\al^h\eta\chi<^{h,\sf}_\tA\vec\al^\sf\vphi$ by assumption on
  $<^{h,\sf}_\tA$. Therefore, $\rs(|{\pi\!\au}|\chi,T)\tgt\G\thfphi
  t:T$.\qed\\
\end{prf}

The previous lemmas assume that there are no size variables in
$\G$. So, to use these lemmas, we need to be able to replace size
variables by constants ({\em aka} eigenvariables). Under this
assumption, we can conclude:

\begin{figure}
  \figrule
    \caption{Algorithm for deciding whether $\G\thfphi t:T$.\label{fig-algo}}
  \normalsize
  \begin{enumerate}
  \item Check whether there is $U$ such that $\G\g\thf t\au U$, where
    $\g$ is an injection from the set of size variables of $\G$, $T$
    and $\vphi$ to the set of constants $\tC$ not occurring in $\G$,
    $T$ and $\vphi$.\\If it fails, then $t$ is not typable in $\G$.
  \item If it succeeds, try to compute $\chi=\mgs\{U\le^?T\g\}$. If it
    fails, $\G\thf t:T$ does not hold.
  \item If it succeeds, then check whether
    $|\upsilon|\chi\tgt\G\g\th^\sf_{\vphi\g} t:U\chi$ where $\upsilon$
    is the unique derivation of $\G\g\thf t\au U$.\\If it succeeds,
    then $\G\thfphi t:T$ holds. Otherwise, $\G\thfphi t:T$ does not
    hold.
\end{enumerate}
  \figrule
\end{figure}

\begin{thm}[Decidability of $\thfphi$]
  Assume that $\tA$ is an algebra such that:
  \begin{itemize}
  \item $\tA$ is monotone;
  \item $\tA$ contains an infinite set of constants $\tC$ such that,
    if $a\lea b$ ($\va\lta^{\sg,\sf}\vb$ resp.) then, for all
    $\tc\in\tC$ and $e\in\tA$, $a\d\lea b\d$
    ($\va\d\lta^{\sg,\sf}\vb\d$ resp.), where $\d$ replaces every
    $\tc$ by $e$;
  \item $\ltf$ is decidable and, for all $\sg\eqf\sf$,
    $\lta^{\sg,\sf}$ is decidable;
  \item the satisfiability of a subtyping problem is decidable;
  \item every satisfiable problem $P$ has a most general
    solution $\mgs(P)$ that is computable.
  \end{itemize}
  Given $\G$, $t$ and $T$, one can decide whether $\G\thfphi t:T$
  by using the algorithm of Figure \ref{fig-algo}.
\end{thm}

\pagebreak
\begin{prf}
  \begin{itemize}
    \item Correctness. Assume that the algorithm succeeds. Then,
      $\G\g\th^\sf_{\vphi\g} t:U\chi$ and $U\chi\le T\g$. By (sub),
      $\G\g\th^\sf_{\vphi\g} t:T\g$. Then, by applying $\d=\g^{-1}$,
      we get $\G\thfphi t:T$.
    \item Completeness. Assume that the algorithm fails in step 1 or 2
      then, by Lemma \ref{lem-complete}, $t$ is not typable in
      $\G\g$. Therefore, it is not typable in $\G$ either. Finally, if
      the algorithm fails in step 3 then, by Lemma
      \ref{lem-complete-decr}, there is no derivation of
      $\G\g\th^\sf_{\vphi\g} t:T\g$. Therefore, there is no derivation
      of $\G\thfphi t:T$ either.\qed
  \end{itemize}
\end{prf}

That the successor algebra satisfies the first two conditions follows
from Lemma \ref{lem-lea}.

\begin{exa}
  To carry on with Example \ref{ex-inf-div}, we now would like to
  check whether $\G\th r:\sN_\al\vphi$ where ${\th}={\thfphi[\div]}$
  and $\vphi=\{(\al,\ts\,x)\}$. We have seen that $\G\th^\div
  r\au\sN_{\ts\,x}$. Hence, $\chi$ is the identity and we are left to
  check that, in every (app) node with $h\eqf\div$, the decreasingness
  condition is satisfied. Here, it amounts to check that, in the (app)
  node for $(\div~(\sub~x~y)~(\ss~y))$, the size annotation for the
  type of $\sub~x~y$, that is $x$, is smaller than $\al\vphi=\ts\,x$,
  which is indeed the case.\qed
\end{exa}

\section{Reducing subtyping problems to size problems}
\label{sec-sub}

For the type inference algorithm we just saw, we assumed the existence
of an algorithm to compute the most general solution of a subtyping
problem. In this section, we show how a subtyping problem can be
reduced to solving constraints in $\o\tA$. As subtyping is
not syntax-directed, we first prove that it is equivalent to a
syntax-directed relation. To this end, we prove that the rules (refl)
and (trans) are redundant, that is, they can be eliminated, following
a proof technique used by Curien and Ghelli in \cite{curien92mscs}:

\begin{thm}
  $T\le U$ iff $T\le_\ra U$, where $\le_\ra$ is defined by the
  rules (size) and (prod) only.
\end{thm}

\begin{prf}
  We first prove that (refl) can be eliminated, hence that
  ${\le}={\le'}$ where $\le'$ is the relation defined by (size),
  (prod) and (trans) only. Indeed, using the reflexivity of $\leai$,
  one can easily prove that $T\le_\ra T$, by induction on $T$.

  We now prove that, in turn, (trans) can be eliminated, hence that
  ${\le}={\le_\ra}$. More precisely, we prove that, if $\pi$ is a
  derivation of $A\le'B$ of height $n$, then $A\le_\ra B$, by
  induction on $n$. We proceed by case on the last rule:
  \begin{itemize}
  \item[(size)] Immediate.
  \item[(prod)] Assume that $U\A V\le'U'\A V'$ ends with (prod). By
    induction hypothesis, $U'\le_\ra U$ and $V\le_\ra V'$. Hence, by
    (prod), $U\A V\le_\ra U'\A V'$.
  \item[(trans)] Assume that $T\le'U$ and $U\le'V$. By induction
    hypothesis, $T\le_\ra U$ and $U\le_\ra V$. If $T\le_\ra U$ ends
    with (size), then $T=\sB_a$, $U=\sB_b$ and $a\leai b$. Therefore,
    $U\le_\ra V$ ends with (size) too, $V=\sB_c$ and $b\leai
    c$. Hence, by transitivity of $\leai$, $T\le_\ra V$. Similarly, if
    $U\le_\ra V$ ends with (size), then $T\le_\ra U$ ends with (size)
    and $T\le_\ra V$. So, we are left with the case where both
    $T\le_\ra U$ and $U\le_\ra V$ ends with (prod):

\begin{center}
\begin{minipage}{9cm}
\begin{ded}
\[\[\pi_{11}\justifies A'\le_\ra A\]\quad\[\pi_{12}\justifies B\le_\ra B'\]
\justifies A\A B\le_\ra A'\A B'\using\mbox{(prod)}\]
\quad\[\[\pi_{21}\justifies A''\le_\ra A'\]\quad\[\pi_{22}\justifies B'\le_\ra B''\]
\justifies A'\A B'\le_\ra A''\A B''\using\mbox{(prod)}\]
\justifies A\A B\le' A''\A B''\using\mbox{(trans)}
\end{ded}
\end{minipage}
\end{center}

\noindent
But $A\A B\le' A''\A B''$ can also be proved as follows:

\begin{center}
\begin{minipage}{9cm}
\begin{ded}
\[\[\pi_{21}\justifies A''\le_\ra A'\]\quad\[\pi_{11}\justifies A'\le_\ra A\]
\justifies A''\le' A\using\mbox{(trans)}\]
\quad\[\[\pi_{12}\justifies B\le_\ra B'\]\quad\[\pi_{22}\justifies B'\le_\ra B''\]
\justifies B\le' B''\using\mbox{(trans)}\]
\justifies A\A B\le' A''\A B''\using\mbox{(prod)}
\end{ded}
\end{minipage}
\end{center}

The derivation heights of $A''\le'A$ and $B\le'B''$ are strictly
smaller than the derivation height of $A\A B\le' A''\A
B''$. Therefore, by induction hypothesis, $A''\le_\ra A$ and $B\le_\ra
B''$. Hence, by (prod), $A\A B\le_\ra A''\A B''$.\qed
  \end{itemize}
\end{prf}

As a consequence, we can prove that a subtyping problem can be reduced
to an equivalent size problem as follows:

\begin{dfn}[Size problem]\label{def-A-sol}
  A size constraint is a pair of size expressions $(a,b)$, written
  $a\le^?b$. A {\em size problem} $P$ is either $\bot$ or a finite set
  of size constraints. It has a solution $\vphi:\tV\a\o\tA$ if
  $P\neq\bot$, $\dom(\vphi)\sle\Var(P)$ and, for all $a\le^?b\in P$,
  $a\vphi\leai b\vphi$. A solution $\vphi$ is {\em finite} if
  $\vphi:\tV\a\tA$. Let $\Sol_{\o\tA}(P)$ ($\Sol_\tA(P)$ resp.) be the
  set of the ({\em finite} resp.) solutions of $P$.
  
  We define the size problem associated to a subtyping problem as follows:
\begin{itemize}
\item $|\vide|=\vide$,
\item $|P\cup Q|=|P|\cup|Q|$ if $|P|\neq\bot$ and $|Q|\neq\bot$,
\item $|\{\sB_a\le^?\sB_b\}|=\{a\le^?b\}$,
\item $|\{U\A V\le^?U'\A V'\}|=|\{U'\le^?U,V\le^?V'\}|$,
\item $|P|=\bot$ otherwise.
\end{itemize}
\end{dfn}

\begin{lem}
$\Sol(P)=\Sol_{\o\tA}(|P|)$.
\end{lem}

\begin{prf}
  We proceed by induction on $P$. We only detail the case where
  $P=\{T\le^?T'\}$:
\begin{itemize}
\item Let $\vphi\in\Sol(P)$. Then, $T\vphi\le_\ra T'\vphi$. If
  $T=\sB_a$, then $T'=\sB_b$ and
  $\vphi\in\Sol_{\o\tA}(\{a\le^?b\})=\Sol_{\o\tA}(|P|)$. Otherwise,
  $T=U\A V$, $T'=U'\A V'$ and
  $\vphi\in\Sol(\{U'\le^?U,V\le^?V'\})$. By induction hypothesis,
  $\vphi\in\Sol_{\o\tA}(|U'\le^?U|)\cap\Sol_{\o\tA}(|V\le^?V'|)=\Sol_{\o\tA}(|P|)$.
\item Let $\vphi\in\Sol_{\o\tA}(|P|)$. If $T=\sB_a$, then $T'=\sB_b$
  and $\vphi\in\Sol(P)$. Otherwise, $T=U\A V$, $T'=U'\A V'$,
  $\vphi\in\Sol_{\o\tA}(|U'\le^?U|)\cap\Sol_{\o\tA}(|V\le^?V'|)$. By
  induction hypothesis,
  $\vphi\in\Sol(U'\le^?U)\cap\Sol(V\le^?V')=\Sol(P)$.\qed
\end{itemize}
\end{prf}

To go further, we need to make more assumptions on the size algebra.

\section{Solving size problems in the successor algebra}
\label{sec-succ}

We have seen in the previous section that solving a subtyping problem
can be reduced to solving inequalities in $\o\tA$. In this
section, we consider a specific size algebra $\tA$, the successor
algebra, and prove that, in this algebra, the solvability of a size
problem is decidable in polynomial time, and that solvable size
problems have a most general solution that can be computed in
polynomial time too.

The relations $\lea$ and $\lta$ of the successor algebra (Definition
\ref{def-succ-alg}) are equivalently defined by the rules of Figure
\ref{fig-succ}. We start by proving basic properties of $\lea$, the
quasi-ordering $\qle$ and its associated equivalence relation $\equiv$
on size substitutions introduced in Definition \ref{def-subtyp-pb}.

\begin{figure}
  \figrule
\caption{Ordering in the successor algebra\label{fig-succ}}
\normalsize
$$\cfrac{}{a\lea a}
\quad
\cfrac{a\lta b}{a\lea b}
\quad
\cfrac{}{a\lta\ts\,a}
\quad
\cfrac{a\lta b\quad b\lta c}{a\lta c}$$
\figrule
\end{figure}

\begin{lem}\label{lem-lea}\hfill
\begin{itemize}
\item $a\lea b$ ($a\lta b$ resp.) iff there is $k\ge 0$ ($k>0$ resp.)
  such that $b=\ts^ka$.
\item $\ts a\lta\ts b$ iff $a\lta b$.
\end{itemize}
\end{lem}

\begin{prf}
\begin{itemize}
\item One can easily check $a\lea\ts^ka$ by induction on $k\ge 0$. We
  have $a\lea a$ by definition. Assume now that $a\lea\ts^ka$. Since
  $\ts^ka\lta\ts^{k+1}a$ holds by definition, we get $a\lea\ts^{k+1}a$
  by transitivity.

  Similarly, once can easily check $a\lta\ts^ka$ by induction on $k\ge
  1$. We have $a\lta\ts a$ by definition. Assume now that
  $a\lta\ts^ka$. Since $\ts^ka\lta\ts^{k+1}a$ by definition, we get
  $a\lta\ts^{k+1}a$ by transitivity.

  We now prove that, if $a\lta b$, then there is $b'$ such that
  $b=\ts b'$ and $a\lea b'$, by induction on the derivation height of
  $a\lta b$. If $b=\ts a$, then this is immediate. Otherwise, there is
  $c$ such that $a\lta c$ and $c\lta b$. By induction hypothesis,
  there is $b'$ such that $b=\ts b'$ and $c\lea b'$. Therefore, $a\lea
  b'$ since $\lea$ is the reflexive closure of $\lta$ and $\lta$ is
  transitive.

  We finally prove that there is $k\ge 0$ whenever $a\lea b$, by
  induction on $b$. If $a=b$, this is immediate. If $a\lta b$, then
  there is $b'$ such that $b=\ts b'$ and $a\lea b'$. By induction
  hypothesis, $b'=\ts^ka$ for some $k\ge 0$. Therefore,
  $b=\ts^{k+1}a$.

\item If $\ts a\lta\ts b$, then $\ts b=\ts^{k+1}\ts a$ for some
  $k$. Therefore, $b=\ts^{k+1}a$. Conversely, if $a\lta b$, then
  $b=\ts^{k+1}a$ for some $k$. Therefore, $\ts b=\ts^{k+1}\ts a$.\qed
\end{itemize}
\end{prf}

It follows that the successor algebra is monotone and also that $\lea$
and $\leai$ are orderings, as well as their pointwise extensions to
substitutions.

\begin{dfn}[Successor and head parts of a substitution]
  \label{def-succ-head-subs}
  To a substitution $\vphi:\tV\a\o\tA$, we associate two unique maps
  $\vphi_s:\tV\a\bN$ and $\vphi_h:\tV\a\tV\cup\tC\cup\{\infty\}$ such
  that, for all $\al$, $\al\vphi=\ts^{\al\vphi_s}\al\vphi_h$ with
  $\al\vphi_s=0$ if $\al\vphi_h=\infty$.
\end{dfn}

\begin{lem}\label{lem-qle-succ}
  $\vphi\qle\psi$ iff there is $\r:\tV\a\tV\cup\tC\cup\{\infty\}$
  such that $\vphi\r\leai\psi$.
\end{lem}

\begin{prf}
  The ``if'' part is immediate. We now prove the ``only if'' part.
  Assume that there is $\t$ such that $\vphi\t\leai\psi$. Let
  $\r=\t_h|_{\Var(\vphi_h)}$, where
  $\Var(\vphi_h)=\bigcup\{\Var(\al\vphi_h)\mid\al\in\dom(\vphi_h)\}$. We
  now check that $\vphi\r\leai\psi$. If $\al\vphi_h\notin\tV$, then
  $\al\vphi\r=\al\vphi\t\leai\al\psi$. Otherwise,
  $\al\vphi\r=\ts^{\al\vphi_s}\al\vphi_h\t_h\leai\ts^{\al\vphi_s+\al\vphi_h\t_s}\al\vphi_h\t_h=\al\vphi\t\leai\al\psi$.\qed
\end{prf}

\begin{lem}\label{lem-inj-ext}
  Let $V$ be a set, and $V_1$ and $V_2$ be subsets of $V$. If
  $\r_1:V_1\a V_2$ and $\r_2:V_2\a V_1$ are injections, then there is
  a permutation $\xi:V\a V$ such that $\xi|_{V_1}=\r_1$.
\end{lem}

\begin{prf}
  By Cantor-Bernstein theorem, $V_1$ and $V_2$ are equipotent. Hence,
  $V_1-V_2$ and $V_2-V_1$ are equipotent as well. Let $\nu$ be any
  bijection from $V_2-V_1$ to $V_1-V_2$, and
  $\xi=\{(\al,\al\r_1)\mid\al\in V_1\}\cup\{(\al,\al\nu)\mid\al\in
  V_2-V_1\}$. The function $\xi$ is a bijection on $V_1\cup V_2$ and
  $\xi|_{V_1}=\r_1$.\qed
\end{prf}

\begin{lem}\label{lem-mgs-permut}
  $\vphi_2\equiv\vphi_1$ iff $\vphi_2=\vphi_1\xi$ for some permutation
  $\xi:\tV\a\tV$.
\end{lem}

\begin{prf}
  If ``if'' part is immediate. We now prove the ``only if'' part. In
  \cite{huet76hdr}, Huet proved this result when $\leai$ is the
  equality. His proof can be adapted to our more general situation
  since $\al\leai\b$ iff $\al=\b$. By assumption and Lemma
  \ref{lem-qle-succ}, there are
  $\r_1,\r_2:\tV\a\tV\cup\tC\cup\{\infty\}$ such that
  $\vphi_1\r_1\leai\vphi_2$ and $\vphi_2\r_2\leai\vphi_1$. By
  stability, we have $\vphi_1\r_1\r_2\leai\vphi_2\r_2$. Hence, by
  transitivity, $\vphi_1\r_1\r_2\leai\vphi_1$. Similarly,
  $\vphi_2\r_2\r_1\leai\vphi_2$.
  
  We now prove that $\r_1$ is an injection from $V_1$ to $V_2$, where
  $V_i=\bigcup\{\Var(\b\vphi_i)\mid\b\in V\}$ and
  $V=\dom(\vphi_1)\cup\dom(\vphi_2)$. Let $\al\in V_1$. Then, there is
  $\b\in V$ such that $\al\in\Var(\b\vphi_1)$. Hence,
  $\b\vphi_1=\ts^k\al$ for some $k\in\bN$. Since
  $\vphi_1\r_1\r_2\leai\vphi_1$, we have
  $\b\vphi_1\r_1\r_2=\ts^k\al\r_1\r_2\leai\b\vphi_1=\ts^k\al$. Therefore,
  $\al\r_1\r_2=\al$ and $\r_1$ is an injection on $V_1$. We now prove that
  $\g=\al\r_1\in V_2$. Since $\vphi_1\r_1\leai\vphi_2$, we have
  $\b\vphi_1\r_1=\ts^k\g\leai\b\vphi_2$. We now prove (*) for all
  $\d\in\tV$, if $\d\vphi_1\neq\infty$, then
  $\d\vphi_2\neq\infty$. Indeed, if $\d\vphi_2=\infty$ then, since
  $\vphi_2\r_2\leai\vphi_1$, we have
  $\d\vphi_2\r_2=\infty\leai\d\vphi_1$ which is not possible since
  $\d\vphi_1\neq\infty$. Applying (*) with $\d=\b$, we get
  $\b\vphi_2=\ts^{k+l}\g$ for some $l$, and $\g\in V_2$.

  Similarly, $\r_2$ is an injection from $V_2$ to $V_1$. So, by Lemma
  \ref{lem-inj-ext}, there is a permutation $\xi:V\a V$ with
  $\xi|_{V_1}=\r_1$. We now prove that, for all $\al$,
  $\al\vphi_1\xi=\al\vphi_2$. If $\al\notin V$, this is
  immediate. Otherwise, we proceed by case on $\al\vphi_1$:
\begin{itemize}
\item $\al\vphi_1=\infty$. Since $\al\vphi_1\r_1\leai\al\vphi_2$, we
  have $\al\vphi_1\xi=\al\vphi_2=\infty$.
\item $\al\vphi_1=\ts^k\b$. Then, $\b\in V_1$ and
  $\al\vphi_1\xi=\ts^k\b\r_1$. Since $\vphi_1\r_1\leai\vphi_2$, we
  have $\ts^k\b\r_1\leai\al\vphi_2$. By (*), we have
  $\al\vphi_2\neq\infty$ since $\al\vphi_1\neq\infty$. So,
  $\al\vphi_2=\ts^{k+l}\b\r_1$ for some $l$. Since
  $\vphi_2\r_2\leai\vphi_1$, we have
  $\al\vphi_2\r_2=\ts^{k+l}\b\r_1\r_2\leai\ts^k\b$. Thus, $l=0$ and
  $\al\vphi_1\xi=\al\vphi_2$.
\item $\al\vphi_1=\ts^k\tc$. Since $\vphi_1\r_1\leai\vphi_2$, we have
  $\ts^k\tc\leai\al\vphi_2$. By (*), we have $\al\vphi_2\neq\infty$
  since $\al\vphi_1\neq\infty$. Hence, $\al\vphi_2=\ts^{k+l}\tc$ for
  some $l$. Since $\vphi_2\r_2\leai\vphi_1$, we have
  $\ts^{k+l}\tc\leai\ts^k\tc$. Therefore, $l=0$ and
  $\al\vphi_1\xi=\al\vphi_2$.\qed
\end{itemize}
\end{prf}

\subsection{Satisfiability}

To check whether a problem is satisfiable, we are going to introduce a
terminating rewrite system that will put the problem into some
normal form whose satisfiability is easy to establish. To do so, we
first need to extend the successor algebra as follows:

\begin{dfn}[Successor-iterator algebra]
  Let $\tB$ be the following multi-sorted algebra:
\begin{itemize}
\item Sorts: $\tA$ interpreted by $\kh$, and $\tN$ interpreted by
  $\w$.
\item Function symbols: $\mt{0}:\tN$ interpreted by $0$,
  $\ts_\tN:\tN\a\tN$ and $\ts:\tA\a\tA$ interpreted by the successor
  function, $\tc:\tA$ for every $\tc\in\tC$, $\ts:\tN\times\tA\a\tA$,
  with $\ts(a,b)$ written $\ts^ab$, interpreted as the iteration of
  the successor function: $(\ts^ab)\mu=b\mu+a\mu$.
\item Variables: the variables $\al,\b,\ldots\in\tV$ are of sort
  $\tA$. In addition, we assume given a set $\tV_\tN$, disjoint from
  $\tV\cup\tC$, of variables $x,y,\ldots$ of sort $\tN$, and an
  injection $\rx:\tV\a\tV_\tN$.
\item ${<_\tB}={\lta}$.
\item ${\le_\tB}={{\lta}\cup{\simeq_\tA}}$ where $\simeq_\tA$ is the
  smallest congruence satisfying the following semantically valid
  equations on terms of sort $\tA$:
\begin{rewc}[~~\simeq_\tA~~]
\ts^{\mt{0}}\al & \al\\
\ts^{\ts_\tN x}\al & \ts(\ts^x\al)\\
\ts^x(\ts\al) & \ts(\ts^x\al)\\
\end{rewc}
\end{itemize}
In the top-extension of $\tB$, $\o\tB$, the symbol $\infty$ is of sort
$\tA$. Let $\Var_s(a)$ be the variables of sort $s$ occurring in
$a$. A problem is {\em constant-free} if it contains no constants
$\tc\in\tC$.
\end{dfn}

Note that, in a multi-sorted algebra, substitutions map a variable of
sort $s$ to a term of sort $s$ (hence a substitution cannot map a
variable of sort $\tN$ to $\infty$), and constraints are pairs of
terms of the same sort. A problem is of sort $s$ if all its
constraints are of sort $s$.

One can easily check that, when oriented from left to right, the
equations defining $\simeq_\tA$ form a confluent and terminating
rewrite system. Hence, every term has a unique normal form and two
equivalent terms have the same normal form. So, wlog, we can always
assume that terms are in normal form, in which case $\simeq_\tA$ is
the equality and $\le_\tB$ is $\lea$.

In the following, we use the letters $e$ and $f$ ($k$ and $l$ resp.)
to denote arbitrary (closed resp.) terms of sort $\tN$. Closed terms
of sort $\tN$ are isomorphic to natural numbers. Hence, we identify
$\ts_\tN\ldots\ts_\tN\mt{0}$ ($k$ times $\ts_\tN$) with $k$, denote
$\ts_\tN\ldots\ts_\tN x$ ($k$ times $\ts_\tN$) by $x+k$, and call a
problem of sort $\tN$ an {\em integer problem}. However, $\ts^k\al$
will not denote $\ts^{\ts_\tN\ldots\ts_N\mt{0}}\al$ ($k$ times
$\ts_\tN$) but its normal form $\ts\ldots\ts\al$ ($k$ times $\ts$).

Given a problem $P$ in $\o\tA$, since $\Sol_{\o\tB}(P)$ may contain
solutions not expressible in $\o\tA$, we consider the following subset
instead:

\begin{dfn}[$\tN$-closed solutions]
  \label{def-B-sol}
  A term $a\in\o\tB$ is $\tN$-closed if $\Var_\tN(a)=\vide$. A
  solution to a problem $P$ is $\tN$-closed if it maps every
  $\al\in\Var(P)$ to an $\tN$-closed term. Let $\Sol_{\o\tB}^\vide(P)$
  ($\Sol_\tB^\vide(P)$ resp.) be the set of all the $\tN$-closed
  (finite resp.) solutions of $P$.
\end{dfn}

\begin{lem}\hfill
  \begin{itemize}
  \item A term of sort $\tA$ belongs to $\o\tA$ iff it is
    $\tN$-closed.
  \item Given a problem $P$ in $\o\tA$, $\Sol_{\o\tA}(P)=\Sol_{\o\tB}^\vide(P)$.
  \end{itemize}
\end{lem}

\begin{prf}
  \begin{itemize}
  \item This is immediate if $a=\infty$. Otherwise,
    $a=\ts\ldots\ts\ts^{x_1}\ldots\ts^{x_n}b$ with
    $b\in\tV\cup\tC$. If $a\in\tA$, then $n=0$ and $a$ is
    $\tN$-closed. Conversely, if $a$ is $\tN$-closed, then $n=0$ and
    $a\in\tA$.
  \item Immediate consequence of the previous property.\qed
  \end{itemize}
\end{prf}

Note that a $\tN$-closed solution maps every variable of sort $\tN$ to
an integer. Hence, for an integer problem $P$,
$\Sol_{\o\tB}^\vide(P)=\Sol_\tB^\vide(P)$ (solutions to integer
problems are always finite) and, given
$\vphi,\psi\in\Sol_\tB^\vide(P)$, $\vphi\qle\psi$ iff
$\vphi\le_\bN\psi$, \ie for all $x\in\Var(P)$, $x\vphi\le_\bN x\psi$.

Now, to a problem in $\o\tB$, we associate a graph as follows:

\begin{dfn}[Graph associated to a problem in $\o\tB$]\label{def-graph}
  Let $\tH=\tV\cup\tV_\tN\cup\tC\cup\{\mt{0}\}$. To a problem $P$ in
  $\o\tB$, we associate a directed graph $G(P)$ on $\tH\cup\{\infty\}$
  with the following labeled edges:
  \begin{itemize}
  \item $x\stackrel{k-l}\longrightarrow y$ for each constraint
    $x+k\le^?y+l\in P$ with $x,y\in\tV_\tN\cup\{\mt{0}\}$;
  \item $\mt{0}\stackrel{0}\longrightarrow y$ for each variable
    $y\in\Var_\tN(P)$;
  \item $\al\stackrel{k-l}\longrightarrow\b$ for each constraint
    $\ts^k\al\le^?\ts^l\b\in P$;
  \item $\al\stackrel\infty\longrightarrow\infty$ for each $\al\in\Var_\tA(P)$;
  \item $\infty\stackrel{0}\longrightarrow\b$ for each constraint
    $\infty\le^?\ts^l\b\in P$;
  \item $\tc\stackrel{0}\longrightarrow\b$ for each constraint
    $\ts^e\tc\le^?\ts^l\b\in P$.
  \end{itemize}
  The {\em weight} of a path $a_1\stackrel{k_1}\longrightarrow\ldots
  \stackrel{k_n}\longrightarrow a_{n+1}$ is $\S_{i=1}^nk_i$, where
  $k+\infty=\infty$. A {\em cycle} (\ie when $a_{n+1}=a_1$) is {\em
    positive} if its weight is $>0$.

  Let $\le_P$ be the smallest quasi-ordering on $\tH$ (we exclude
  $\infty$) such that $a\le_P b$ iff there is a path from $a$ to $b$
  in $G(P)$.

  A triple $(\al,\tc,\td)$ such that $\tc\le_P\al$, $\td\le_P\al$ and
  $\tc\neq\td$, is called {\em incompatible}.
\end{dfn}

For instance, the graph of the problem
$P=\{\tc\le^?\al,\ts\al\le^?\b,\b\le^?\al\}$ is:

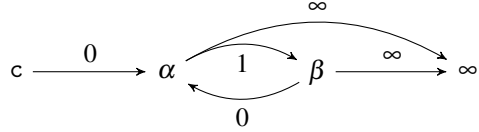
\begin{center}
\begin{tikzpicture}[->,auto,node distance=2cm,>=stealth',shorten >=1pt]
  \node (c) {$\tc$};
  \node (alpha) [right of=c] {$\al$};
  \node (beta) [right of=alpha] {$\b$};
  \node (infty) [right of=beta] {$\infty$};

  \path
  (c) edge node {0} (alpha)
  (alpha) edge [bend left] node [below] {1} (beta)
          edge [bend left] node {$\infty$} (infty)
  (beta) edge [bend left] node [below] {0} (alpha)
         edge node {$\infty$} (infty);
\end{tikzpicture}
\end{center}

If we replace $\al$ by $\rx_\al\in\tV_\tN$, $\b$ by $\rx_\b\in\tV_\tN$
and $\tc$ by $\mt{0}$, we get the integer problem
$I(P)=\{0\le^?\rx_\al,\rx_\al+1\le^?\rx_\b,\rx_\b\le^?\rx_\al\}$ whose
graph is:

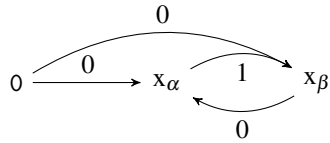
\begin{center}
\begin{tikzpicture}[->,auto,node distance=2cm,>=stealth',shorten >=1pt]
  \node (c) {$\mt{0}$};
  \node (alpha) [right of=c] {$\rx_\al$};
  \node (beta) [right of=alpha] {$\rx_\b$};

  \path
  (c) edge node {0} (alpha) edge [bend left] node {0} (beta)
  (alpha) edge [bend left] node [below] {1} (beta)
  (beta) edge [bend left] node [below] {0} (alpha);
\end{tikzpicture}
\end{center}

Following Pratt \cite{pratt77}, an integer problem $P$ has an integer
solution iff $G(P)$ has no positive cycles, which can be decided in
polynomial time ``e.g., by forming the max/+ transitive closure of the
graph and searching for a self-edge with a positive label''.

In the graph of $I(P)$, the cycle
$\rx_\al\stackrel{1}\a\rx_\b\stackrel{0}\a\rx_\al$ has weight $1$ and
thus is positive. So, $I(P)$ has no integer solution. On the other
hand, $P$ can be solved by taking $\al=\b=\infty$.

Next, we introduce a data structure used to transform an arbitrary
problem into a problem in normal form using the rules of Figure
\ref{fig-rules1}:

\begin{dfn}[Configuration]
  A term is admissible if it contains at most one variable.
  A constraint $a\le^?b$ is admissible if both $a$ and $b$ are admissible.
  
  A configuration $C$ is $\bot$ or a tuple $(C_0,C_1,C_2,C_3,C_4)$ with:
  \begin{itemize}
  \item $C_0\sle\tV$,
  \item $C_1\sle\tV$,
  \item $C_2$ is a finite map from $\tV$ to $\tC$,
  \item $C_3$ is a set of admissible constraints of sort $\tN$,
  \item $C_4$ is a set of admissible constraints of sort $\tA$,
  \item $C_0$, $C_1$, $\dom(C_2)$ and $\Var(C_4)$ are pairwise disjoint,
  \item $\Var_\tN(C_3)=\{\rx_\al\mid\al\in\dom(C_2)\}$,
  \item $\Var_\tN(C_4)\sle\{\rx_\al\mid\al\in\dom(C_2)\}$.
  \end{itemize}
\noindent
Let $\Sol_{\o\tB}^\vide(C)=\Sol_{\o\tB}^\vide(\pi(C))$ and
$\Var(C)=\Var(\pi(C))$, where $\pi(\bot)=\bot$ and
$\pi(C_0,\ldots,C_4)$ is the union of:
  \begin{itemize}
  \item $\{\al\le^?\infty\mid\al\in C_0\}$,
  \item $\{\infty\le^?\al\mid\al\in C_1\}$,
  \item $\{\al\le^?\ts^{\rx_\al}\tc\mid(\al,\tc)\in C_2\}\cup\{\ts^{\rx_\al}\tc\le^?\al\mid(\al,\tc)\in C_2\}$,
  \item $C_3\cup C_4$.
  \end{itemize}
  \noindent
  $C$ is {\em normal} if there is no $D$ such that $C\leadsto D$ where
  $\leadsto$ is defined in Figure \ref{fig-rules1}.

  Finally, given $C$ and $\psi$, let:
  \begin{itemize}
  \item $\s_0(C,\psi)=\{(\al,\al\psi)\mid\al\in C_0\}$,
  \item $\s_1(C)=\{(\al,\infty)\mid\al\in C_1\}$,
  \item $\s_2(C,\psi)=\{(\al,\ts^{\rx_\al\psi}\tc)\mid(\al,\tc)\in C_2\}$,
  \item $\s_{3,4}(C,\psi)=\{(\al,\al\psi)\mid\al\in\Var(C_3\cup C_4)\}$,
  \item $\s_{4\tA}(C,\psi)=\{(\al,\al\psi)\mid\al\in\Var_\tA(C_4)\}$.
  \end{itemize}
\end{dfn}

$C_0$ records the variables with no constraints, $C_1$ records the
variables that must be set of $\infty$, $C_2$ records the variables
that must be set to a value of the form $\ts^k\tc$, $C_3$ contains the
constraints on integer variables, and $C_4$ contains all the other
constraints.

Note that Figure \ref{fig-rules1} describes an infinite set of rules
since $a$ stands for an arbitrary size expression of sort $\tA$, $e$
and $f$ for arbitrary size expressions of sort $\tN$, $k$ for an
arbitrary natural number, $\al$ for an arbitrary size variable of sort
$\tA$, $\tc$ and $\td$ for arbitrary constants, and $P\uplus Q$ for an
arbitrary set with two disjoint parts, $P$ and $Q$.

\begin{itemize}
\item[$(\_\infty)$] removes the constraints of the form $a\le^?\infty$
  that are always satisfied, and records in $C_0$ variables not
  occurring elsewhere.
\item[$(\infty\al_1)$] detects variables that must be set to $\infty$
  because they belong to a positive cycle.
\item[$(\infty\al_2)$] detects variables $\al$ that must be set to
  $\infty$ because some constraints imply that it should otherwise be
  set to a term of the form $\ts^k\tc$ and some other constraints that
  it should be set to a term of the form $\ts^l\td$ with $\tc\neq\td$.
\item[$(\infty\tc)$] detects an unsatisfiable constraint of the form
  $\infty\le^?\ts^e\tc$.
\item[$(\tc\td)$] detects an unsatisfiable constraint of the form
  $\ts^e\tc\le^?\ts^f\td$ with $\tc\neq\td$.
\item[$(\tc\tc)$] replaces a constraint of the form
  $\ts^e\tc\le^?\ts^f\tc$ by the integer constraint $e\le^?f$.
\item[$(\al\tc)$] replaces a constraint of the form
  $\ts^k\al\le^?\ts^e\tc$ by recording in $C_2$ that $\al$ must be set
  to a term of the form $\ts^{\rx_\al}\tc$, propagating it in other
  constraints, and recording in integer constraints the fact that
  $\rx_\al+k\le^?e$.
\end{itemize}

The rule $(\infty\al_2)$ is not necessary for deciding the
satisfiability of a problem. It is included here because it is useful
to compute a most general solution in next section.

\begin{figure}
  \figrule
\caption{Rules for computing the normal form of a problem\label{fig-rules1}}
\normalsize
\begin{center}
  $\begin{array}{@{}l@{~}r@{~~}c@{~~}l@{}}
    (\_\infty)& C_0,C_1,C_2,C_3,C_4\uplus\{a\le^?\infty\}
    &\leadsto& C_0\cup(\Var_\tA(a)-\Var(C_4)),C_1,C_2,C_3,C_4\\
    (\infty\al_1)& C_0,C_1,C_2,C_3,C_4\uplus Q
      &\leadsto& C_0,C_1\!\cup\Var(Q),C_2,\\
      &&& \quad C_3,C_4\{(\al,\infty)\mid\al\in\Var(Q)\}\\
      &\mbox{if}&Q& \mbox{is constant-free and $G(Q)$ is a positive cycle}\\
    (\infty\al_2)& C_0,C_1,C_2,C_3,C_4
    &\leadsto& C_0,C_1\cup\{\al\},C_2,C_3,C_4\{(\al,\infty)\}\\
    &&&\mbox{if}~\tc\le_{C_4}\al,\td\le_{C_4}\al,\tc\neq\td\\    
    (\infty\tc)& C_0,C_1,C_2,C_3,C_4\uplus\{\infty\le^?\ts^e\tc\} &\leadsto& \bot\\
    (\tc\td)& C_0,C_1,C_2,C_3,C_4\uplus\{\ts^e\tc\le^?\ts^f\td\}
      &\leadsto& \bot~~\mbox{if $\tc\neq\td$}\\
    (\tc\tc)& C_0,C_1,C_2,C_3,C_4\uplus\{\ts^e\tc\le^?\ts^f\tc\}
    &\leadsto& C_0,C_1,C_2,C_3\cup\{e\le^?f\},C_4\\
    (\al\tc)& C_0,C_1,C_2,C_3,C_4\uplus\{\ts^k\al\le^?\ts^e\tc\}
      &\leadsto& C_0,C_1,C_2\cup\{(\al,\tc)\},\\
       &&& \quad C_3\cup\{\rx_\al+k\le^?e\},C_4\{(\al,\ts^{\rx_\al}\tc)\}\\
    



    

    

\end{array}$
\end{center}
\figrule
\end{figure}

\pagebreak
\begin{lem}\label{lem-rules1-prop}\hfill
  \begin{enumerate}
  \item
    $\Sol_{\o\tB}^\vide(C)\!=\!{\{\s_0(C,\vphi)\cup\s_1(C)\cup\s_2(C,\psi)\cup\s_{3,4}(C,\psi)\!\mid\!\vphi\,\mbox{$\tN$-closed},\psi\in\Sol_{\o\tB}^\vide(C_3\cup C_4)\}}$.
  \item For all problems $P$ in $\o\tA$,
    $(\vide,\vide,\vide,\vide,P)$ is a configuration and
    $\Sol_{\o\tA}(P)=\Sol_{\o\tB}^\vide(\vide,\vide,\vide,\vide,P)$.
  \item In a configuration, every term of sort $\tA$ is of the form
    $\infty$, $\ts^\al$ or $\ts^e\tc$.
  \item If $C$ is a configuration and $C\leadsto D$, then:
    \begin{enumerate}
    \item $D$ is a configuration.
    \item If $D\neq\bot$, then $\Var(C)\sle\Var(D)$.
    \item Correctness: if $\vphi\in\Sol_{\o\tB}^\vide(D)$, then
      $\vphi|_{\Var(C)}\in\Sol_{\o\tB}^\vide(C)$.
    \item Completeness: if $\psi\in\Sol_{\o\tB}^\vide(C)$, then
      $\psi=\vphi|_{\Var(C)}$ for some $\vphi\in\Sol_{\o\tB}^\vide(D)$.
    \end{enumerate}
  \item The relation $\leadsto$ terminates.
  \item If $(\vide,\vide,\vide,\vide,P)\leadsto^*C\neq\bot$, then
    $\Var(C)=\Var(P)\cup\Var(C_3)$.
  \end{enumerate}
\end{lem}

\begin{prf}
  \begin{enumerate}
  \item Let
    $S(C)\!=\!{\{\s_0(C,\vphi)\!\cup\!\s_1(C)\!\cup\!\s_2(C,\psi)\!\cup\!\s_{3,4}(C,\psi)\!\mid\!\vphi\,\mbox{$\tN$-closed},\psi\in\Sol_{\o\tB}^\vide(C_3\!\cup
    C_4)\}}$.  One can easily check that
    $S(C)\sle\Sol_{\o\tB}^\vide(C)$. Assume now that
    $\vphi\in\Sol_{\o\tB}^\vide(C)$. Then,
    $\vphi=\s_0(C,\vphi)\cup\s_1(C)\cup\s_2(C,\psi)\cup\s_{3,4}(C,\psi)$
    where $\psi=\vphi|_{\Var(C_3\cup C_4)}$. Indeed, if $\al\in C_1$,
    then $\infty\le^?\al\in\pi(C)$. Hence, $\al\vphi=\infty$. Now,
    if$(\al,\tc)\in C_2$, then $\pi(C)$ contains
    $\al\le^?\ts^{\rx_\al}\tc$ and $\ts^{\rx_\al}\tc\le^?\al$. Hence,
    $\al\vphi=\ts^{\rx_\al\vphi}\tc$ and $\rx_\al\vphi=\rx_\al\psi$
    since $\{\rx_\al\mid\al\in\dom(C_2)\}\sle\Var_\tN(C_3)$.

  \item One can easily check that $(\vide,\vide,\vide,\vide,P)$ is a
    configuration. The fact that
    $\Sol_{\o\tA}(P)=\Sol_{\o\tB}^\vide(\vide,\vide,\vide,\vide,P)$
    directly follows from the previous property.

  \item Straightforward.
    
  \item
    \begin{enumerate}
    \item One can easily check that all the conditions defining what
      is a configuration are preserved by each rule. In particular,
      $(\al\tc)$ replaces $\al$ by $\ts^{\rx_\al}\tc$, hence every
      term of $D$ is admissible if every term of $C$ so is.
    \item Straightforward.
    \item Straightforward.
    \item We only detail the following cases:
      \begin{itemize}
      \item Rule ($\al\tc$). We have
        $(\ts^k\al)\psi\leai(\ts^e\tc)\psi=\ts^{e\psi}\tc$. So,
        $\al\psi\neq\infty$ and
        $(\ts^k\al)\psi=\ts^k(\al\psi)\lea\ts^{e\psi}\tc$. By Lemma
        \ref{lem-lea}, there is $l$ such that
        $\ts^{e\psi}\tc=\ts^l\ts^k(\al\psi)$. Hence, there is $m$ such
        that $\al\psi=\ts^m\tc$ and $e\psi=l+k+m$. Let now
        $\vphi=\psi\cup\{(\rx_\al,m)\}$. We have
        $\al\vphi=\al\psi=\ts^m\tc=(\ts^{\rx_\al}\tc)\vphi$ and
        $(\rx_\al+k)\vphi=m+k\le l+k+m=e\psi=e\vphi$. Therefore,
        $\vphi\in\Sol_{\o\tB}^\vide(D)$ and $\vphi|_{\Var(C)}=\psi$.
  
      \item Rule ($\infty\al_1$). We first prove that, if
        $\al_1\stackrel{k_1}\longrightarrow\ldots
        \stackrel{k_n}\longrightarrow\al_{n+1}$ is a path in $G(Q)$,
        $\psi\in\Sol_{\o\tB}^\vide(Q)$ and $k=\S_{i=1}^nk_i\ge 0$
        ($k<0$ resp.), then $\ts^k\al_1\psi\leai\al_{n+1}\psi$
        ($\al_1\psi\leai\ts^{-k}\al_{n+1}\psi$ resp.) (*), by
        induction on $n$. If $n=1$, this is immediate. We now prove it
        for $n+1$.
        \begin{itemize}
        \item Case $k\ge 0$. By induction hypothesis,
          $\ts^k\al_1\psi\leai\al_{n+1}\psi$.

          \begin{itemize}
          \item Case $k_{n+1}\ge 0$. Then,
            $\ts^{k_{n+1}}\al_{n+1}\psi\leai\al_{n+2}\psi$.
            \begin{itemize}
            \item Case $k+k_{n+1}\ge 0$. By monotony and transitivity,
              $\ts^{k+k_{n+1}}\al_1\psi\leai\al_{n+2}\psi$.
            \item Case $k+k_{n+1}<0$. Impossible.
            \end{itemize}

          \item Case $k_{n+1}<0$. Then,
            $\al_{n+1}\psi\leai\ts^{-k_{n+1}}\al_{n+2}\psi$ and, by
            transitivity, $\ts^k\al_1\psi\leai\ts^{-k_{n+1}}\al_{n+2}\psi$.
            \begin{itemize}
            \item Case $k+k_{n+1}\ge 0$. Since $-k_{n+1}\le k$,
              $\ts^{k+k_{n+1}}\al_1\psi\leai\al_{n+2}\psi$.
            \item Case $k+k_{n+1}<0$. Since $k<-k_{n+1}$,
              $\al_1\psi\leai\ts^{-k-k_{n+1}}\al_{n+2}\psi$.
            \end{itemize}
          \end{itemize}

        \item Case $k<0$. Symmetric to previous case.
        \end{itemize}

        Assume now that $Q$ is constant-free and $G(Q)$ is a positive
        cycle. If $G(Q)$ contains $\infty$, then $\al\psi=\infty$ for
        all $\al\in\Var(Q)$. Otherwise, $G(Q)$ is
        $\al_1\stackrel{k_1}\longrightarrow\ldots
        \stackrel{k_n}\longrightarrow\al_{n+1}=\al_1$. Hence,
        $\ts^k\al_1\psi\leai\al_1\psi$ with
        $k=\S_{i=1}^nk_i>0$. Therefore, $\al_1\psi=\infty$ and
        $\al\psi=\infty$ for all $\al\in\Var(Q)$.

      \item Rule ($\infty\al_2$). We first prove that (a) for any
        problem $P$, if $\b\le_P\al$ by a path of length $n$,
        $\vphi\in\Sol_{\o\tB}^\vide(P)$ and $\b\vphi=\infty$, then
        $\al\vphi=\infty$, by induction on $n$. If $n=0$, this is
        immediate. Otherwise, there is $\ts^p\b\le^?\ts^q\g\in P$ with
        $\g\le_P\al$ by a path of length $n-1$. Since
        $\vphi\in\Sol_{\o\tB}^\vide(P)$ and $\b\vphi=\infty$, we have
        $\infty\leai\ts^q\g\vphi$. Therefore $\g\vphi=\infty$ and, by
        induction hypothesis, $\al\vphi=\infty$.

        We now prove that (b) if $\b\le_P\al$ by a path of length $n$,
        $\vphi\in\Sol_{\o\tB}^\vide(P)$ and $\b\vphi=\ts^k\tc$ for
        some $k$, then either $\al\vphi=\infty$ or $\al\vphi=\ts^i\tc$
        for some $i$, by induction on $n$. If $n=0$, this is
        immediate. Otherwise, there is $\ts^p\b\le^?\ts^q\g\in P$ with
        $\g\le_P\al$ by a path of length $n-1$. Since
        $\vphi\in\Sol_{\o\tB}^\vide(P)$ and $\b\vphi=\ts^k\tc$, we
        have $\ts^{p+k}\tc\leai\ts^q\g\vphi$. If $\g\vphi=\infty$
        then, by (a), $\al\vphi=\infty$. Otherwise, $\g\vphi=\ts^l\tc$
        for some $l$ and, by induction hypothesis, either
        $\al\vphi=\infty$ or $\al\vphi=\ts^i\tc$ for some $i$.

        Hence, if $(\al,\tc,\td)$ is incompatible in $C_4$ and
        $\vphi\in\Sol_{\o\tB}^\vide(C_4)$, then $\al\vphi=\infty$.
      \end{itemize}
    \end{enumerate}
  \item Every rule decreases the number of constraints in $C_4$ except
    rule $(\infty\al_2)$. In $(\infty\al_2)$, this number is unchanged
    but the number of variables decreases. Since the number of
    variables in $C_4$ never increases, the system terminates.
  \item Straightforward.\qed
  \end{enumerate}
\end{prf}

The properties 4(c) and 4(d) give
$\Sol_{\o\tB}^\vide(C)=\{\vphi|_{\Var(C)}\mid\vphi\in\Sol_{\o\tB}^\vide(D)\}$
whenever $C\leadsto D$.

\begin{dfn}[Affine problem]
  A constraint is {\em affine} if it is of sort $\tN$, of the form
  $\ts^k\al\le^?\ts^l\b$ or of the form $\ts^e\tc\le^?\ts^l\b$. A
  problem is affine if all its constraints are affine.
\end{dfn}

\begin{lem}
  In any normal configuration $C\neq\bot$, $C_4$ is an affine problem
  with no positive cycles and no incompatible triples.
\end{lem}

\begin{prf}
  By Lemma \ref{lem-rules1-prop}, every term of sort $\tA$ occurring
  in $C$ is of the form $\infty$, $\ts^k\al$ or $\ts^e\tc$. Now, $C_4$
  cannot contain a constraint of the form:
\begin{itemize}
\item $a\le^?\infty$ because of rule $(\_\infty)$,
\item $\infty\le^?\ts^l\b$ because of rule $(\infty\al_1)$,
\item $\infty\le^?\ts^f\td$ because of rule $(\infty\tc)$,
\item $\ts^k\al\le^?\ts^f\td$ because of rule $(\al\tc)$,
\item $\ts^e\tc\le^?\ts^f\td$ because of rules $(\tc\tc)$ and $(\tc\td)$.
\end{itemize}
Therefore, a constraint in $C_4$ can only be either of the form
$\ts^k\al\le^?\ts^l\b$ or of the form
$\ts^e\tc\le^?\ts^l\b$. Moreover, $G(C_4)$ cannot have positive cycles
because of rule $(\infty\al_1)$, and cannot have incompatible triples
because of rule $(\infty\al_2)$.\qed\\
\end{prf}

Since affine problems of sort $\tA$ are always satisfiable (by setting
their variables to $\infty$), we can conclude:

\begin{thm}[Satisfiability]\label{thm-sat}
  The satisfiability of a size problem in the successor algebra is
  decidable in polynomial time wrt. the number of symbols by the
  algorithm of Figure \ref{fig-sat}.
\end{thm}

\begin{prf}
  Let $|P|$ be the number of symbols in $P$. Constructing $G(P)$
  requires at most $\sharp\Var(P)+\sharp P$ steps, where $\sharp X$ is
  the cardinal of $X$. But $\Var(P)\le 2\sharp P$ since there are at
  most 2 variables per constraint, and $2\sharp P\le|P|$ since every
  constraint is of size 2 at least. Therefore, constructing $G(P)$
  requires at most $3|P|/2$ steps.

  Whether there is a positive cycle in a graph is decidable in
  polynomial time \cite{pratt77}. Whether there is an incompatible
  triple in a graph can be done in polynomial time too. Hence, whether
  a rule can be applied is decidable in polynomial time. Now, since
  $\leadsto$ terminates, the algorithm describes a computable
  function.

  We now prove that it is correct and complete. If $C=\bot$ then, by
  completeness, $P$ is unsatisfiable. Otherwise,
  $C=(C_0,C_1,C_2,C_3,C_4)$. If $G(C_3)$ has a positive cycle then, by
  completeness, $P$ is unsatisfiable. Otherwise, let
  $\vphi_3\in\Sol_{\o\tB}^\vide(C_3)$. Then, one can easily check that
  $\vphi=\vphi_3\cup\{(\al,\infty)\mid\al\in\Var(C_1)\cup\Var(C_4)\}\cup\{(\al,\ts^{\rx_\al\vphi_3}\tc)\mid(\al,\tc)\in
  C_2\}\in\Sol_{\o\tB}^\vide(C)$. Therefore, by correctness,
  $\vphi|_{\Var(P)}\in\Sol_{\o\tB}(P)=\Sol_{\o\tA}(P)$ and $P$ is
  satisfiable.

  Finally, to prove that the complexity for computing $C$ is
  polynomial, it suffices to show that the number of rewrite steps and
  the size $|C|=|\pi(C)|$ of intermediate configurations $C$ are
  polynomially bounded by $|P|$.

  By definition of $\leadsto$,
  $\Var(C)\sle\Var(P)\cup\{\rx_\al\mid\al\in\Var(P)\}$ and
  $\sharp\Var(C)\le 2\sharp\Var(P)\le 2|P|$. So, after the termination
  proof, the number of rewrite steps is $\le\sharp P\times
  2|P|\le|P|^2$.
  
  Let $\|C\|_\infty$ be the maximum size of a constraint in
  $\pi(C)$. No rule but $(\al\tc)$ can make $\|C\|_\infty$
  increase. $\|C\|_\infty$ can be increased by at most $2$ for each
  replacement of a variable $\al$ by $\ts^{\rx_\al}\tc$. However,
  there cannot be more than two such replacements in a constraint
  since, after two such replacements, there is no variable of sort
  $\tA$ anymore. Therefore,
  $\|C\|_\infty\le\|P\|_\infty+4\le|P|+4$. Now, $\sharp\pi(C)\le
  5\sharp P\le 5|P|/2$ since $\sharp C_0+\sharp C_1+\sharp C_2\le
  2\sharp\Var(P)\le 4\sharp P$ and $\sharp C_3+\sharp C_4\le\sharp
  P$. Therefore, $|C|\le\|C\|_\infty\times\sharp C\le(|P|+4)\times
  5|P|/2$.\qed\\
\end{prf}

\begin{figure}
\figrule
\caption{Algorithm for deciding the satisfiability of a problem $P$ in the successor algebra.\label{fig-sat}}
\normalsize
\begin{enumerate}
  \item Compute a normal form $C$ of $(\vide,\vide,\vide,\vide,P)$ wrt
    the rules of Figure \ref{fig-rules1}.
  \item If $C=\bot$, then $P$ is not satisfiable. Otherwise,
    $C=(C_0,C_1,C_2,C_3,C_4)$.
  \item If $C_3$ has a positive cycle, then $P$ is not
    satisfiable. Otherwise, $P$ is satisfiable.
\end{enumerate}
\figrule
\end{figure}


Our procedure can be related to the one described in
\cite{barthe05tlca} where, like many works on type inference, the
authors consider constrained types. But they do not bring out the
properties of the size algebra and, in particular that, in the
successor algebra, satisfiable sets of constraints have a most general
solution as we shall see in next section.

\begin{exa}\label{ex-config-bottom}
  Let $P=\{\tc\le^?\al,\ts\al\le^?\b,\b\le^?\al,\td\le^?\b\}$. We
  have $(\vide,\vide,\vide,\vide,P)$

  \noindent
  $\leadsto(\vide,\{\al\},\vide,\vide,\{\tc\le^?\infty,\infty\le^?\b,\b\le^?\infty,\td\le^?\b\})$,
  by $(\infty\al_2)$ since $\tc\le_P\al$ and $\td\le_P\al$;

  \noindent
  $\leadsto(\vide,\{\al,\b\},\vide,\vide,\{\tc\le^?\infty,\infty\le^?\td\})$, by $(\infty\al_1)$ since $\infty\stackrel{0}\a\b\stackrel\infty\a\infty$ is positive;

  \noindent
  $\leadsto\bot$, by $(\infty\tc)$.\qed
\end{exa}

\begin{exa}\label{ex-config-1}
  Let $P=\{\al\le^?\ts\tc,\b\le^?\al\}$. We
  have $(\vide,\vide,\vide,\vide,P)$

  \noindent
  $\leadsto(\vide,\vide,\{(\al,\tc)\},\{\rx_\al\le^?1\},\{\b\le^?\ts^{\rx_\al}\tc\})$,
  by $(\al\tc)$;

  \noindent
  $\leadsto(\vide,\vide,\{(\al,\tc),(\b,\tc)\},\{\rx_\al\le^?1,\rx_\b\le^?\rx_\al\},\vide)$,
  by $(\al\tc)$ again.  This is a normal form and the graph of
  $\{\rx_\al\le^?1,\rx_\b\le^?\rx_\al\}$ has no positive cycle, so it
  is satisfiable (the solutions for $(\rx_\al,\rx_\b)$ are $(0,0)$,
  $(1,0)$ and $(1,1)$).\qed
\end{exa}

\subsection{Computing the most general solution}
\label{sec-mgs}

We now turn to the problem of whether, in the successor algebra
$\o\tA$, a satisfiable problem $P$ has a most general solution and, if
so, how to compute it.

Let $\mgs_{\o\tA}(P)$ ($\mgs_\tA(P)$ resp.) be the set of most general
(finite resp.) solutions of $P$, and $\mgs_{\o\tB}^\vide(C)$
($\mgs_\tB^\vide(C)$ resp.) be the set of most general (finite resp.)
$\tN$-closed solutions of $C$.

We first prove a refinement of Lemma \ref{lem-qle-succ} to
$\tN$-closed solutions of a configuration:

\begin{lem}\label{lem-qle-succ-iter-closed}
  Given $\vphi,\psi\in\Sol_{\o\tB}^\vide(C)$, $\vphi\qle\psi$ iff
  there is $\r:\tV\a\tV\cup\tC\cup\{\infty\}$ such that
  $\dom(\r)\sle\Var_\tA(C)$ and, for all $\al\in
  C_0\cup\Var(C_3)\cup\Var_\tA(C_4)$, $\al\vphi\r\leai\al\psi$.
\end{lem}

\begin{prf}
  \begin{itemize}
  \item[$\A$] By Lemma \ref{lem-qle-succ}, there is
    $\r:\tV\cup\tV_\tN\a\tV\cup\tV_\tN\cup\tC\cup\{\mt{0},\infty\}$
    such that $\vphi\r\leai\psi$. Since $\vphi$ and $\psi$ are
    $\tN$-closed, we also have $\vphi(\r|_\tV)\leai\psi$. Indeed, if
    $\al\in\tV_\tN$, then
    $\al\vphi(\r|_\tV)=\al\vphi=\al\vphi\r\leai\al\psi$. Let now
    $\al\notin\Var(C)$. Then,
    $\al(\r|_\tV)=\al\vphi(\r|_\tV)\leai\al\psi=\al$. Therefore,
    $\al(\r|_\tV)=\al$ and $\dom(\r|_\tV)\sle\Var_\tA(C)$.
  \item[$\lA$] After Lemma \ref{lem-qle-succ}, it is enough to prove
    that, for all $\al\in\tV\cup\tV_\tN$, $\al\vphi\r\leai\al\psi$.
    By assumption, the property holds if $\al\in
    C_0\cup\Var(C_3)\cup\Var_\tA(C_4)$. If $\al\in C_1$, then
    $\al\vphi=\infty=\al\psi$ and $\al\vphi\r\leai\al\psi$. If
    $(\al,\tc)\in C_2$, then $\al\vphi=\ts^{\rx_\al\vphi}\tc$,
    $\al\psi=\ts^{\rx_\al\psi}\tc$. Since $\rx_\al\in\Var(C_3)$ and
    $\vphi$ is $\tN$-closed, we have
    $\rx_\al\vphi=\rx_\al\vphi\r\leai\rx_\al\psi$. Therefore,
    $\al\vphi\r\leai\al\psi$. Since $\Var_\tN(C_4)\sle\Var(C_3)$, we
    are left with the case where $\al\notin\Var(C)$. But, in this
    case, $\al\vphi=\al\psi=\al\r=\al$ since $\dom(\vphi)$,
    $\dom(\psi)$ and $\dom(\r)$ are all included in $\Var(C)$.\qed
  \end{itemize}
\end{prf}

We now prove that the most general solutions of a problem $P$ in
$\o\tA$ can be obtained from the most general $\tN$-closed solutions
of the normal form of $(\vide,\vide,\vide,\vide,P)$.

\begin{lem}\label{lem-sol-config}
  Assume that $(\vide,\vide,\vide,\vide,P)\leadsto^*C$.
  \begin{itemize}
  \item\label{lem-sol-config-correct}
    Correctness: if $\vphi\in\mgs_{\o\tB}^\vide(C)$, then
    $\vphi|_{\Var(P)}\in\mgs_{\o\tA}(P)$.
  \item\label{lem-sol-config-complete}
    Completeness: if $\psi\in\mgs_{\o\tA}(P)$, then there is
    $\vphi\in\mgs_{\o\tB}^\vide(C)$ such that $\vphi|_{\Var(P)}=\psi$.
  \end{itemize}
\end{lem}

\begin{prf}
   Note that $\Var(C)=\Var(P)\cup\Var(C_3)$.
  \begin{itemize}
  \item Let $\vphi\in\mgs_{\o\tB}^\vide(C)$. By correctness of
    $\leadsto$, $\vphi|_{\Var(P)}\in\Sol_{\o\tA}(P)$. Let now
    $\psi\in\Sol_{\o\tA}(P)$. By completeness of $\leadsto$, there is
    $\vphi'\in\Sol_{\o\tB}^\vide(C)$ such that
    $\psi=\vphi'|_{\Var(P)}$. Since $\vphi=\mgs(C)$,
    $\vphi\qle\vphi'$. By Lemma \ref{lem-qle-succ-iter-closed},
    $\vphi\r\leai\vphi'$ for some $\r:\tV\a\tV\cup\tC\cup\{\infty\}$
    such that $\dom(\r)\sle\Var_\tA(C)=\Var(P)$. Therefore, for all
    $\al\in\tV\cup\tV_\tN$,
    $\al\vphi|_{\Var(P)}\r\leai\al\vphi'|_{\Var(P)}$ and
    $\vphi|_{\Var(P)}\qle\vphi'|_{\Var(P)}=\psi$.
  \item Let $\psi\in\mgs_{\o\tA}(P)$. By completeness of $\leadsto$,
    there is $\vphi\in\Sol_{\o\tB}^\vide(C)$ such that
    $\psi=\vphi|_{\Var(P)}$. Assume now that there is
    $\vphi'\in\Sol_{\o\tB}^\vide(C)$ such that
    $\vphi\not\qle\vphi'$. By correctness of $\leadsto$,
    $\vphi'|_{\Var(P)}\in\Sol_{\o\tA}(P)$. Since $\psi=\mgs(P)$,
    $\psi=\vphi|_{\Var(P)}\qle\vphi'|_{\Var(P)}$, that is, there is
    $\r$ such that, $\vphi|_{\Var(P)}\r\leai\vphi'|_{\Var(P)}$. Since
    $\vphi\not\qle\vphi'$, there is $x$ such that
    $x\vphi\r\not\leai x\vphi'$. Since
    $\vphi|_{\Var(P)}\r\leai\vphi'|_{\Var(P)}$, $x=\rx_\b$ for some
    $\b\in\Var(P)$. By definition of $\Sol_{\o\tB}^\vide(C)$, there is
    $\tc$ such that $\b\vphi=\ts^{x\vphi}\tc$ and
    $\b\vphi'=\ts^{x\vphi'}\tc$. Hence,
    $\ts^{x\vphi}\tc\leai\ts^{x\vphi'}\tc$ and
    $x\vphi\not\leai x\vphi'$. Contradiction.\qed
  \end{itemize}
\end{prf}

We now prove that the most general $\tN$-closed solutions of
$(C_0,C_1,C_2,C_3,C_4)$ can be obtained from the most general
$\tN$-closed solutions of $C_3\cup C_4$.

\begin{lem}\label{lem-sol34}
  Let $C=(C_0,C_1,C_2,C_3,C_4)$ be a configuration.
  \begin{itemize}
  \item
    Correctness: if $\psi\in\mgs_{\o\tB}^\vide(C_3\cup C_4)$ and, for
    all $\al\in\Var_\tA(C_4)$, $\Var(\al\psi)\cap
    C_0=\vide$,\footnote{Thanks to Lemma \ref{lem-mgs-permut}, this
      condition can always be satisfied by applying some permutation
      to $\psi$.} then
    $\s_1(C)\cup\s_2(C,\psi)\cup\s_{3,4}(C,\psi)\in\mgs_{\o\tB}^\vide(C)$.
  \item
    Completeness: if $\vphi\in\mgs_{\o\tB}^\vide(C)$, then
    $\vphi|_{\Var(C_3\cup C_4)}\in\mgs_{\o\tB}^\vide(C_3\cup C_4)$.
  \end{itemize}
\end{lem}

\begin{prf}
  \begin{itemize}    
  \item Let $\psi'=\s_1(C)\cup\s_2(C,\psi)\cup\s_{3,4}(C,\psi)$ and
    $\vphi\in\Sol_{\o\tB}^\vide(C)$. We have
    $\vphi_{3,4}=\vphi|_{\Var(C_3\cup
    C_4)}\linebreak\in\Sol_{\o\tB}^\vide(C_3\cup C_4)$. Since
    $\psi\in\mgs_{\o\tB}^\vide(C_3\cup C_4)$,
    $\psi\qle\vphi_{3,4}$. By applying Lemma
    \ref{lem-qle-succ-iter-closed} on $(\vide,\vide,\vide,C_3,C_4)$,
    there is $\r:\tV\a\tV\cup\tC\cup\{\infty\}$ such that
    $\dom(\r)\sle\Var_\tA(C_4)$ and, for all
    $\al\in\Var(C_3)\cup\Var_\tA(C_4)$,
    $\al\psi\r\leai\al\vphi_{3,4}$. Then, let
    $\r'=\{(\al,\al\vphi)\mid\al\in
    C_0\}\cup\{(\al,\al\r)\mid\al\in\Var_\tA(C_4)\}$. We prove that
    $\psi'\qle\vphi$ by using Lemma \ref{lem-qle-succ-iter-closed}. We
    have $\dom(\r')\sle\Var_\tA(C)$ by definition. If $\al\in C_0$,
    then $\al\psi'\r'=\al\r'=\al\vphi$ by definition. If
    $\al\in\Var(C_3)$, then $\al\psi'\r'=\al\psi\r'=\al\psi\r$ because
    $\psi$ is $\tN$-closed, and
    $\al\psi\r\leai\al\vphi_{3,4}=\al\vphi$. If $\al\in\Var_\tA(C_4)$,
    then $\al\psi'\r'=\al\psi\r'=\al\psi\r$ since $\Var(\al\psi)\cap
    C_0=\vide$ by assumption, and
    $\al\psi\r\leai\al\vphi_{3,4}=\al\vphi$.

  \item We first check that $\vphi|_{C_0}$ maps variables to variables
    and is injective. Let $\al\in C_0$ and
    $\vphi'=\vphi|_{\Var(C)-\{\al\}}$. Then,
    $\vphi'\in\Sol_{\o\tB}(C)$ too since, by definition of
    configuration, $\al\notin\Var(C_i)$ for every $i>0$. Hence,
    $\vphi\qle\vphi'$, that is, there is $\r$ such that
    $\al\vphi\r\leai\al\vphi'=\al$. Therefore, $\al\vphi$ is a
    variable $\g$. Assume now that $\g=\b\vphi$ for some $\b\in
    C_0$. Then,
    $\vphi''=\vphi|_{\Var(C)-\{\al,\b\}}\in\Sol_{\o\tB}(C)$
    too. Hence, $\vphi\qle\vphi'$, that is, there is $\r'$ such that
    $\g\r'\leai\al\vphi'=\al$ and $\g\r'\leai\b\vphi'=\b$. Therefore,
    $\al=\g\r'=\b$.

    So, by taking in Lemma \ref{lem-inj-ext} $V=\tV$, $V_1=C_0$,
    $V_2=\vphi(C_0)$, $\r_1=\{(\al,\al\vphi)\mid\al\in C_0\}$ and
    $\r_2=\{(\al\vphi,\al)\mid\al\in C_0\}$ (the inverse of $\rho_1$),
    there is a permutation $\xi:\tV\a\tV$ such that
    $\xi|_{C_0}=\vphi|_{C_0}$. By Lemma \ref{lem-mgs-permut},
    $\vphi\xi^{-1}$ is a mgs of $C$ too. So, wlog, we can assume that
    $\vphi|_{C_0}$ is the identity.
    
    We now prove that $\vphi_{3,4}=\vphi|_{\Var(C_3\cup
      C_4)}\in\mgs_{\o\tB}^\vide(C_3\cup C_4)$. Let
    $\psi\in\Sol_{\o\tB}(C_3\cup C_4)$. By Lemma \ref{lem-rules1-prop}
    (1),
    $\psi'=\s_1(C)\cup\s_2(C,\psi)\cup\s_{3,4}(C,\psi)\in\Sol_{\o\tB}^\vide(C)$. Hence,
    $\vphi\qle\psi'$. By Lemma \ref{lem-qle-succ-iter-closed}, there
    is $\r:\tV\a\tV\cup\tC\cup\{\infty\}$ such that
    $\dom(\r)\sle\Var_\tA(C_4)$ and, for all $\al\in
    C_0\cup\Var(C_3)\cup\Var_\tA(C_4)$, $\al\vphi\r\leai\al\psi'$. For
    all $\al\in\Var(C_3)\cup\Var_\tA(C_4)$,
    $\al\vphi_{3,4}\r=\al\vphi\r\leai\al\psi'=\al\psi$. Therefore, by
    Lemma \ref{lem-qle-succ-iter-closed}, $\vphi_{3,4}\qle\psi$.\qed\\
  \end{itemize}
\end{prf}

Next, we prove that, for all affine problems $P$ with no incompatible
triples (like $C_3\cup C_4$ in a normal configuration $C$), the set of
finite $\tN$-closed solutions of $P$ is in bijection with the set of
finite $\tN$-closed solutions of:

\begin{dfn}[Integer problem associated to an affine problem]
  Given an affine problem $P$, let $I(P)$ be the integer problem
  obtained by replacing in $P$ every constraint $\ts^k\al\le^?\ts^l\b$
  by $\rx_\al+k\le^?\rx_\b+l$, and every constraint
  $\ts^e\tc\le^?\ts^l\b$ by $e\le^?\rx_\b+l$.
\end{dfn}

\begin{lem}\label{lem-sol-affine}
  If $P$ is an affine problem with no incompatible triples, then:
\begin{enumerate}
\item\label{lem-ip-p} there is a strictly monotone map $\psi\to\acute\psi$
  from $(\Sol_\tB^\vide(I(P)),\qle)$ to ${(\Sol_\tB^\vide(P),\qle)}$;
\item\label{lem-p-ip} there is a monotone map $\vphi\to\grave\vphi$
  from $(\Sol_\tB^\vide(P),\qle)$ to $(\Sol_\tB^\vide(I(P)),\qle)$;
\item\label{lem-ip-p-ip} for all $\psi\in\Sol_\tB^\vide(I(P))$,
  $\check\psi=\psi$;
\item\label{lem-p-ip-p} for all $\vphi\in\Sol_\tB^\vide(P)$, there is
  $\r:\tV\a\tV\cup\tC$ such that $\vphi=\h\vphi\r$, hence
  $\h\vphi\qle\vphi$;
\item\label{lem-sol-affine-correct}
  correctness: if $\psi\in\mgs_\tB^\vide(I(P))$, then
  $\acute\psi\in\mgs_\tB^\vide(P)$;
\item\label{lem-sol-affine-complete}
  completeness: if $\vphi\in\mgs_\tB^\vide(P)$, then
  $\grave\vphi\in\mgs_\tB^\vide(I(P))$.
\end{enumerate}
\end{lem}

\begin{prf}
  Let $\simeq_P$ be the symmetric and transitive closure of $\le_P$
  and $\eta:{{\tH}/{\simeq_P}}\a\tH$ be any function such that, for
  all equivalence classes $X$, $\eta(X)\in X$ ($\tH$ and $\le_P$ are
  introduced in Definition \ref {def-graph}). Such a function always
  exists because equivalence classes are non-empty. Because $P$ has no
  incompatible triples, an equivalence class modulo $\simeq_P$ cannot
  contain two different constants. Hence, we can assume that
  $\eta(X)=\tc$ iff $\tc\in X$.

  Given $\psi\in\Sol_\tB^\vide(I(P))$, let $\acute\psi=\{(x,x\psi)\mid
  x\in\Var_\tN(P)\}\cup\{(\al,\ts^{\rx_\al\psi}\al^*)\mid\al\in\Var_\tA(P)\}$
  where $\al^*=\eta([\al]_P)$ and $[\al]_P$ is the equivalence class
  of $\al$ modulo $\simeq_P$.

  Given $\vphi\in\Sol_\tB^\vide(P)$, let $\grave\vphi=\{(x,x\vphi)\mid
  x\in\Var_\tN(P)\}\cup\{(\rx_\al,\al\vphi_s)\mid\al\in\Var_\tA(P)\}$
  ($\vphi_s$ is introduced in Definition \ref{def-succ-head-subs}).
  
\begin{enumerate}
\item We first check that $\acute\psi\in\Sol_\tB^\vide(P)$ whenever
  $\psi\in\Sol_\tB^\vide(P)$, that is, $\acute\psi$ satisfies every
  constraint of $P$. This is immediate for constraints of sort
  $\tN$. Otherwise, since $P$ is affine, there are two cases. If
  $\ts^k\al\le^?\ts^l\b\in P$, then $\al^*=\b^*$ and
  $\rx_\al\psi+k\lea\rx_\b\psi+l$. Hence,
  $(\ts^k\al)\acute\psi=\ts^{k+\al\psi}\al^*\lea\ts^{l+\b\psi}\b^*=(\ts^l\al)\acute\psi$. If
  $\ts^e\tc\le^?\ts^l\b\in P$, then $\b^*=\tc$ and
  $e\psi\lea\rx_\b\psi+l$. Therefore,
  $(\ts^e\tc)\acute\psi=\ts^{e\psi}\tc\lea\ts^{\b\psi+l}\b^*=(\ts^l\b)\acute\psi$.

  Next, one can easily check that $\psi\to\acute\psi$ is injective
  ($\psi_1=\psi_2$ whenever $\acute{\psi_1}=\acute{\psi_2}$) and monotone wrt.
  $\lea$ ($\acute\vphi\lea\acute\psi$ whenever $\vphi\lea\psi$) and thus wrt.
  $\qle$. Therefore, $\psi\to\acute\psi$ is strictly monotone wrt. $\qle$.

\item We first check that $\grave\vphi\in\Sol_\tB^\vide(I(P))$
  whenever $\vphi\in\Sol_\tB^\vide(P)$. If $\ts^k\al\le^?\ts^l\b\in
  P$, then
  $(\ts^k\al)\vphi=\ts^{\al\vphi_s+k}\al\vphi_h\lea(\ts^l\b)\vphi=\ts^{\b\vphi_s+l}\b\vphi_h$. So,
  $\al\vphi_h=\b\vphi_h$ and
  $\rx_\al\grave\vphi+k\lea\rx_\b\grave\vphi+l$. Assume now that
  $\ts^e\tc\le^?\ts^l\b\in P$. Then,
  $(\ts^e\tx)\vphi=\ts^{e\vphi_s}\tc\lea(\ts^l\b)\vphi=\ts^{\b\vphi_s+l}\b\vphi_h$. So,
  $\tc=\b\vphi_h$ and $e\grave\vphi\lea\rx_\b\grave\vphi+l$.

  We now check that $\vphi\to\grave\vphi$ is monotone. Let
  $\vphi_1,\phi_2\in\Sol_\tB^\vide(P)$ such that
  $\vphi_1\qle\vphi_2$. Hence, there is $\r:\tV\a\tC\cup\tV$ such that
  $\vphi\r\lea\psi$. Therefore, $\grave\vphi\lea\grave\psi$.

\item Immediate.
  
\item Let $\r$ the map from $\tV$ to $\tV\cup\tC$ such that, if
  $\al^*\in\tV$, then $\al^*\r=\al\vphi_h$. The map $\r$ is well
  defined since $\vphi_h$ is invariant by $\simeq_P$: if
  $\al\simeq_P\b$, then $\al\vphi_h=\b\vphi_h$. Now, one can easily
  check that $\vphi=\h\vphi\r$. If $\al^*=\tc$, then there is a
  constraint $\ts^k\tc\le^?\ts^l\al\in P$. Since
  $\vphi\in\Sol_\tB^\vide(P)$, $\al\vphi_h=\tc$ and
  $\al\h\vphi\r=\ts^{\al\vphi_s}\al^*\r=\al\vphi$. Otherwise,
  $\al\h\vphi\r=\ts^{\al\vphi_s}\al^*\r=\ts^{\al\vphi_s}\al\vphi_h=\al\vphi$.
  
\item Let $\vphi\in\Sol_\tB^\vide(P)$. By \ref{lem-p-ip},
  $\grave\vphi\in\Sol_\tB^\vide(I(P))$ and $\psi\qle\grave\vphi$. By
  \ref{lem-ip-p}, $\acute\psi\qle\h\vphi$. By \ref{lem-p-ip-p},
  $\h\vphi\qle\vphi$. Therefore, $\acute\psi\qle\vphi$.
  
\item Let $\psi\in\Sol_\tB^\vide(I(P))$. By \ref{lem-ip-p},
  $\acute\psi\in\Sol_{\o\tB}^\vide(P)$ and $\vphi\qle\acute\psi$. By
  \ref{lem-p-ip}, $\grave\vphi\qle\check\psi$. By \ref{lem-ip-p-ip},
  $\check\psi=\psi$. Therefore, $\grave\vphi\qle\psi$.\qed
\end{enumerate}
\end{prf}

\begin{lem}\label{lem-int-smallest}
  Every satisfiable integer problem has a smallest $\tN$-closed
  solution that can be computed in polynomial time.
\end{lem}

\newcommand\Z{\o\bZ_{\max}}

\begin{prf}
  Let $P$ be a satisfiable integer problem whose variables are
  $x_1,\ldots,x_n$. We first prove that $P$ is equivalent to a
  problem in the dioid $(\Z^{n\times n},\oplus,\otimes)$ where
  $\Z=\bZ\cup\{\pm\infty\}$, $\oplus=\max$ and $\otimes=+$
  both applied component wise \cite{baccelli92book}.

  Wlog. we can assume that $P$ contains no constraints of the form
  $\mt{0}+k\le\mt{0}$ (since $P$ is satisfiable, these constraints are
  always satisfied and thus can be removed). Hence, $P$ contains only
  constraints of the form $x_i+k\le^?x_j$, $\mt{0}+k\le^?x_j$ or
  $x_i+k\le^?\mt{0}$, that is, in the syntax of $(\Z,\oplus,\otimes)$,
  $k\otimes x_i\le x_j$, $k\le x_j$ or $x_i\le-k$.

  Given a problem $P$, let $a_{ij}=\sup\{k\in\Z\mid x_j+k\le^?x_i\in
  P\}$, $b_{ij}=\sup(\{0\}\cup\{k\in\Z\mid\mt{0}+k\le^?x_i\in P\})$
  (we add $0$ because solutions must be non-negative) and
  $c_{ij}=\inf\{-k\in\Z|x_i+k\le^?\mt{0}\in P\}$ with, as usual,
  $\sup\vide=-\infty$ and $\inf\vide=+\infty$. Note that, in $b$ and
  $c$, every column is the same ($b_{ij}$ and $c_{ij}$ do not depend
  on $j$).
  
  We now prove that, if $\psi\in\Sol_{\o\tB}^\vide(P)$, then there is
  $x\in\Z^{n\times n}$ such that $(a\otimes x)\oplus b\le x\le c$ and,
  for all $j$, $x_{ij}=x_i\psi$ (the columns of $x$ are equal). For
  all $i$ and $j$, the set of inequations $\{k\otimes x_i\le x_j\mid
  x_i+k\le^?x_j\in P\}$ is equivalent to $a_{ji}\otimes x_i\le x_j$
  since $k\le a_{ji}$ and $(-\infty)\otimes x_i=-\infty\le
  x_j$. Hence, $\{a_{ji}\otimes x_i\le x_j\mid i\in\{1,\ldots,n\}\}$
  is equivalent to $\bigoplus_{i=1}^na_{ji}\otimes x_i\le x_j$. By
  taking $x_{il}=x_i$ for all $l$, we therefore get $(a\otimes
  x)_{jl}\le x_{jl}$. Similarly, for all $j$ and $l$, $\{k\le
  x_j\mid\mt{0}+k\le^?x_j\in P\}\cup\{0\le x_j\}$ (implicit in $P$) is
  equivalent to $b_{jl}\le x_{jl}$. Therefore, $(a\otimes x)\oplus
  b\le x$. Finally, for all $i$, $\{x_i\le-k\mid x_i+k\le^?\mt{0}\in
  P\}$ is equivalent to $x_{il}\le c_{il}$ for all $l$, that is, $x\le
  c$.

  Because we proceeded by equivalence, we also have the converse: if
  $(a\otimes x)\oplus b\le x\le c$, then
  $\psi_l\in\Sol_{\o\tB}^\vide(P)$ where $\psi_l$ is the substitution
  such that $x_i\psi_l=x_{il}$ ($l$-th column of $x$).
  
  By Theorem 4.75 in \cite{baccelli92book}, $(a\otimes x)\oplus b\le
  x$ has $a^*\otimes b$ as smallest solution, where
  $a^*=\bigoplus_{k\in\bN}a^k$, $a^{k+1}=a^k\otimes a$, and $a^0$ is
  the matrix with $0$ on the diagonal and $-\infty$ everywhere
  else. Since $P$ is satisfiable, $G(P)$ has no positive
  cycles. Hence, $a^*=\bigoplus_{k=0}^na^k$ \cite{cuninghame79book}
  (Theorem 3.20 in \cite{baccelli92book}). Therefore,
  $\psi\in\Sol_{\o\tB}^\vide(P)$ iff $a^*\otimes b\le c$, and the
  smallest solution of $P$ is the function $\psi$ such that
  $x_i\psi=\bigoplus_{k=1}^na^*_{ik}\otimes b_{k1}$, which can be
  computed in polynomial time.\qed\\
\end{prf}

\begin{figure}
\figrule
\caption{Algorithm computing a most general solution in the successor algebra.\label{fig-mgs}}
\normalsize
\begin{enumerate}
  \item Apply the algorithm of Figure \ref{fig-sat}.
  \item Compute the most general $\tN$-closed solution $\psi$ of
    $C_3\cup I(C_4)$ using Lemma \ref{lem-int-smallest}.
  \item Compute $\acute\psi$ defined in Lemma \ref{lem-sol-affine}.
  \item Return $\s_1(C)\cup\s_2(C,\acute\psi)\cup\s_{4\tA}(C,\acute\psi)$.
\end{enumerate}
\figrule
\end{figure}

Therefore, we can now conclude:

\begin{thm}
  In the successor algebra, any satisfiable size problem has a most
  general solution that can be computed in polynomial time following
  the algorithm of Figure \ref{fig-mgs}.
\end{thm}

\begin{prf}
  Correctness. By Lemma \ref{lem-int-smallest},
  $\psi\in\mgs_\tB^\vide(I(C_3\cup C_4))$. By Lemma
  \ref{lem-sol-affine} (\ref{lem-sol-affine-correct}),
  $\acute\psi\in\mgs_\tB^\vide(C_3\cup
  C_4)\sle\mgs_{\o\tB}^\vide(C_3\cup C_4)$. By Lemma \ref{lem-sol34}
  (1),
  $\vphi=\s_1(C)\cup\s_2(C,\acute\psi)\cup\s_{3,4}(C,\acute\psi)\in\mgs_{\o\tB}^\vide(C)$. By
  Lemma \ref{lem-sol-config} (1),
  $\vphi|_{\Var(P)}=\s_1(C)\cup\s_2(C,\acute\psi)\cup\s_{4\tA}(C,\acute\psi)\in\mgs_{\o\tA}(P)$.

  Complexity. After Theorem \ref{thm-sat}, $C_3\cup C_4$ is of
  polynomial size wrt. the size of $P$. The computation of $I(C_3\cup
  C_4)$ is linear. After Lemma \ref{lem-int-smallest}, the computation
  of $\psi$ is polynomial. After Lemma \ref{lem-sol-affine}
  (\ref{lem-ip-p}), the computation of $\acute\psi$ is
  polynomial. Therefore, the algorithm of Figure \ref{fig-mgs} is
  polynomial.\qed
\end{prf}

\begin{exa}
  In Example \ref{ex-config-1}, we have seen that the normal form of $(\vide,\vide,\vide,\vide,P)$ where $P=\{\al\le^?\ts\tc,\b\le^?\al\}$ is $(\vide,\vide,\{(\al,\tc),(\b,\tc)\},C_3,\vide)$ with
  $C_3=\{\rx_\al\le^?1,\rx_\b\le^?\rx_\al\}$. Following Lemma
  \ref{lem-int-smallest}, by taking $x_1=\rx_\al$ and $x_2=\rx_\b$,
  the corresponding max-linear system is $(a\otimes x)\oplus b\le x\le
  c$ where $a_{11}=\sup\{k\mid\rx_\al+k\le^?\rx_\al\in
  C_3\}=\sup\vide=-\infty$, $a_{12}=\sup\{\rx_\b+k\le^?\rx_\al\in
  C_3\}=\sup\{0\}=0$, $a_{21}=\sup\{k\mid\rx_\al+k\le^?\rx_\b\in
  C_3\}=\sup\vide=-\infty$, $a_{22}=\sup\{k\mid\rx_\b+k\le^?\rx_\b\in
  C_3\}=\sup\vide=-\infty$, $b_1=\sup(\{0\}\cup\{k\mid
  k\le^?\rx_\al\in C_3\})=\sup\{0\}=0$, $b_2=\sup(\{0\}\cup\{k\mid
  k\le^?\rx_\al\in C_3\})=\sup\{0\}=0$,
  $c_1=\inf\{k\mid\rx_\al\le^?k\in C_3\}=\inf\{1\}=1$ and
  $c_2=\inf\{k\mid\rx_\b\le^?k\in C_3\}=\inf\vide=+\infty$. To
  summarize, we have:\linebreak
  $a=\left(\begin{array}{cc}-\infty&0\\-\infty&-\infty\end{array}\right)$,
    $b=\left(\begin{array}{c}0\\0\end{array}\right)$ and
      $c=\left(\begin{array}{c}1\\+\infty\end{array}\right)$.
         One can easily check that, if $x=\left(\begin{array}{c}\rx_\al\\\rx_\b\end{array}\right)$, then $(a\otimes x)\oplus b=\left(\begin{array}{c}\rx_\b\oplus 0\\0\end{array}\right)$, hence that $(a\otimes x)\oplus b\le x\le c$ is equivalent to $\rx_\b\oplus 0\le\rx_\al\le 1$ and $0\le\rx_\b\le+\infty$, which is $C_3$. Now, $a^0=\left(\begin{array}{cc}0&-\infty\\-\infty&0\end{array}\right)$ and $a^2=\left(\begin{array}{cc}-\infty&-\infty\\-\infty&-\infty\end{array}\right)$. Hence, $a^*=a^0\oplus a=\left(\begin{array}{cc}0&0\\-\infty&0\end{array}\right)$ and $a^*\otimes b=\left(\begin{array}{c}0\\0\end{array}\right)$. So, the smallest solution of $C_3$ is $\acute\psi=\{(\rx_\al,0),(\rx_\b,0)\}$ and the smallest solution of $P$ is $\s_1(C)\cup\s_2(C,\acute\psi)\cup\s_{4\tA}(C,\acute\psi)=\{(\al,\tc),(\b,\tc)\}$.\qed
\end{exa}

\section{Conclusion}

We have presented a general and modular termination criterion for the
combination of $\b$-reduction and user-defined rewrite rules, based on
the use of type-checking with size-annotated types approximating a
semantic notion of size defined by the annotations given to
constructor symbols. This extends to rewriting-based function
definitions and more general notions of size, an approach initiated by
Hughes, Pareto and Sabry for function definitions based on a fixpoint
combinator and case analysis \cite{hughes96popl}.

First, we have shown that these termination conditions can be reduced
to solving problems in the quasi-ordered algebra used for size
annotations. Then, we have shown that the successor algebra (successor
symbol with arbitrary constants) enjoys nice properties: decidability
of the satisfiability of sets of inequalities (in polynomial time),
and existence and computability of a most general solution for
satisfiable problems (in polynomial time too). As a consequence, we
have a complete algorithm for checking the termination conditions in
the successor algebra.

We have implemented a simple heuristic that turns this termination
criterion into a fully automated termination prover for higher-order
rewriting called HOT \cite{hot}, which tries to detect size-preserving
functions and, following \cite{abel02jfp}, to find a lexicographic
ordering on arguments. Combined with other (non-)termination
techniques \cite{jouannaud91lics,blanqui00rta,blanqui02tcs}, HOT won
the 2012 international competition of termination provers \cite{tc}
for higher-order rewriting against THOR \cite{thor} and WANDA
\cite{wanda}. It could be improved by replacing the lexicographic
ordering by the size-change principle \cite{lee01popl,hyvernat14lmcs},
and using abstract interpretation techniques for annotating function
symbols \cite{telford99bctcs,chin01hosc}. A more complete (and perhaps
more efficient) implementation would be obtained by encoding
constraints into a SAT problem and send it to state-of-art SAT solvers
\cite{fuhs07sat,benamram08tacas,codish11jar}.

A natural following is to study other size algebras like the
max-successor algebra (\ie the successor algebra extended with a
$\mt{max}$ operator), the plus algebra (\ie the successor algebra
extended with addition) or their combination, the max-plus
algebra. Indeed, the richer the size algebra is, the more precise the
typing of function symbols is, and the more functions can be proved
terminating.

Following \cite{blanqui06lpar-sbt}, it is also possible to consider
full Presburger arithmetic \cite{presburger29} and handle conditional
rewrite rules, by extending the system with explicit quantifiers and
constraints on size variables, in the spirit of HM(X)
\cite{sulzmann01flops}. Simplification of constraints is then an
important issue in practice \cite{pottier01ic}.

We have presented this criterion in Church' simply typed $\l$-terms
but, following \cite{blanqui05mscs}, it should be possible to extend
it to richer type systems with polymorphic and dependent
types. Similarly, we considered matching modulo $\al$-congruence only
but, following \cite{blanqui16tcs}, it should be possible to extend it
to rewriting modulo some equational theory and to rewriting on
$\b$-normal forms with matching modulo $\b\eta$ as used in Klop's
combinatory reduction systems \cite{klop93tcs} or Nipkow's
higher-order rewrite systems \cite{mayr98tcs}.

Another interesting extension would be to consider size-annotated
types in the computability path ordering \cite{blanqui15lmcs},
following Kamin and L\'evy's extension of Dershowitz' recursive path
ordering \cite{dershowitz79focs,kamin80note}, and Borralleras and
Rubio's extension of Jouannaud and Okada's higher-order recursive path
ordering \cite{jouannaud99lics,borralleras01lpar}.

\medskip

{\bf Acknowledgments.} I would like to thank Christophe Raffalli for a
short but useful discussion on max-plus algebra, and Nachum
Dershowitz, Jean-Pierre Jouannaud and Sylvain Schmitz for their
comments on the introduction and the conclusion. I also want to thank
very much the anonymous referees for their very careful reading and
the numerous remarks and suggestions they made. This greatly helped me
to improve the article.

\small
\renewcommand\ss\latexss
\bibliographystyle{jfp}

\end{document}